\UseRawInputEncoding

\documentclass{article}

\usepackage{amsmath}
\usepackage{graphicx}
\usepackage{color}
\usepackage{multirow}
\usepackage{algorithm}
\usepackage{algorithmic}
\usepackage{enumitem}
\usepackage{authblk}
\usepackage{xcolor}
\usepackage{lineno}
\usepackage{dsfont}

\usepackage[version=4]{mhchem}
\usepackage{siunitx}

\title{Self-configuring feedback loops for sensorimotor control}

\author{Sergio Verduzco-Flores}
\author{Erik De Schutter}
\affil{Computational Neuroscience Unit, OIST}

\date{}

\begin{document}

\maketitle

\begin{abstract}
    How dynamic interactions between nervous system regions in mammals performs 
    online motor control remains an unsolved problem. 
    In this paper we show that feedback control is a simple, yet
    powerful way to understand the neural dynamics of sensorimotor control.
    We make our case 
    using a minimal model comprising spinal cord, sensory and motor 
    cortex, coupled by long connections that are plastic. It succeeds in 
    learning how to perform reaching movements of a planar arm with 6 muscles
    in several directions from scratch. The model satisfies biological 
    plausibility constraints, like neural implementation, transmission delays,
    local synaptic learning and continuous online learning.  
    Using differential Hebbian plasticity the model can go 
    from motor babbling to reaching arbitrary targets in less than 10 minutes
    of {\it in silico} time. Moreover, independently of the learning mechanism,
    properly configured feedback control has many emergent properties: 
    neural populations in motor cortex show directional tuning and oscillatory 
    dynamics, the spinal cord creates convergent force fields that add linearly,
    and movements are ataxic (as in a motor system without a cerebellum).
\end{abstract}


\section{Introduction}

\subsection{The challenge}
Neuroscience has made great progress in decoding how cortical regions perform specific brain functions like primate vision 
\cite{kaas_primate_2003,ballard_hierarchical_2021} and rodent navigation 
\cite{chersi_cognitive_2015,moser_spatial_2017}.
Conversely, the evolutionary much older motor control system still poses
fundamental questions, despite a large body of experimental work. 
This is because, in mammals, in addition to areas in cortex like premotor and motor areas and to some degree sensory and parietal ones, many {\it extracortical regions} have important and unique functions: basal ganglia, thalamus, cerebellum, pons, brain stem nuclei like the red nucleus and spinal cord \cite{eccles_physiology_1981, loeb_major_2015}. 
These structures are highly interconnected by fast conducting axons and all 
show strong dynamic activity changes, related to the ongoing dynamics of the 
performed motor act. Clinical and lesion studies have confirmed the necessity 
of each of these regions for normal smooth motor control of arm reaching 
\cite{shadmehr_computational_2005,arber_connecting_2018}.

Fully understanding motor control will thus entail understanding
the simultaneous function and interplay of all brain regions involved.
Little by little, new experimental techniques will allow us to monitor
more neurons, in more regions, and for longer periods 
\cite[e.g.]{tanaka_thalamocortical_2018}. But to make sense
of these data computational models must step up to the task of integrating all those
regions to create a functional neuronal machine.

Finally, relatively little is known about the neural basis of motor
development {\it in infants} \cite{hadders-algra_early_2018}. Nevertheless, a full understanding of primate motor control will not only require explanation of how these brain regions complement and interact with each other but also how this can be learned during childhood.

With these challenges in mind we recently developed a motor control framework based
on differential Hebbian learning \cite{verduzco-flores_differential_2022}. 
A common theme in physiology is the control of {\it homeostatic variables} (e.g. blood
glucose levels, body temperature, etc.) using negative feedback mechanisms 
\cite{woods_homeostasis_2007}.
From a broad perspective, our approach considers the musculoskeletal system as
an extension of this homeostatic control system: movement aims to make the
external environment conducive to the internal control of homeostatic variables
(e.g. by finding food, or shelter from the sun).

Our working hypothesis (see
\cite{verduzco-flores_differential_2022})
is that control of homeostatic variables requires a feedback controller that
uses the muscles to produce a desired set of sensory perceptions. The
motosensory loop, minimally containing motor cortex, spinal cord, and sensory
cortex may implement that feedback controller. To test this hypothesis we
implemented a relatively complete model of the sensorimotor loop (figure
\ref{fig:architecture}), using the learning rules in 
\cite{verduzco-flores_differential_2022} to produce 2D arm reaching. 
The activity of the neural populations and the movements they produced showed
remarkable consistency with the experimental observations that we describe next.

\subsection{Relevant findings in motor control} \label{sub:relevant}
Before describing our modeling approach, we summarize some of the relevant experimental data that will be important to understanding the results. We focus on three related issues: 1) the role of the spinal cord in movement, 2) the nature of representations in motor cortex, and 3) muscle synergies, and how the right pattern of muscle activity is produced.

For animals to move, spinal motoneurons must activate the skeletal muscles.
In general, descending signals from the corticospinal tract do not activate the 
motoneurons directly, but instead provide input to a network of excitatory and 
inhibitory interneurons \cite{bizzi_new_2000, lemon_descending_2008, 
arber_motor_2012, asante_differential_2013, alstermark_circuits_2012,
jankowska_spinal_2013, wang_deconstruction_2017, ueno_corticospinal_2018}.
Learning even simple behaviors involves long-term plasticity, both at the spinal
cord (SC) circuit, and at higher regions of the motor hierarchy 
\cite{wolpaw_adaptive_1983, grau_learning_2014, meyer-lohmann_dominance_1986,wolpaw_complex_1997,norton_acquisition_2018}.
Despite its obvious importance, there are comparatively few attempts to elucidate the nature of the SC computations, and the role of synaptic plasticity.

The role ascribed to SC is closely related to the role assumed from motor cortex,
particularly M1. One classic result is that M1 pyramidal neurons of macaques activate preferentially when the hand is moving in a particular direction. When the preferred directions of a large population of neurons are added as vectors, a population vector appears, which points close to the hand's direction of motion 
\cite{georgopoulos_relations_1982,georgopoulos_neuronal_1986}.
This launched the hypothesis that M1 represents kinematic, or other high-level
parameters of the movement, which are transformed into movements in concert with the SC.
This hypothesis mainly competes with the view that M1 represents muscle forces. Much
research has been devoted to this issue 
\cite[e.g.]{kakei_muscle_1999,truccolo_primary_2008,kalaska_intention_2009,georgopoulos_local_2007,harrison_towards_2013,
tanaka_modeling_2016,morrow_prediction_2003,todorov_direct_2000}.
 
Another important observation is that the preferred directions of motor neurons cluster
around one main axis. As shown in \cite{scott_dissociation_2001}, this suggests that
M1 is mainly concerned with 
dynamical aspects of the movement, rather than representing  its kinematics.

A related observation is that the preferred directions in M1 neurons experience
random drifts that overlap learned changes  
\cite{rokni_motor_2007,padoa-schioppa_neuronal_2004}.
This leads to the hypothesis that M1 is a redundant network that is constantly using feedback error signals to capture the {\it task-relevant dimensions}, placing the configuration of synaptic weights in an {\it optimal manifold}.

A different perspective for studying motor cortex is to focus on how it can
produce movements, rather than describing its activity 
\cite{shenoy_cortical_2013}.
One specific proposal is that motor cortex has a collection of pattern generators,
and specific movements can be created by combining their activity
\cite{shenoy_cortical_2013,sussillo_neural_2015}.
Experimental support for this hypothesis came through the surprising finding of
rotational dynamics in motor cortex activity \cite{churchland_neural_2012},
suggesting that oscillators with 
different frequencies are used to produce desired patterns.
This begs the question of how the animal chooses its desired patterns
of motion.

Selecting a given pattern of muscle activation requires {\it planning}.
Motor units are the final actuators in the motor system, but they number in the tens 
of thousands, so planning movements in this space is unfeasible. 
A low-dimensional representation of desired limb configurations 
(such as the location of the hand in Euclidean coordinates) is better.
Movement generation likely involves a coordinate transformation,  from the 
endpoint coordinates (e.g. hand coordinates) into actuator coordinates (e.g. 
muscle lengths), from which motor unit activation follows directly. Even 
using pure engineering methods, as for robot control, computing this 
coordinate transformation is very challenging. 
For example, this must overcome kinematic redundancies, as when many 
configurations of muscle lengths put the hand in the same location.

The issue of coordinate transformation is central for motor control 
\cite{shadmehr_computational_2005,schoner_reaching_2018,valero-cuevas_mathematical_2009};
{\it motor primitives and muscle synergies} are key concepts in this discussion.
Representing things as combinations of elementary components is a fundamental 
theme in applied mathematics. For example, linear combinations of basis vectors
can represent any vector, and linear combinations of wavelets can approximate any
smooth function \cite{keener_principles_1995}. In motor control, this idea arises
in the form of motor primitives. Motor primitives constitute a set of basic
motions, such that that any movement can be decomposed into them 
\cite{giszter_motor_2015,mussa-ivaldi_motor_2000,bizzi_computations_1991}.
This is closely related to the concept of synergies. The term ``synergy'' may mean
several things \cite{kelso_synergies_2009, bruton_synergies_2018},
but in this paper we use it to denote a pattern of muscle activity arising as a coherent unit. Synergies may be composed of motor primitives, or they may be the motor primitives themselves.

A promising candidate for motor primitives comes in the form of convergent force
fields, which have been observed for the hindlimbs of 
frogs and rats
\cite{giszter_convergent_1993,mussa-ivaldi_linear_1994},
or in the forelimbs of monkeys \cite{yaron_forelimb_2020}.
In experiments where the limb is held at a particular location, local stimulation
of the spinal cord will cause a force to the limb's endpoint. The collection 
of these force vectors for all of the limb endpoint's positions forms a force 
field, and these force fields have two important characteristics: 1) they 
have a unique fixed point, and 2) simultaneous stimulation of two spinal cord 
locations produces a force field which is the sum of the force fields from 
stimulating the two locations independently. It is argued that movement 
planning may be done in terms of force fields, since they can produce 
movements that are resistant to perturbations, and also permit a solution to 
the problem of coordinate transformation with redundant 
actuators 
\cite{mussa-ivaldi_motor_2000}.

The neural origin of synergies, and whether they are used by the motor system is a matter
of ongoing debate \cite{tresch_case_2009,de_rugy_are_2013,bizzi_neural_2013}.
To us, it is of interest that single spinal units found in the mouse
\cite{levine_identification_2014} and monkey \cite{takei_neural_2017}
spinal cord (sometimes called Motor Synergy Encoders, or MSEs) 
can reliably produce specific patterns of motoneuron activation.

\subsection{Model concepts}
We believe that it is impossible to understand the complex dynamical system
in biological motor control without the help of computational modeling.
Therefore we set out to build a minimal model that could eventually 
control an autonomous agent, while still satisfying biological plausibility 
constraints. 

Design principles and biological-plausibility constraints for neural network modeling have
been proposed before \cite{pulvermuller_biological_2021, oreilly_six_1998, richards_deep_2019}. 
Placing emphasis on the motor system, we compiled
 a set of characteristics that cover the majority of these constraints. Namely: 

\begin{itemize}
    \item {\bf Spanning the whole sensorimotor loop.} 
    \item {\bf Using only neural elements. 
    Learning their connection strengths is part of the model.}
    \item {\bf Learning does not rely on a training dataset. It is
    instead done by synaptic elements using local information.}
    \item {\bf Learning arises from continuous-time interaction with a
    continuous-space environment.} 
    \item {\bf There is a clear vision on how the model integrates with the rest of
    the brain in order to enact more general behavior.}
\end{itemize} 

Our aim is hierarchical control of homeostatic variables, with the spinal cord 
and motor cortex at the bottom of this hierarchy. 
At first glance, spinal plasticity poses a conundrum, 
because it changes the effect of corticospinal inputs. Cortex is playing
a piano that keeps changing its tuning. A solution comes when we consider the 
corticospinal loop (e.g. the long-loop reflex) as a negative control system, 
where the spinal cord activates the effectors to reduce an error. 
The role of cortex is to produce perceptual variables that are controllable, 
and can eventually improve homeostatic regulation. In this regard,
our model is a variation of Perceptual Control Theory 
\cite{powers_feedback:_1973, powers_behavior:_2005}, but if the desired value
of the controller is viewed as a prediction, then this approach resembles active
inference models \cite{adams_predictions_2013}. Either way, the
goal of the system is to reduce the difference between the desired and the
perceived value of some variable.

If cortex creates representations for perceptual variables, the sensorimotor
loop must be configured so those variables can be controlled. This happens when
the error in those variables activates the muscles in a way that brings the
perceived value closer to the desired value. In other words, we must find the
input-output structure of the feedback controller implicit in the long-loop
reflex. We have found that this important problem can be solved by the
differential Hebbian learning rules introduced in 
\cite{verduzco-flores_differential_2022}. We favor the hypothesis that this 
learning takes place is in the connections from motor cortex to interneurons and
brainstem. Nevertheless, we show that all our results are valid if learning
happens in the connections from sensory to motor cortex.

In the Results section we will describe our model, its variations, and how it
can learn to reach. Next we will show that many phenomena described above
are present in this model. These phenomena
emerge from having a properly configured neural feedback controller with a
sufficient degree of biological realism. This means that even if the synaptic
weights of the connections are set by hand and are static, the
phenomena still
emerge, as long as the system is configured to reduce errors. In short, we show
that a wealth of phenomena in motor control can be explained simply by feedback
control in the sensorimotor loop, and that this feedback control can be
configured in a flexible manner by the learning rules presented in
\cite{verduzco-flores_differential_2022}. 

\section{Results}

\subsection{A neural architecture for motor control} \label{sub:architecture}

\begin{figure}
    \centering
    \includegraphics[width=0.9\columnwidth]{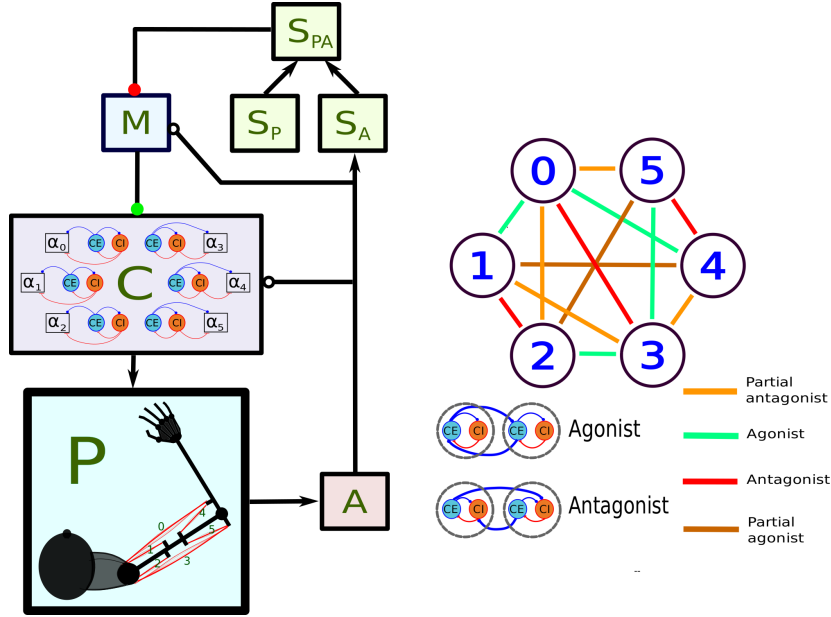}
    \caption{Main components of the model. In the left panel each box stands for a
    neural population, except for {\bf P}, which represents the arm and
    the muscles. Arrows indicate static connections, open circles
    show {\it input correlation} synapses, and the two colored circles show 
    possible locations of synapses with the learning rule in
    \cite{verduzco-flores_differential_2022}. In the {\it spinal learning} model
    the green circle connections  are plastic, and the red circle connections
    are static. In the {\it cortical learning} model the red circle connections
    are plastic, whereas the green circle connections are static. In the
    {\it static network} all connections are static.
    {\bf A}: afferent population. ${\bf S_A}$:
    Somatosensory cortex, modulated by afferent input. ${\bf S_P}$:
    somatosensory cortex, prescribed pattern. ${\bf S_{PA}}$: population
    signaling the difference between ${\bf S_P}$ and ${\bf S_A}$. ${\bf M}$:
    primary motor cortex. ${\bf C}$: spinal cord. Inside the ${\bf C}$ box
    the circles represent the excitatory (${\bf CE}$) and inhibitory 
    (${\bf CI}$) interneurons, organized into six pairs. The interneurons in
    each pair innervate an alpha motoneuron (${\bf \alpha}$), each of which
    stimulates one of the six muscles in the arm, numbered from 0 to 5. The
    trios consisting of ${\bf CE}$, ${\bf CI}$, ${\bf \alpha}$ units are
    organized into agonists and antagonists, depending on whether their 
    $\alpha$ motoneurons cause torques in similar or opposite directions.
    These relations are shown in the right-side panel. }
    \label{fig:architecture}
\end{figure}

The model in this paper contains the main elements of the long-loop reflex, applied
to the control of a planar arm using six muscles. The left panel of figure
\ref{fig:architecture} shows the architecture of the model, which contains 74
firing rate neurons organized in 
six populations. This architecture resembles a feedback controller that makes the
activity in a neural population $S_A$ approach the activity in a
different population $S_P$.

The six firing-rate neurons (called {\it units} in this paper) in $S_A$ represent 
a region of somatosensory cortex, and its inputs consist of the static
gamma (II) afferents. In steady state, activity of the II afferents is monotonically 
related to muscle length \cite{mileusnic_mathematical_2006}, which in turn 
can be used to prescribe hand location.
Other afferent signals are not provided to $S_A$ in the interest of
simplicity.

$S_P$ represents a different cortical layer of the same somatosensory region as $S_A$, 
where a ``desired'' or ``predicted'' activity has been caused by
brain regions not represented in the model. Each firing rate neuron in $S_A$ 
has a corresponding unit in $S_P$, and they represent
the mean activity at different levels of the same microcolumn 
\cite{mountcastle_columnar_1997}.
$S_{PA}$ is a region (either in sensory or motor cortex) that conveys
the difference between activities in $S_P$ and $S_A$, which is the error signal
to be minimized by negative feedback control.

Population $A$ represents sensory thalamus and dorsal parts of the spinal cord. 
It contains 18 units with logarithmic
activation functions, each receiving an input from a muscle afferent.
Each muscle provides proprioceptive
feedback from models of the Ia, Ib, and II afferents. In rough terms,
Ia afferents provide information about contraction velocity, and Ib
afferents signal the amount of tension in the muscle and tendons.

Population $M$ represents motor cortex. Ascending inputs to $M$ arise from
population $A$, and use a variation of the {\it input  correlation} learning 
rule \cite{porr_strongly_2006}, where the $S_{PA}$ inputs act as a learning signal. 
The input correlation rule enhances the stability of the controller.
More details are presented in Methods.
The $S_{PA}$ inputs to $M$ can either be static, or use a learning rule
to be described below.

To represent positive and negative values, both $M$ and 
$S_{PA}$ use a ``dual representation'', where each error signal is represented by
two units. Let $e_i = s_P^i - s_A^i$ be the error associated with the $i$-th muscle. 
One of the two
$S_{PA}$ units representing $e_i$ is a monotonic function of $\text{max}(e_i, 0)$, whereas the
other unit increases according to $\text{max}(-e_i, 0)$. These opposing inputs, along with mutual
inhibition between the two units creates dynamics where sensorimotor events cause both excitatory
and inhibitory responses, which agrees with experimental observations 
\cite{shafi_variability_2007,steinmetz_distributed_2019,najafi_excitatory_2020},
and allows transmitting ``negative'' values using excitatory
projections.
Dual units in $M$ receive the same inputs, but with the opposite sign.

Plasticity mechanisms within the sensorimotor loop should specify which muscles
contract in order to reduce an error signaled by $S_{PA}$. We suggest that this
plasticity could take place in the spinal cord and/or motor cortex. To show that
our learning mechanisms work regardless of where the learning takes place, we
created two main configurations of the model. In the first configuration, called the
``spinal learning'' model, a ``spinal'' network $C$ 
transforms the $M$ outputs into muscle stimulation. $C$ learns to transform
sensory errors into appropriate motor commands using a differential Hebbian
learning rule \cite{verduzco-flores_differential_2022}. In this configuration
the error input to each $M$ unit comes from one of the $S_{PA}$ activities.
A second configuration, called the ``cortical learning'' model, has
``all-to-all'' connections from $S_{PA}$ to $M$ using the differential Hebbian
rule, whereas the connections from $M$ to $C$ use appropriately patterned static
connections. Both configurations are basically the same model; the difference is
that one configuration has our learning rule on the inputs to $C$, whereas the
other has it on the inputs to $M$ (figure \ref{fig:architecture}).

While analyzing our model we reproduced several experimental
phenomena (described below). Interestingly, these phenomena did not arise 
because of the learning rules. To make this explicit we created a third
configuration of our model, called the ``static network''. This configuration
does not change the weight of any synaptic connection during the simulation. The
initial weights were hand-set to approximate the optimal solution
everywhere (see Methods). We will show that all emergent phenomena in the paper
are also present in the static network.

We explain the idea behind the differential Hebbian rule as
applied in the connections from $M$ to $C$. $C$ contains $N$ interneurons,
whose activity vector we denote as $\mathbf{c} = [c_1, \dots, c_N ]$. The input
to each of these units is an $M$-dimensional vector 
$\mathbf{e} = [e_1, \dots, e_M]$. Each unit in $C$ has an output
$ c_i = \sigma \left( \sum_j \omega_{ij}e_j \right)$, where $\sigma(\cdot)$
is a positive sigmoidal function. The inputs are assumed to be errors,
and to reduce them we want $e_j$ to activate $c_i$ when $c_i$ can reduce $e_j$.
One way this could happen is when the weight $\omega_{ij}$ from 
$e_j$ to $c_i$ is proportional to the negative of their sensitivity derivative:
\begin{equation} \label{eq:sensitivity}
    \omega_{ij} \propto - \frac{\partial e_j}{\partial c_i}.
\end{equation}

Assuming a monotonic relation between the motor commands and the errors,
relation \ref{eq:sensitivity} entails that errors will
trigger an action to cancel them, with some caveats considered in 
\cite{verduzco-flores_differential_2022}.
Synaptic weights akin to equation \ref{eq:sensitivity} can be obtained using
a learning rule that extracts correlations between the derivatives of $c_i$
and $e_j$ (see Methods). Using this rule, the commands coming from population
$C$ can eventually move the
arm so that $S_A$ activity resembles $S_P$ activity.

$C$ is organized to capture the most basic motifs of spinal cord connectivity 
using a network where balance between excitation and inhibition is crucial
\cite{berg_balanced_2007,berg_when_2019,goulding_inhibition_2014}.
Each one of six $\alpha$ motoneurons stimulate one muscle, and is stimulated by one
excitatory ($CE$), and one inhibitory ($CI$) interneuron. $CE$ and $CI$
stimulate one another, resembling the classic Wilson-Cowan model 
\cite{cowan_wilsoncowan_2016}. The
trios composed of $\alpha, CE$, and $CI$ neurons compose a group that controls
the activation of one muscle, with $CE$ and $CI$ receiving convergent inputs from
$M$. This resembles the premotor network model in
\cite{petersen_premotor_2014}.
($\alpha, CE, CI$) trios are connected to other trios following the
agonist-antagonist motif that
is common in the spinal cord 
\cite{pierrot-deseilligny_circuitry_2005}. This means that $CE$ units project to the
$CE$ units of agonists, and to the $CI$ units of antagonists (figure 1, right
panel).  When the agonist/antagonist relation is not strongly defined, muscles can be
``partial'' agonists/antagonists, or unrelated.

Connections from $A$ to $C$ (the ``short-loop reflex'') use the input
correlation learning rule, analogous to the connections from $A$ to $M$.

Direct connections from $M$ to alpha motoneurons are not necessary for the model
to reach, but they were introduced in new versions because in higher
primates these connections are present for distal joints
\cite{lemon_descending_2008}.
Considering that bidirectional plasticity has been observed in corticomotoneural
connections \cite{nishimura_spike-timing-dependent_2013}, we chose to endow them
with the differential
Hebbian rule of \cite{verduzco-flores_differential_2022}.

Because timing is essential to support the conclusions of this paper,
every connection has a transmission delay, and all firing rate neurons are modeled with
ordinary differential equations.

All the results in this paper apply to the 3 configurations described above
(spinal learning, cortical learning, and static network). To emphasize the
robustness and potential of the learning mechanisms, in the Appendix we 
introduce two variations of the spinal learning model (in the {\it Variations of
the spinal learning model} section). 
All results in the paper also apply to those
two variations. In one of the variations (the ``synergistic'' network) each spinal
motoneuron stimulates two muscles rather than one. 
In the second variation (the ``mixed errors'' network) the inputs from $S_{PA}$ to $M$
are not one-to-one, but instead come from a matrix that combines multiple error
signals as the input to each $M$ unit.

Since most results apply to all configurations, and since results could depend
on the random initial weights, we report simulation results using 3 means and 3
standard deviations 
$(m_1 \pm \sigma_1| m_2  \pm \sigma_2| m_3  \pm \sigma_3)$, 
with the
understanding that these 3 value pairs correspond respectively to the
spinal learning, motor learning, and static network models.
The statistics come from 20 independent simulations with different initial
conditions.

A reference section in the Appendix (the {\it Comparison of the 5
configurations} section) summarizes the basic traits of all
different model configurations (including the two variations of the spinal
learning model), and compiles all their numerical results.

For each configuration, a single simulation was used to produce all the
representative plots in different sections of the paper.

\subsection{The model can reach by matching perceived and desired sensory
activity}
\label{sub:can_reach}

Reaches are performed by specifying an $S_P$ pattern equal 
to the $S_A$ activity when the hand is at the target.
The acquisition of these $S_P$ patterns is not in the scope of this paper (but
see \cite{verduzco-flores_differential_2022}).

We created a set of random targets by sampling uniformly from the space
of joint angles. Using this to set a different pattern in $S_P$ every 40 seconds, we 
allowed the arm to move freely during 16 $S_P$ target
presentations. To encourage exploratory movements we used noise and
two additional units described in the Methods.

\begin{figure}
    \centering
    \includegraphics[width=0.9\columnwidth]{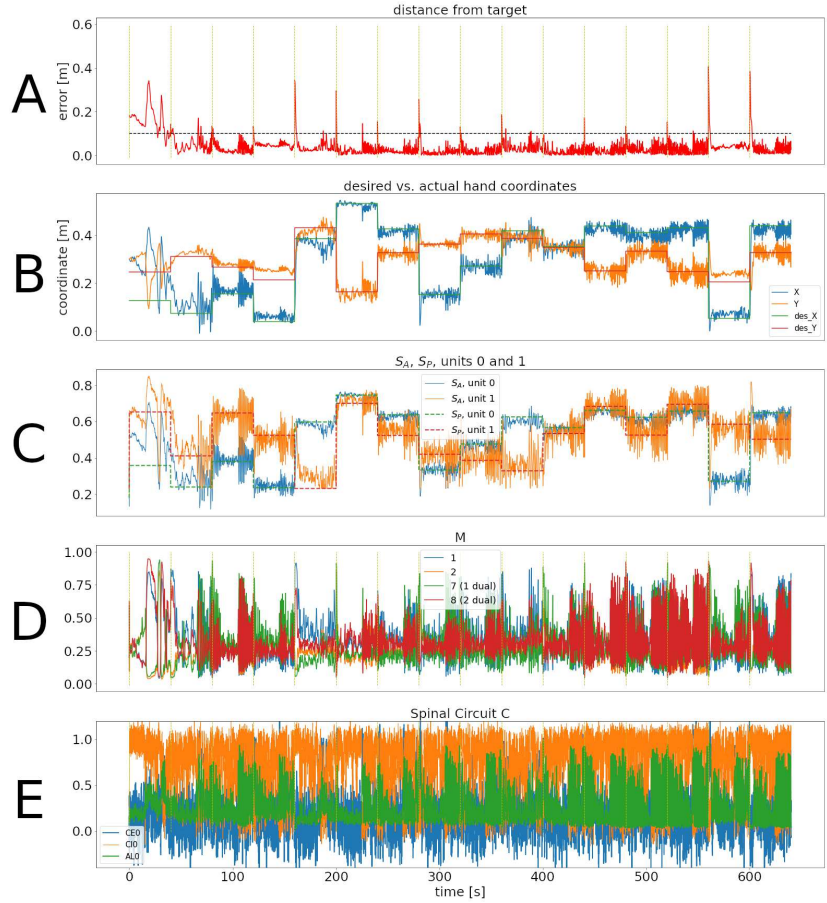}
    \caption{Representative training phase of a simulation for
    the spinal learning model. 
    {\bf A}: Distance between the target and the
    hand through 640 seconds of simulation, corresponding to 16 reaches to 
    different targets. 
    The horizontal dotted line corresponds to 10 centimeters. 
    The times when $S_P$ changes are indicated with a vertical, dotted yellow line. 
    Notice that the horizontal time axis is the same for all
    panels of this figure.
    The average error can be seen to decrease through the first two reaches.
    {\bf B}: Desired versus actual hand coordinates through the training phase.
    The straight lines denote the desired X (green) and Y (red) coordinates of the
    hand. The noisy orange and blue lines show the actual coordinates of the
    hand.
    {\bf C}: Activity of units 0 and 1 in $S_P$ and $S_A$. This panel shows that
    the desired values in the $S_P$ units (straight dotted lines) start
    to become tracked by the perceived values.
    {\bf D}: Activity of $M$ units 1, 2, and their duals. Notice that even when
    the error is close to zero the activity in the $M$ units does not disappear.
    {\bf E}: Activity of the $CE,CI, \alpha$ trio for muscle 0. The intrinsic
    noise in the units causes ongoing activity. Moreover, the inhibitory
    activity (orange line) dominates the excitatory activity (blue line).}
    \label{fig:training_phase}
\end{figure}

All model configurations were capable of reaching.
To decide if reaching was learned in a trial we took the
average distance between the hand and the target (the 
{\it average error}) during the
last 4 target presentations. Learning was achieved when this error was smaller
than 10 cm.

The system learned to reach in 99 out of 100 trials (20 for each
configuration). One simulation with the spinal learning model
had an average 
error of 14 cm during the last 4 reaches of training.
To assess the speed of learning we recorded the average number of target 
presentations required before the error  became less than 10 cm for the first
time. This average number of failed reaches before the first success was:
$(1.8 \pm 2| 1.2 \pm .9| 0 \pm 0)$.

Figure \ref{fig:training_phase}A shows the  error through 16 successive reaches 
(640 seconds of {\it in-silico} time) in a typical case for 
the spinal learning model.
A supplementary video shows the arm's movements during this simulation.
Figures similar to figure \ref{fig:training_phase} can be seen for all
configurations in the Appendix
(in the {\it Supplementary figures} section).

In figure \ref{fig:training_phase}A the error increases each time a new
target was presented (yellow vertical lines), but as learning continues 
it was consistently reduced below 10 centimeters.

Panel B also shows the effect of learning, as the hand's Cartesian coordinates
eventually track the target coordinates whenever they change.
This is also reflected as the activity in $S_A$ becoming similar
to the activity in $S_P$ (panel C).

Panels D and E of figure \ref{fig:training_phase} show the activity of a few
units in population $M$ and population $C$ during the 640
seconds of this training phase.
During the first few reaches, $M$ shows
a large imbalance between the activity of units and their duals, reflecting larger
errors. Eventually these activities balance out, leading to
a more homogeneous activity that may increase when a new target appears.
M1 activation patterns that produce no movement are called the
{\it null-space activity} \cite{kaufman_cortical_2014}. In our case, this
includes patterns where $M$ units have the same activity as their duals.
This, together with the noise and oscillations intrinsic to the
system cause the activity in $M$ and $C$ to never disappear.

In panel E the noise in the $C$ units becomes evident. It can also be seen that
inhibition dominates excitation (due to $CE$ to $CI$ connections), which promotes
stability in the circuit.

We tested whether any of the novel elements in the model were superfluous.
To this end, we removed each of the elements individually and checked if the
model could still learn to reach. In conclusion, removing
individual elements generally deteriorated performance, but the factor that
proved essential for all configurations with plasticity was the differential
Hebbian learning in the connections from $M$ to $C$ or from $S_{PA}$ to $M$.
For details, see the the Appendix section titled {\it The model fails when
elements are removed}.

\subsection{Center-out reaching 1: The reach trajectories present traits of cerebellar ataxia}
\label{sub:center_out1}
In order to compare our model with experimental data, after the
training phase we began a standard center-out reaching task. Switching 
to this task merely
consisted of presenting the targets in a different way, but for the sake of smoother
trajectories we removed the noise from the units in $C$ or $M$. 

\begin{figure}
    \centering
    \includegraphics[width=1.0\columnwidth]{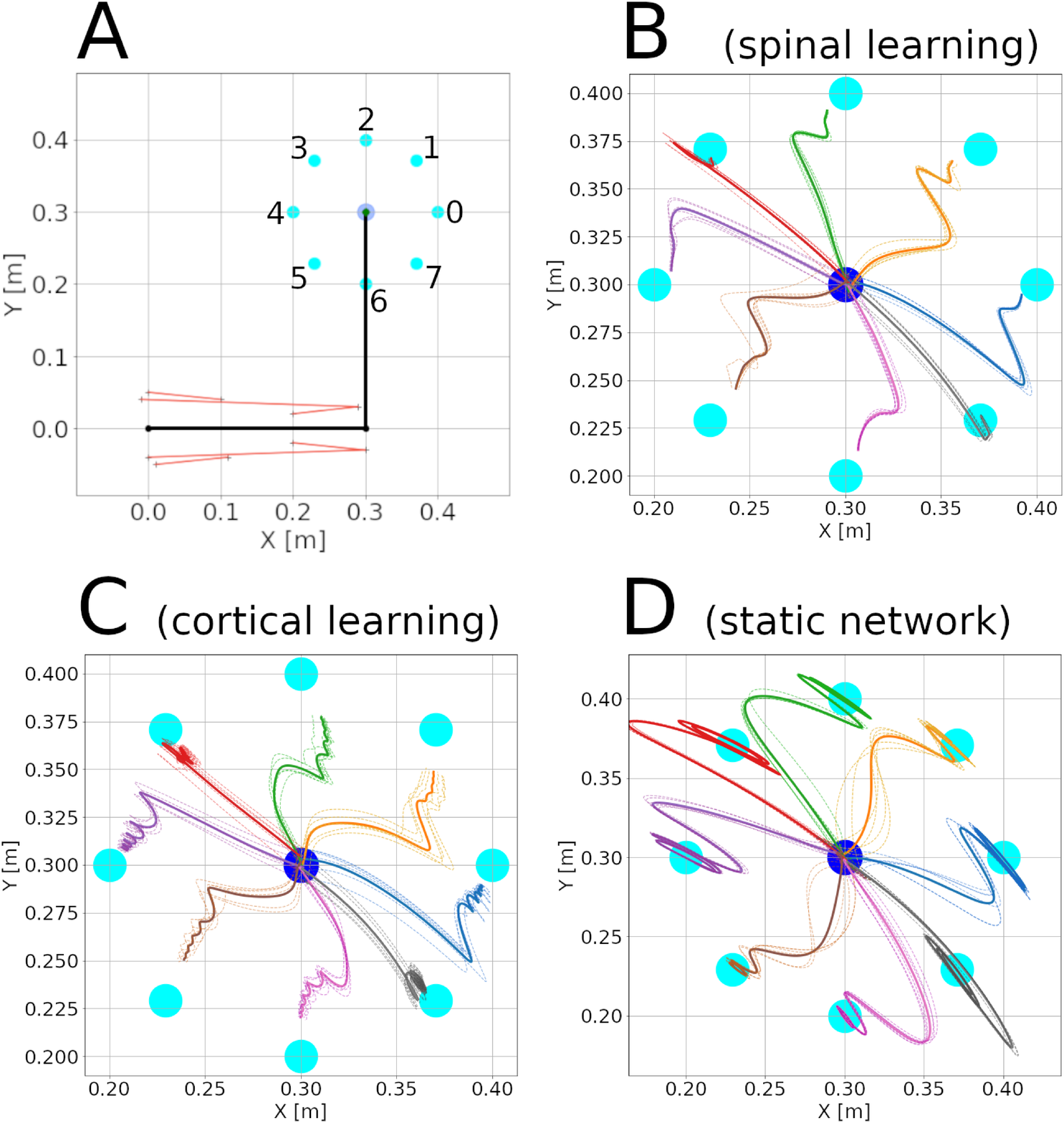}
    \caption{Center-out reaching. {\bf A}: The arm at its resting
	    position, with hand coordinates (0.3, 0.3) meters, where a center target 
        is located.
    Eight peripheral targets (cyan dots) were located on a circle around the center 
    target, with a 10 cm radius. The muscle lines, connecting the muscle insertion
    points, are shown in red. The shoulder is at the origin, whereas the elbow
    has coordinates (0.3, 0). Shoulder insertion points remain fixed.
    {\bf B-F}: Hand trajectories for all reaches in the 3 configurations.
    The trajectory's color indicates the target. Dotted lines show individual
    reaches, whereas thick lines indicate the average of the 6 reaches.
    }
    \label{fig:center_out_1A}
\end{figure}

Figure \ref{fig:center_out_1A}A shows the 8 peripheral targets around
a hand rest position. Before reaching a peripheral target, a reach to the
center target was performed, so the whole experiment was a single
continuous simulation controlled by the $S_P$ pattern.
 
Peripheral targets were selected at random, each appearing
6 times. This produced 48 reaches (without counting reaches to the center),
each one lasting 5 seconds. 
Panels B through D of figure \ref{fig:center_out_1A}
show the trajectories followed by the hand in the 3 configurations. During these
48 reaches the average distance between the hand and the target was
$(3.3 \pm .01| 2.9 \pm .001| 2.9 \pm .0003)$ centimeters.

Currently our system has neither cerebellum nor visual information.
Lacking a ``healthy" model to make quantitative
comparisons, we analyzed and compared them to data from cerebellar
patients.

For the sake of stability and simplicity, our system is configured to perform
slow movements. Fast and slow reaches are different in cerebellar patients 
\cite{bastian_cerebellar_1996}.
Slow reaches undershoot the target, follow longer hand paths, and show
movement decomposition (joints appear to move one at a time).
In figure \ref{fig:center_out_1A} the trajectories begin close to the 135-degree
axis, indicating a slower response at the elbow joint. 
With the parameters used, the spinal learning and cortical learning
models tend to undershoot the target, whereas in the static network the hand can
oscillate around the target.

The traits of the trajectories can be affected by many hyperparameters in the
model, but the dominant factor seems to be the gain in the control loop.
Our model involves delays, activation latencies, momentum, and interaction
torques. Unsurprisingly, increasing the gain leads to oscillations along
with faster reaching. On the other hand, low gain leads to slow, stable reaching
that often undershoots the target. Since we do not have a cerebellum to 
overcome this trade off, the gain was the only hyperparameter that was manually 
adjusted for all configurations (See Methods). 
In particular, we adjusted the slope of the $M$ and $S_A$ units so the system
was stable, but close to the onset of oscillations. Gain was allowed
to be a bit larger in the static network so oscillations could be observed.
The {\it Supplementary figures} section in the Appendix shows more examples of 
configurations with higher gain.

The shape of the trajectory also depends on the target. Different reach directions
cause different interaction forces, and encounter different levels of viscoelastic 
resistance from the muscles.

\begin{figure}
    \centering
    \includegraphics[width=0.9\columnwidth]{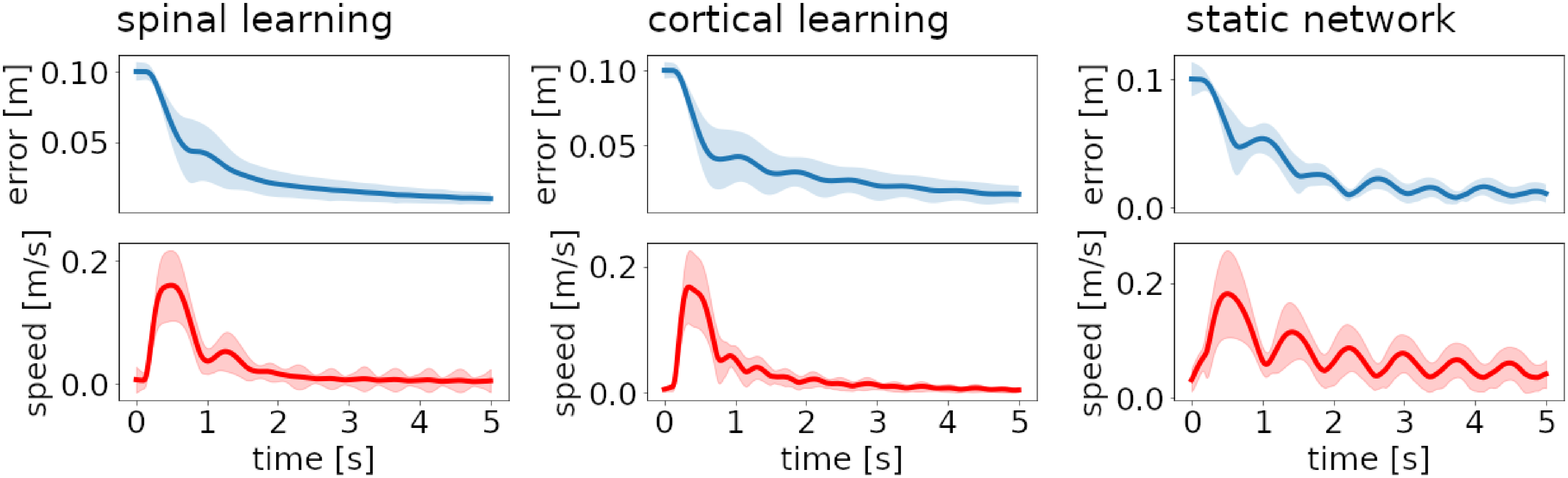}
    \caption{Distance to target and reach velocity through time for the 3
    configurations. Thick lines show the average over 48 reaches (8 targets, 6
    repetitions). Filled stripes show standard deviation.
    For the spinal and cortical learning configurations (left and
    center plots) the hand initially moves quickly to the target, but the
    direction is biased, so it needs to gradually correct the error from this
    initial fast approach; most of the variance in error and
    velocity appears when these corrections cause small-amplitude
    oscillations. In the case of the static network (right plots) oscillations 
    are ongoing, leading to a large variance in velocity.
    }
    \label{fig:center_out_1B}
\end{figure}

Figure \ref{fig:center_out_1B} reveals that the approach to the target is initially fast,
but gradually slows down. Healthy subjects usually present a bell-shaped velocity
profile, with some symmetry between acceleration and deceleration. This symmetry
is lost with cerebellar ataxia
\cite{becker_multi-joint_1991,gilman_kinematic_1976}.

We are not aware of center-out reaching studies for cerebellar patients
in the dark, but \cite{day_influence_1998} does examine reaching in these
conditions. Summarizing its findings:
\begin{enumerate}
    \item Movements were slow.
    \item The endpoints had small variation, but they had constant errors.
    \item Longer, more circuitous trajectories, with most changes in direction
          during the last quarter.
    \item Trajectories to  the same target showed variations.
\end{enumerate}

From figures \ref{fig:center_out_1A} and \ref{fig:center_out_1B} we can observe
constant endpoint errors when the gain is low, in the spinal and cortical
learning models. Circuitous trajectories with a
pronounced turn around the end of the third quarter are also observed.
Individual trajectories can present variations. A higher gain, as in the static
network on the right plots, can increase these
variations (illustrated in the {\it Supplementary figures} of The Appendix.)

\subsection{Center-out reaching 2: Directional tuning and preferred directions}

To find whether directional tuning could arise during learning, we analyzed 
the $M$ population activity for the 48 radial reaches described in the previous
subsection.

\begin{figure}
    \centering
    \includegraphics[width=0.9\columnwidth]{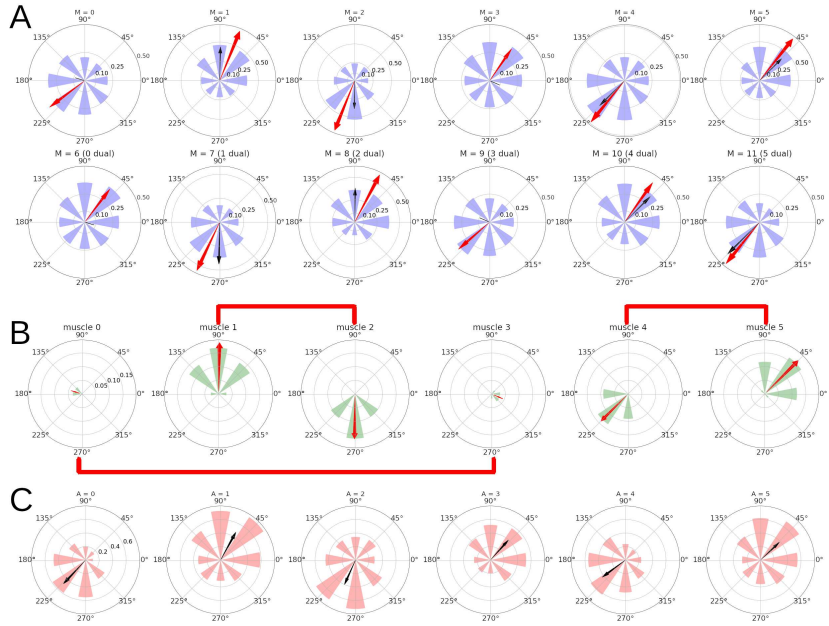}
    \caption{Directional tuning of the units in $M$ for
    a simulation with the spinal learning model.
	    {\bf A}: Average firing rate per target, and preferred direction (see Methods)
	    for each of the 12 units in $M$. 
        Each polar plot corresponds to a single unit, and each of the
    8 purple wedges corresponds to one of the 8 targets. The length of a wedge
    indicates the mean firing rate when the hand was reaching the corresponding
    target. The red arrow indicates the direction and relative magnitude of the
    PD vector. The black arrow shows the predicted PD vector, in this case just the
    corresponding arrows from panel B.
    {\bf B}: For each muscle and target, a wedge shows the muscle's length at
    rest position minus the length at the target, divided by the rest position length.
    The red arrow comes from the sum of the wedges taken as vectors, and represents
    the muscle's direction of maximum contraction. Plots corresponding
    to antagonist muscles are connected by red lines.
    {\bf C}: Average activity of the 6 $A$ units indicating muscle tension. 
    The black arrows come from the sum of wedges taken as vectors,
    showing the relation between muscle tension and preferred direction.
    }
    \label{fig:center_out_2A}
\end{figure}

For each of the 12 units in $M$, figure \ref{fig:center_out_2A}A shows the mean
firing rate of the unit when reaching each of the 8 targets. The red arrows
show the Preferred Direction (PD) vectors that arise from these distributions
of firing rates. 
For the sake of exposition, figure \ref{fig:center_out_2A} 
shows data for the simpler case of one-to-one connectivity between $S_{PA}$ and $M$
in the spinal learning model, but these results generalize to the
case when each $M$ unit receives a linear combination of the $S_{PA}$ activities
(the ``mixed errors'' variation presented in the {\it Variations of the spinal
learning model} section of the Appendix.)

We found that 
$(11.8 \pm .4| 12 \pm 0| 12 \pm 0)$ units were
significantly tuned to reach direction ($p < 0.001$, bootstrap test), with PD vectors
of various lengths.
The direction of the PD vectors is not mysterious. Each $M$ unit controls the length
error of one muscle. 
Figure \ref{fig:center_out_2A}B shows that the required
contraction length depends on both the target and the muscle. 
The PD vectors of units 0-5 point to the targets that require the most
contraction of their muscle. Units 6-11 are the duals of 0-5, and their PD
is in the opposite direction. Figure \ref{fig:center_out_2A}C
shows that the PD may also be inferred from the muscle activity,
reflected as average tension. 

In the case when each $M$ unit receives a linear combination of $S_{PA}$ errors,
its PD can be predicted using a linear combination of the ``directions of maximum
contraction'' shown in figure
\ref{fig:center_out_2A}B, using the same weights as the $S_{PA}$ inputs.
When accounting for the length of the PD vectors, this can predict the PD
angle with a coefficient of determination 
$R^2 \approx (.74 \pm.18| .88 \pm .14| .86 \pm .01)$.

As mentioned in the Introduction, the PDs of motor cortex neurons tend
to align in particular directions \cite{scott_dissociation_2001}. This is almost 
trivially true for this model, since the PD vectors are mainly produced by 
linear combinations of the vectors in figure \ref{fig:center_out_2A}B.

\begin{figure}
    \centering
    \includegraphics[width=0.9\columnwidth]{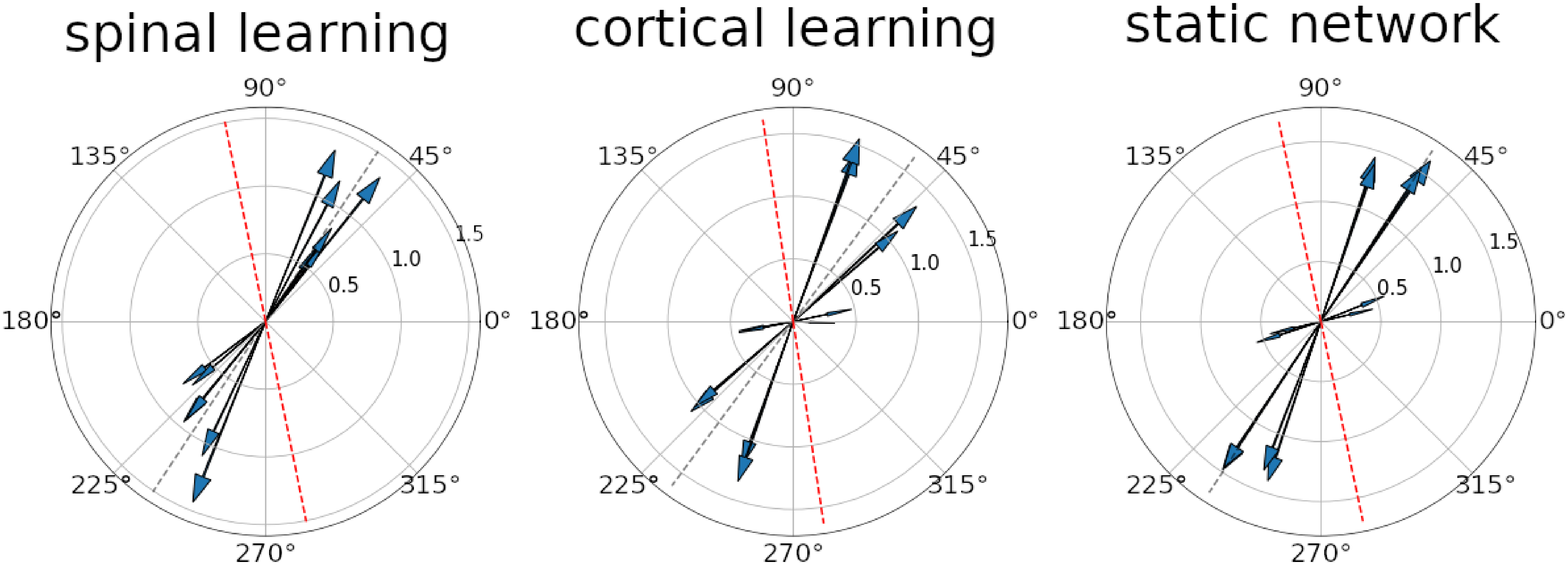}
    \caption{Preferred direction vectors for the 12 $M$ units.
    In all three plots the arrows denote the direction and
    magnitude of the preferred direction (PD) for an individual unit. The gray 
    dotted lines shows the main axis of the distribution. The red dotted
    lines are a 45-degree rotation of the gray line, for comparison with 
    \cite{scott_dissociation_2001}. It can be seen that all configurations
    display a strong bimodality, especially when considering the units with a
    larger PD vector. The axis where the PD vectors tend
    to aggregate is in roughly the same position for the 3 configurations.
    }
    \label{fig:center_out_2B}
\end{figure}

Figure \ref{fig:center_out_2B} shows the PD for all the $M$
units in a representative simulation for each of the configurations.
In every simulation the PD distribution showed significant
bimodality ($ p < 0.001$). The main axis of the PD distribution (see Methods)
was $(59 \pm 7| 52 \pm 2| 54 \pm .5)$ degrees.

To compare with \cite{scott_dissociation_2001} we rotate this line 45 degrees
so the targets are in the same position relative to the shoulder
(e.g. \cite{lillicrap_preference_2013} Fig. 1, \cite{kurtzer_primate_2006}
Fig. 1). This places the average main axes above in a range
between 99 and 104 degrees, comparable to the 117 degrees in 
\cite{scott_dissociation_2001}.

The study in \cite{lillicrap_preference_2013} suggested that a rudimentary
spinal cord feedback system should be used to understand {\it why} the PD 
distribution arises. Our model is the first to achieve this.

The PD vectors are not stationary, but experience random fluctuations that become
more pronounced in new environments 
\cite{rokni_motor_2007, padoa-schioppa_neuronal_2004}. 
The brain is constantly remodeling itself, without
losing the ability to perform its critical operations 
\cite{chambers_stable_2017}.
Our model is continuously learning, so
we tested the change in the PDs by setting 40 additional center-out reaches
(no intrinsic noise) after the previous experiment,
once for each configuration.

To encourage changes we set 10 different targets instead of 8. 
After a single trial for each configuration the change in angle
for the 12 PD vectors had means and standard deviations of
$(3.3 \pm 2.4| 4.9 \pm 2.1| .3 \pm .2)$ degrees. Larger changes
(around 7 degrees) could be observed in the ``mixed errors'' variation of the
model, presented in the Appendix ({\it Variations of the spinal learning model}
section).
We also measured the change in the preferred directions of the muscles,
obtained as in figure \ref{fig:center_out_2A}C. This yielded differences and
standard deviations
$(3.8 \pm 2.1| 6.4 \pm 2.9| .2 \pm .2)$ degrees.

The average distance between hand and target during the 40 reaches
was $(3| 3.6| 2.9)$ cm, showing that the hand was still moving 
towards the targets, although with different errors due to their new 
locations. 

\subsection{Center-out reaching 3: Rotational dynamics}

Using a dynamical systems perspective, \cite{shenoy_cortical_2013}
considers that the muscle activity $\mathbf{m}(t)$ (a vector function of time)
arises from the cortical activity vector $\mathbf{r}(t)$ after it is
transformed by the downstream circuitry:
\begin{equation} \label{eq:downstream}
    \mathbf{m}(t) = G[\mathbf{r}(t)].
\end{equation}
It is considered that the mapping $G[\cdot]$ may consist of sophisticated controllers,
but for the sake of simplicity this mapping is considered static, omitting spinal cord plasticity.
The cortical activity arises from a dynamical system:
\begin{equation} \label{eq:dynamical}
    \tau \mathbf{\dot{r}}(t) = h(\mathbf{r}(t)) + \mathbf{u}(t),
\end{equation}
where $\mathbf{u}(t)$ represents inputs to motor cortex from other areas, and
$h(\cdot)$ is a function that describes how the state of the system evolves.

A difficulty associated with equation \ref{eq:dynamical} is explaining how
$\mathbf{r}(t)$ generates a desired muscle pattern
$\mathbf{m}(t)$ when the function $h(\cdot)$ represents the dynamics of a
recurrent neural network. One possibility is that M1 has intrinsic oscillators
of various frequencies, and they combine their outputs to shape
the desired pattern. This prompted the search for oscillatory activity in
M1 while macaques performed center-out reaching motions. A brief oscillation
(in the order of 200 ms, or 5 Hz) was indeed found in the population activity
\cite{churchland_neural_2012}, and the model in \cite{sussillo_neural_2015}
was able to reproduce this result, although this was done in the
open-loop version of equations
\ref{eq:downstream} and \ref{eq:dynamical}, where $\mathbf{u}(t)$ contains no
afferent feedback (this is further commented in the Supplemental Discussion).

Recently it was shown that the oscillations in motor cortex can arise 
when considering the full sensorimotor loop, without the need of recurrent 
connections in motor cortex \cite{kalidindi_rotational_2021}.
A natural question is whether our model can also
reproduce the oscillations in \cite{churchland_neural_2012} without requiring 
M1 oscillators or recurrent connections.
 
The analysis in \cite{churchland_neural_2012} is centered around
measuring the amount of rotation in the M1 population activity. The first step
is to project the M1 activity vectors onto their first six principal
components. These six components are then rotated so the evolution of the
activity maximally resembles a pure rotation. These rotated components are
called the ``jPCA vectors''. The amount of variance in the M1 activity explained
by the first two jPCA vectors is a measure of rotation.
The Methods section provides more details of this procedure.

Considering that we have a low-dimensional, non-spiking, slow-reaching model, we
can only expect to qualitatively replicate the essential result in 
\cite{churchland_neural_2012}, which is most of the variance being contained in
the first jPCA plane.

\begin{figure} \label{fig:jpca}
    \centering
    \includegraphics[width=0.9\columnwidth]{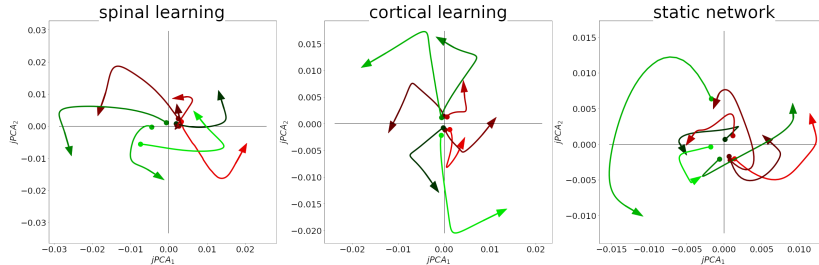}
    \caption{Rotational dynamics in the $M$ population in a
    representative simulation for all configurations. Each plot shows the first
    two jPCA components during 0.25 seconds, for each of the 8 
    conditions/targets. Traces are 
    colored according to the magnitude of their initial
    $jPCA_1$ component, from smallest (green) to largest (red).
    }
    \label{fig:center_out_3}
\end{figure}

We replicated the jPCA analysis, with adjustments to account for the smaller number
of neurons, the slower dynamics, and the fact that there is no delay period
before the reach (See Methods). 
The result can be observed in figure \ref{fig:center_out_3}, 
where 8 trajectories are seen in the plots. Each
trajectory is the average activity of the 12 $M$ units when reaching to one of the
8 targets, projected onto the jPCA plane. The signature of a
rotational structure in these plots is that most trajectories circulate in a
counterclockwise direction. Quantitatively, the first jPCA plane (out of six)
captures $(.42 \pm .04| .42 \pm .04| .46 \pm .03)$
of the variance.

With this analysis we show that our model does not require intrinsic 
oscillations in motor cortex to produce rotational dynamics, in agreement
with \cite{kalidindi_rotational_2021} and \cite{dewolf_spiking_2016}.

\subsection{The effect of changing the mass}
Physical properties of the arm can change, not only as the arm grows, but also
when tools or new environments come into play. As a quick test of whether the 
properties in this paper are robust to moderate changes, we 
changed the mass of the arm and forearm from 1 to 0.8 kg and ran one simulation
for each of the 5 configurations.

\begin{figure}
    \centering
    \includegraphics[width=1.0\columnwidth]{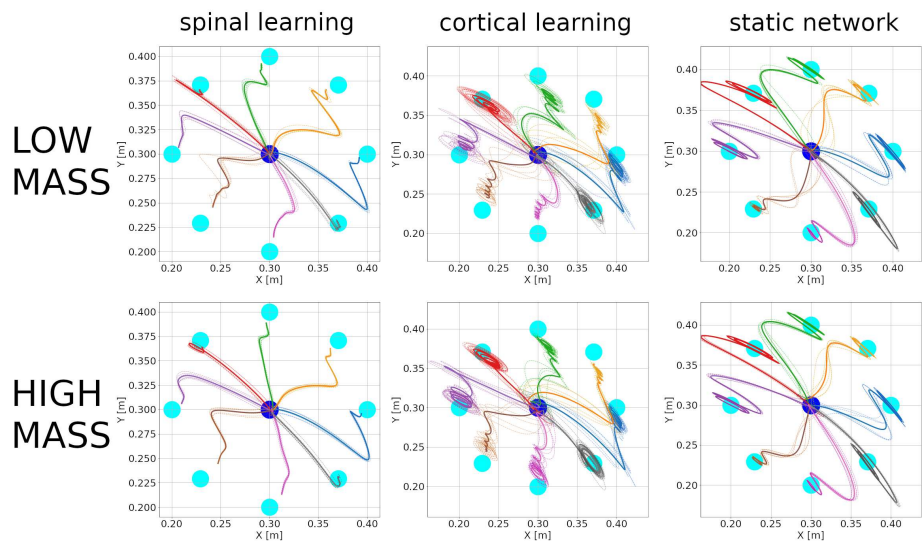}
    \caption{Hand trajectories with low mass (0.8 kg, top 3 
    plots) and high mass (1.2 kg, bottom 3 plots) for the 3 configurations.
    Plots are as in figure \ref{fig:center_out_1A}.
    The spinal learning model and the static network show qualitatively similar
    trajectories compared to those in figure \ref{fig:center_out_1A}. In 
    contrast, the
    cortical learning model began to display considerable endpoint oscillations
    for several targets after its mass was reduced. These oscillations persist
    after the mass has been increased.
    }
    \label{fig:altered}
\end{figure}

With a lighter arm the average errors during center-out reaching were
$(2.5| 3.2| 3)$ cm. The hand trajectories with a reduced mass can
be seen in the top 3 plots of figure \ref{fig:altered}.
We can observe that the spinal learning model slightly reduced its mean error,
whereas the cortical learning model increased it. This can be understood by
noticing that a reduction in mass is akin to an increase in gain. 
The spinal learning model with its original gain was below the threshold of
oscillations at the endpoint, and a slight mass decrease did not change this.
The cortical learning model with the original gain was already oscillating
slightly, and an increase in gain increased the oscillations.


In the same simulation, after the center-out reaching was completed, we once more
modified the mass of the arm and forearm, from 0.8 to 1.2 kg, after which we began
the center-out reaching again. This time the center-out reaching errors were
$(2.4| 3.3| 2.9)$ cm. The hand trajectories for this high mass
condition are in the bottom 3 plots in figure \ref{fig:altered}.
It can be seen that the spinal learning and cortical learning models retained
their respectively improved and decreased performance, whereas the static
network performed roughly the same for all mass conditions. A tentative
explanation is that with reduced mass the synaptic learning rules tried to 
compensate for faster movements with weights that effectively increased the gain
in the loop. After the mass was increased these weights did not immediately
revert, leading to similar trajectories after the increase in mass.

The results of the paper still held after our mass manipulations.
For all configurations, PD vectors could be predicted with a coefficient of 
determination between .74 and .92; All units in $M$ were significantly tuned
to direction; the main axis of the PD distribution ranged between 56 and 61
degrees, and the first jPCA plane captured between 33\% and 58\% of the variance.

\subsection{Spinal stimulation produces convergent direction fields}
\label{sub:convergent}

Due to the viscoelastic properties of the muscles, the mechanical system
without active muscle contraction will have a fixed point with lowest
potential energy at the arm's rest position.
Limited amounts of muscle contraction shift the position of
that fixed point. 
This led us to question whether this could produce convergent force fields,
which as discussed before are candidate motor primitives, and have
been found experimentally.

To simulate local stimulation of an isolated spinal cord we removed all
neuronal populations except for those in $C$, and applied inputs to
the individual pairs of $CE, CI$ units projecting to the same motoneuron.
Doing this for different starting positions of the hand, and recording
its initial direction of motion, produces a {\it direction field}.
A direction field maps each initial hand location to a vector
pointing in the average direction of the force that initially moves the hand.

\begin{figure} 
    \centering
    \includegraphics[width=0.9\columnwidth]{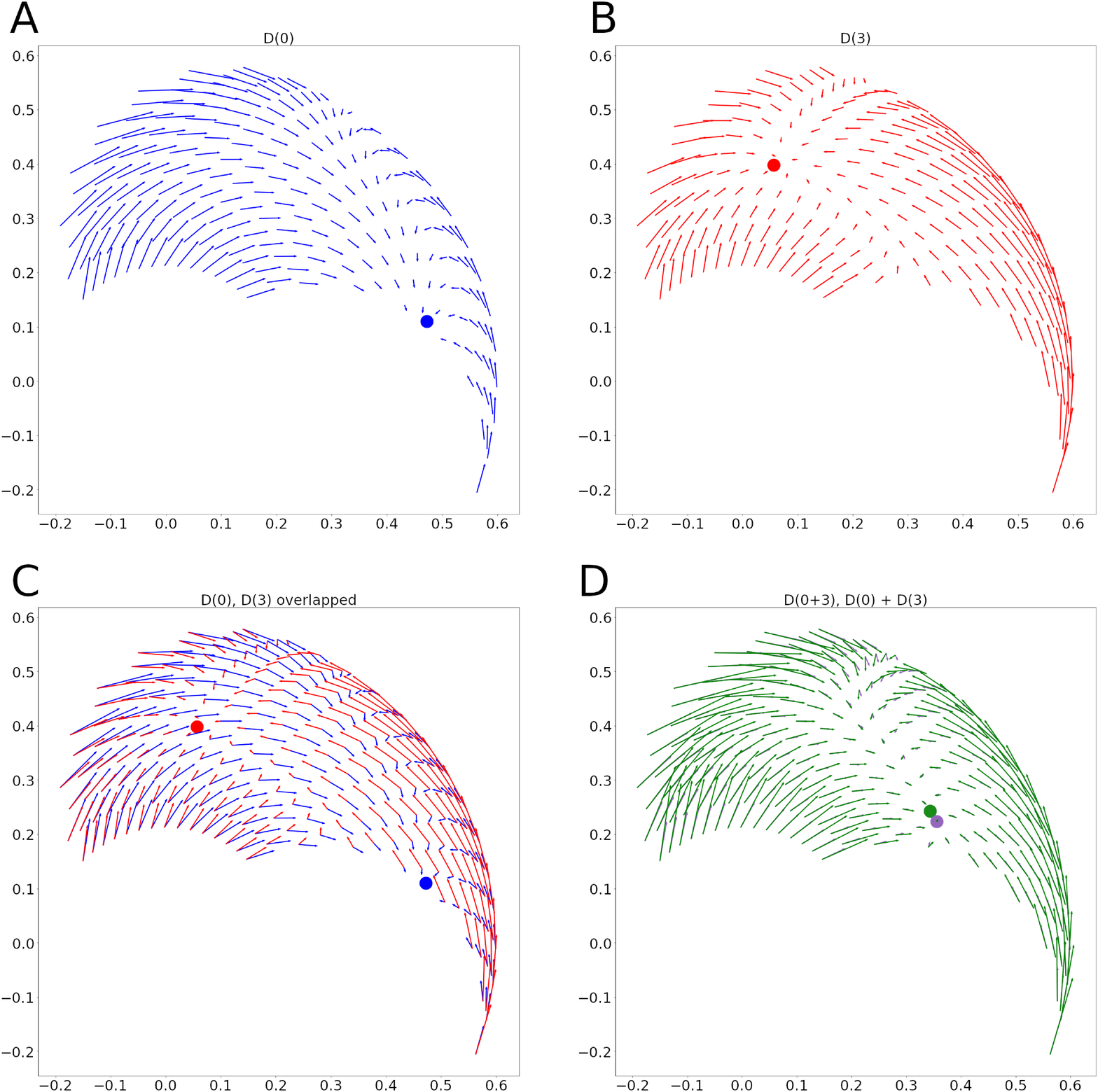}
    \caption{Two sample direction fields and their linear addition
    for circuit $C_1$.
    {\bf A}: Direction Field (DF) from stimulation of the 
    interneurons for muscle 0 (biarticular biceps).  The
    approximate location of the fixed point is shown with a blue dot.
    {\bf B}: DF from stimulation of muscle 3 (biarticular triceps) interneurons.
    A red dot shows the fixed point.
    {\bf C}: Panels A and B overlapped.
    {\bf D}: In green, the DF from stimulating the interneurons 
    for muscles 0 and 3 together. In purple, the sum of the DFs from panels
    A and B. Dots show the fixed points.
    The average angle between the green and purple vectors is ~4 degrees.
    } \label{fig:direction_fields}
\end{figure}

The first two panels of figure \ref{fig:direction_fields} show the result of
stimulating individual E-I pairs in $C$, which will indeed produce 
direction fields with different fixed points. 
 
We found that these direction fields add approximately linearly 
(Fig. \ref{fig:direction_fields}D). 
More precisely, let $D(a+b)$ be the direction field from stimulating spinal 
locations $a$ and $b$ simultaneously, and $\alpha_{a+b}(x,y)$ be the angle of
$D(a+b)$ at hand coordinates $(x,y)$. Using similar definitions for
$D(a),D(b), \alpha_a(x,y), \alpha_b(x,y)$, we say the direction fields
add linearly if

$\alpha_{a+b}(x,y) = \alpha_a(x,y) + \alpha_b(x,y) , \ \forall (x,y)$.

We define the mean angle difference between $D(a+b)$ and $D(a)+D(b)$ as
\begin{equation}
    \gamma_{a,b} = \sum_{x,y} \frac{\alpha_{a+b}(x,y) - \Big( \alpha_a(x,y) +
    \alpha_b(x,y) \Big)}{N_s},
\end{equation}
where $N_s$ is the number of $(x,y)$ sample points. We found that when averaged
over the  15 ($C_1$) or 144 ($C_2$) possible $(a,b)$ pairs, 
the mean of $\gamma_{a,b}$ was 
13.5 degrees.

Randomly choosing two possibly different pairs $(a,b)$ and $(c,d)$ for the
stimulation locations leads to a mean angle difference of 
37.6 degrees between
the fields $D(a+b)$ and $D(c)+D(d)$. A bootstrap test showed that 
these angles are significantly larger ($p < 0.0001$) than in the previous case
where $(a,b)=(c,d)$. 

The resting field is defined as the direction field when no units are stimulated.
Removing the resting field from $D(a+b), D(a)$, and $D(b)$ does not alter 
these results.

Recent macaque forelimb experiments \cite{yaron_forelimb_2020} show that the
magnitude of the vectors in the $D(a+b)$ fields is larger than expected from 
$D(a) + D(b)$ (supralinear summation). We found no evidence for this effect, 
suggesting that it depends on mechanisms beyond those present in our model.

\section{Discussion}

\subsection{Summary of findings and predictions}
We have presented a model of the long loop reflex
with a main assumption: negative feedback configured with two
differential Hebbian learning rules. One novel rule sets the
loop's input-output structure, and the other rule (input correlation) promotes
stability. We showed that this model can make arm reaches by trying to perceive
a given afferent pattern.

Our study made two main points:
\begin{enumerate}
    \item Many experimental phenomena emerge from a feedback controller
    with minimally-complete musculoskeletal and neural models
    (emphasis is placed on the balance between excitation and inhibition).
    \item Even if the feedback controller has multiple inputs and outputs, its
    input-output structure can be flexibly configured by a differential Hebbian
    learning rule, as long as errors are monotonic.
\end{enumerate}

The first main point above was made using a feedback control network with
no learning (called the static network in the Results). We showed that in 
this static network:
1) reaching trajectories are similar to models of cerebellar ataxia, 2)
motor cortex units are tuned to preferred directions, 3) those preferred
directions follow a bimodal distribution, 4) motor cortex units present
rotational dynamics, 5) reaching is still possible when mass is altered, 
and 6) spinal stimulation produces convergent direction fields.

The second main point was made using two separate models, both using the
same differential Hebbian learning rules, but applied at different locations.
The spinal learning model presents the hypothesis that the spinal cord learns
to adaptively configure the input-output structure of the feedback controller.
The cortical learning model posits that configuring this structure
could instead be a function of motor cortex; this would not disrupt our
central claims. These two models should not be considered
as incompatible hypotheses. Different elements performing
overlapping functions are common in biological systems \cite{edelman_degeneracy_2001}.

Two variations of the spinal learning model in the Appendix show that this learning 
mechanism is quite flexible, opening the doors for certain types of synergies,
and for more complex errors (that still maintian the constraint of
monotonicity).

We list some properties of the model, and possible implications:
\begin{itemize}
    \item Basic arm reaching happens through negative feedback, trying
    to perceive a target value set in cortex. Learning the input-output 
    structure of the feedback controller may require spinal cord
    plasticity.
    \begin{itemize}
        \item Cerebellar patients should not be able to adapt to tasks that
        require fast reactions, as negative feedback alone cannot compensate
        for delays in the system \cite{sanguineti_cerebellar_2003}.
        On the other hand, they should be able to learn tasks that
        require remapping afferent inputs to movements. One example is
        \cite{richter_adaptive_2004}, where cerebellar patients learned to
        move in a novel dynamic environment, but their movements were less
        precise than those of controls.
    \end{itemize}
    \item The shape of reaches is dominated by mechanical
    and viscoelastic properties of the arm and muscles.
        \begin{itemize}
            \item Unfamiliar viscous forces as in \cite{richter_adaptive_2004}
            should predictably alter
            the trajectory (Fig. \ref{fig:center_out_1A}) for cerebellar patients,
            who should not be able to adapt unless they move slowly and are 
            explicitly compensating.
        \end{itemize}
    \item Preferred Directions (PDs) in motor cortex happen because muscles
    need to contract more when reaching in certain directions.
    \begin{itemize}
        \item The PD distribution should align with the directions where the
        muscles need to contract to reduce the error. These directions depend
        on which error $S_{PA}$ is encoding. If the error is not related to
        reaching (e.g. related to haptic feedback), a different PD distribution
        may arise after overtraining.
        \item Drift in the PD vectors comes from the ongoing adaptation, and it should
        not disrupt performance.
    \end{itemize}
    \item The oscillations intrinsic to delayed feedback control
    after the onset of a target are sufficient to explain the quasi-oscillations
    observed in motor cortex
    \cite{churchland_neural_2012,kalidindi_rotational_2021}.
    \item Convergent force fields happen naturally in musculoskeletal systems when
    there is balance in the stimulation between agonists and antagonists. Linear addition
    of force fields is a result of forces/torques adding linearly.
\end{itemize}

Since our relatively simple model reproduces
these phenomena, we believe it constitutes a good null hypothesis for them. 
But beyond explaining experimental observations, this model makes inroads into
the hard problem of how the central nervous system (CNS) can generate effective
control signals, recently dubbed the ``supraspinal pattern formation" problem
\cite{bizzi_motor_2020}. From our perspective, the CNS does
not need to generate precise activation patterns for muscles and synergies; it
needs to figure out which perceptions need to change. It is subcortical
structures that learn the movement details. The key to make such a model work
is the differential Hebbian learning framework in 
\cite{verduzco-flores_differential_2022}, which handles the final credit assignment problem.

We chose not to include a model of the cerebellum at this stage. Our model
reflects the brain structure of an infant baby who can make clumsy reaching
movements. At birth the cerebellum is incomplete and presumably not functional.
It requires structured input from spinal cord and cortex to establish correct
synaptic connections during postnatal development and will contribute to smooth
reaching movements at a later age.

Encompassing function, learning, and experimental phenomena in a single simple
model is a promising start towards a more integrated computational neuroscience.
We consider that such models have  the potential to steer complex 
large-scale models so they can also achieve learning and functionality from scratch.

\section{Methods}

Simulations were run in the Draculab simulator \cite{verduzco-flores_draculab:_2019}.
All the parameters from the equations in this paper are presented in the
Appendix. 
Parameters not shown can be obtained from Python dictionaries in the source code.
This code can be downloaded from: \\
{\small
\verb+https://gitlab.com/sergio.verduzco/public_materials/-/tree/master/adaptive_plasticity+}

\subsection{Unit equations}
With the exception of the $A$ and $S_P$ populations, the activity $u_i$ of
any unit in figure \ref{fig:architecture} has dynamics:

\begin{align} \label{eq:unit}
    \tau_u \dot{u_i} &= \sigma(I) - u_i, \\
    \sigma(I) &= \frac{1}{1 + \exp(\beta(I-\eta))}, \label{eq:sig}
\end{align}
where $\tau$ is a time constant, $\beta$ is the slope of the sigmoidal
function, $\eta$ is its threshold, and $I = \sum_j \omega_{ij} u_j(t-\Delta t_j)$ 
is the sum of delayed inputs times their synaptic weights.

Units in the $CE,CI$ populations (in the spinal learning model) or
in $M$ (in the cortical learning model) had an additional noise term, which turned
equation \ref{eq:unit} into this Langevin equation:
\begin{equation} \label{eq:langevin}
    du_i(t) = \frac{1}{\tau_u} \left( \sigma(I) - u_i(t) \right)
    + \varsigma dW(t),
\end{equation}
where $W(t)$ is a Wiener process with unit variance, and $\varsigma$ is a parameter
to control the noise amplitude. This equation was solved using the Euler-Maruyama
method. All other unit equations were integrated using the forward Euler method. 
The equations for the plant and the muscles were integrated with SciPy's 
(\verb+https://scipy.org/+) explicit Runge-Kutta 5(4) method.

Units in the $A$ population use a rectified logarithm activation function,
leading to these dynamics for their activity:
\begin{equation} \label{eq:log}
    \tau_a \dot{a} = \log ([1+ I - T]_+) - a,
\end{equation}
where $\tau_a$ is a time constant, $I$ is the scaled sum of inputs,
$T$ is a threshold, and $[x]_+ = \max(x,0)$ is the "positive part" function.

\subsection{Learning rules}
The learning rule for the connections from $M$ to $CE,CI$ units 
in the spinal learning model was first
described in \cite{verduzco-flores_differential_2022}. It has an equation:
\begin{equation} \label{eq:rga21}
\dot{\omega}_{ij}(t) = -\Big(\ddot{e}_j(t) - \langle \ddot{e}(t) 
\rangle \Big) \Big(\dot{c}_i(t-\Delta t) - 
\langle \dot{c}(t-\Delta t) \rangle \Big).
\end{equation}
In this equation, $e_j(t)$ represents the activity of the $j$-th unit in $M$ at time
$t$, and $\ddot{e}_j(t)$ is its second derivative. Angle brackets denote averages,
so that $\langle \ddot{e} \rangle \equiv \frac{1}{N_M} \sum_k \ddot{e}_k$, where
$N_M$ is the number of $M$ units. $\dot{c}_i(t)$ is the derivative of the 
activity for the
postsynaptic unit, and $\Delta t$ is a time delay ensuring that the rule captures the
proper temporal causality. 
In the Supplementary Discussion of the Appendix we elaborate on how
such a learning rule could be present in the spinal cord.

The learning rule in \ref{eq:rga21} was also fitted with soft weight-bounding to prevent
connections from changing sign, and multiplicative normalization was used to
control the magnitude of the weights by ensuring two requirements: (1)
all weights from projections of the same $M$ unit should add to $w_{sa}$,
(2) all weights ending at the same $C$ unit should add to $w_{sb}$. With this,
the learning rule adopted the form:
\begin{equation} \label{eq:w_norm}
\dot{\omega}_{ij} = -\alpha \omega_{ij} \Big( -\Delta  + 
\lambda \left[ (0.5(\zeta_{sa}+\zeta_{sb}) -
1\right] \Big),
\end{equation}
In this equation
$\alpha$ is a constant learning rate, $\Delta$ is the right-hand side expression
of equation \ref{eq:rga21}, and $\lambda$ is a scalar parameter.
The value $\zeta_{sa}$ is 
$w_{sa}$ divided by the sum of outgoing weights from the $j$-th $M$ unit, and 
$\zeta_{sb}$ is $w_{sb}$ divided by the sum of incoming $M$
weights on $c_i$. This type of normalization is meant to reflect the 
competition for resources among synapses, both at the presynaptic and 
postsynaptic level.

The synapses in the connections from $A$ to $M$ and from $A$ to
$C$ used the input correlation rule \cite{porr_strongly_2006}:
\begin{equation} \label{eq:inp_corr}
    \dot{w} =  \alpha_{IC} w I_A \dot{I}_{PA},
\end{equation}  
where $I_A$ is the scaled sum of inputs from the $A$ population, $\alpha_{IC}$
is the learning rate, $I_{PA}$ is the scaled sum of inputs from $S_{PA}$
or $M$, and $\dot{I}_{PA}$ is its derivative.
Unlike the original input correlation rule, this rule uses soft weight bounding
to avoid weights changing signs. Moreover, the sum of the weights was kept
close to a $\omega_{s}$ value. In practice
this meant dividing the each individual $w$ value by the sum of weights from
$A$-to-$M$ (or $A$-to-$C$) connections, and multiplying times $\omega_s$ at each update.
In addition, weight clipping was used to keep individual weights below a
value $\omega_{max}$.

The learning rule in the cortical learning model was the same, but
the presynaptic units were in $S_{PA}$, and the postsynaptic units in $M$.

\subsection{Exploratory mechanism} \label{sub:exploration}
Without any additional mechanisms the model risked getting stuck in a fixed
arm position before it could learn. We included two mechanisms to permit exploration
in the system. We describe these two mechanisms as they were applied to the
spinal learning model and its two variations. The description below also applies
to the case of the cortical learning model, with the $M$ units (instead of the
$C$ units) receiving the noise and extra connections.

The first exploratory mechanism consists of intrinsic noise in the
$CE$ and $CI$ interneurons, which causes low-amplitude oscillations in the arm. We
have observed that intrinsic oscillations in the $CE,CI$ units are also effective to allow learning (data not shown),
but the option of intrinsic noise permits the use of simple sigmoidal units in $C$,
and contributes to the discussion regarding the role of noise in neural computation.

The second mechanism for exploration consists of an additional unit, called $ACT$. 
This unit acted similarly to a leaky integrator of the
total activity in $S_{PA}$, reflecting the total error. If the leaky integral of
the $S_{PA}$ activity crossed a threshold, then $ACT$ would send a signal to all
the $CE$ and $CI$ units, causing adaptation. The adaptation consisted of an
inhibitory current that grew depending on the accumulated previous activity.

To model this, $CE$ and $CI$ units received an extra input $I_{adapt}$.
When the input from the $ACT$ unit was larger than 0.8, and $I_{adapt} < 0.2$,
the value of $I_{adapt}$ would be set to $(u_i^{slow})^2$. This is the square
of a low-passed filtered version of $u_i$. More explicitly,
\begin{equation} \label{eq:low-pass}
    \tau_{slow} \dot{u}_i^{slow} = u_i - u_i^{slow}. 
\end{equation}
If the input from $ACT$ was smaller than 0.8, or $I_{adapt}$ became larger
than 0.2, then $I_{adapt}$ would decay towards zero:
\begin{equation} \label{eq:decay}
    \tau_{slow}\dot{I}_{adapt} = -I_{adapt}.
\end{equation}

With this mechanism, if the arm got stuck then error would accumulate, leading
to adaptation in the spinal interneurons. This would cause the most active
interneurons to receive the most inhibition, shifting the ``dominant''
activities, and producing larger-amplitude exploratory oscillations.

When a new target is presented, $ACT$ must reset its own activity back to a low value.
Given our requirement to fully implement the controller using neural elements, we needed
a way to detect changes in $S_P$. A unit denominated $CHG$ can detect these changes
using synapses that react to the derivative of the activity in $S_P$ units.
$CHG$ was connected to $ACT$ in order to reset its activity.

More precisely, when inputs from $CHG$ were larger than 0.1, the activity of $ACT$
had dynamics:
\begin{equation}
    \dot{a}(t) = -40 a(t).
\end{equation}
Otherwise it had these dynamics:
\begin{align} \label{eq:act1}
    \dot{a}(t) &= a(t) \big( \sigma(I) - \theta_{ACT} \big), \text{ if }
    \sigma(I) < \theta_{ACT}, \\ \label{eq:act2}
    \tau_{ACT} \dot{a}(t) &= \big( \sigma(I) - \theta_{ACT} \big) \big[1 - a(t) +
    \gamma \dot{\sigma}(I) \big], \text{ otherwise.}
\end{align}
As before, $\sigma(\cdot)$ is a sigmoidal function, and $I$ is the scaled sum
of inputs other than $CHG$. When $\sigma(I)$ is smaller than a threshold
$\theta_{ACT}$ the value of $a$ actually decreases, as this error is deemed
small enough. When $\sigma(I) > \theta_{ACT}$ the activity increases, but
the rate of increase is modulated by a rate of increase
$\dot{\sigma}(I) \equiv \sigma(I) - \sigma(\tilde{I})$, where $\tilde{I}$ is
a low-pass filtered version of $I$. $\gamma$ is a constant parameter.

$CHG$ was a standard sigmoidal unit receiving inputs from $S_P$, with each
synaptic weight obeying this equation:
\begin{equation} \label{eq:chg_syn}
    \omega_j(t) = \alpha | \dot{s}_j(t) | - \omega_j(t),
\end{equation}
where $s_j$ represents the synapse's presynaptic input.

\subsection{Plant, muscles, afferents}
The planar arm was modeled as a compound double pendulum, where both
the arm and forearm were cylinders with 1 kg. of mass. No gravity was
present, and a moderate amount of viscous friction was added at each joint
(3 $\frac{N\ m \ s}{rad}$). The derivation and validation of the double
pendulum's equations can be consulted in a Jupyter notebook included with
Draculab's source code (in the \verb+tests+ folder).

The muscles used a standard Hill-type model, as described in
\cite{shadmehr_computational_2005}, Pg. 99.
The muscle's tension $T$ obeys:
\begin{equation} \label{eq:hill1}
    \dot{T} = \frac{K_{SE}}{b} \left[ g \cdot I + K_{PE} \Delta x + b \dot{x}
    - \left(1 + \frac{K_{PE}}{K_{SE}} \right) T \right],
\end{equation}
where $I$ is the input, $g$ an input gain, $K_{PE}$ the parallel elasticity
constant, $K_{SE}$ the series elasticity constant, $b$ is the damping
constant for the parallel element, $x$ is the length of the muscle, and
$\Delta x = x - x_1^* - x_2^*$. In here, $x_1^*$ is the resting length of the series
element, whereas $x_2^*$ is the resting length of the parallel element.
All resting lengths were calculated from the steady state when the hand was
located at coordinates (0.3, 0.3).

We created a model of the Ia and II afferents using simple
structural elements. This model includes,
for each muscle one dynamic nuclear bag fiber, and one static bag fiber.
Both of these fibers use the same tension equation as the muscle, but
with different parameters. For the static bag fiber:
\begin{equation} \label{eq:hill2}
    \dot{T}^{s} = \frac{K_{SE}^{s}}{b^{s}} \left[
    K_{PE}^{s} \Delta x + b^{s} \dot{x}
    - \left(1 + \frac{K_{PE}^{s}}{K_{SE}^{s}} \right) T^{s} \right].
\end{equation}

The dynamic bag fiber uses the same equation, with the $s$ superscript
replaced by $d$. No inputs were applied to the static or dynamic bag fibers,
so they were removed from these equations. The rest lengths of the static and
dynamic bag fibers where those of their corresponding muscles times factors
$l_0^s, l_0^d$, respectively.

The Ia afferent output is proportional to a linear combination of the
lengths for the serial elements in both  dynamic and static bag fibers.
The II output has two components, one proportional
to the length of the serial element, and one approximately proportional to 
the length of the parallel element, both in the static bag fiber. In practice
this was implemented through the following equations:
\begin{equation} \label{eq:Ia}
    I_a = g_{I_a} \left[ \left(\frac{f_s^{I_a}}{K_{SE}^s} \right) T^s +
    \left(\frac{1-f_s^{I_a}}{K_{SE}^d} \right) T^d \right],
\end{equation}
\begin{equation} \label{eq:II}
    II = g_{II} \left[ \left(\frac{f_s^{II}}{K_{SE}^s} \right) T^s +
    \left(\frac{1-f_s^{II}}{K_{PE}^s} \right) \left(
    T^s - b^s \dot{x} \right) \right].
\end{equation}
In here, $g_{I_a}$ and $g_{II}$ are gain factors. $f_s^{I_a}$ and $f_s^{II}$ 
are constants determining the fraction of $I_a$ and $II$ output that comes from
the serial element.

The model of the Golgi tendon organ producing the Ib outputs was taken
from \cite{lin_neural_2002}. First, a rectified tension was obtained as:
\begin{equation} \label{eq:gto1}
    r = g_{I_b} \log ( T^+ / T_0 + 1).
\end{equation}
$g_{I_b}$ is a gain factor, $T_0$ is a constant that can further alter the 
slope of the tension, and $T^+ = \max(T, 0)$ is the tension, half-rectified.
The $I_b$ afferent output followed dynamics:
\begin{equation} \label{eq:gto2}
    \tau_{I_b} \dot{I}_b = r - I_b.
\end{equation}

\subsection{Static connections} \label{sub:connections}
In all cases the connections to $S_A$ used one-to-one connectivity with the $A$
units driven by the II afferents, whereas connections from $A$ to $M$ and $C$
used all-to-all projections from the units driven by the Ia and Ib afferents.
Projections from $S_A$ to $S_{PA}$ used one-to-one excitatory connections to
the first 6 units, and inhibitory projections to the next 6 units. Projections
from $S_P$ to $S_{PA}$ used the opposite sign from this.

Connections from $S_{PA}$ to $M$ were one-to-one, so
the $j$-th unit in $S_{PA}$ only sent a projection to unit $j$ in $M$.
A variation of this connectivity is presented in the Appendix
(See {\it Variations of the spinal learning model}).

We now explain how we adjusted the synaptic weights of the static network.
To understand the projections from $M$ to $C$ and to the alpha motoneurons it is
useful to remember that each $CE, CI, \alpha$ trio is associated with one
muscle, and the $M$ units also control the error of a single muscle. This error
indicates that the muscle is longer than desired. Thus, the $M$
unit associated with muscle $i$ sent excitatory projections to the 
$CE$ and $\alpha$ units associated with muscle $i$, and to the $CI$ units of the
antagonists of $i$. Additionally, weaker projections were sent to the $CE,
\alpha$ units of muscle $i$'s agonists. Notice that only excitatory connections
were used.

The reverse logic was used to set the connections from $A$ to $C$ and $M$. If
muscle $i$ is tensing or elongating, this can predict an increase in the error
for its antagonists, which is the kind of signal that the input correlation rule
is meant to detect. Therefore, the $Ib$ afferent (signaling tension) of muscle
$i$ sent an excitatory signal to the $CI$ unit associated with muscle $i$, and
to the $CE, \alpha$ units associated with $i$'s antagonists. Moreover, this $Ib$
afferent also sent an excitatory projection to the dual of the $M$ unit
associated with muscle $i$. Connections from $Ia$ afferents (roughly signaling
elongation speed) followed the same pattern, but with slightly smaller 
connection strengths.

\subsection{Rotational dynamics}
\label{sub:rotational_meth}
We explain the method to project the activity of $M$ onto the jPCA plane.
For all units in $M$ we considered the activity during a 0.5 seconds sample
beginning 50 milliseconds after the target onset. 
Unlike \cite{churchland_neural_2012}, we did not apply PCA preprocessing, since
we only have 12 units in $M$.
Let $m_{i,j,k,t}$ be the
activity at time $t$ of the unit $i$ in $M$, when reaching at target $j$ for the
$k$-th repetition. By $m_{i,j,\langle k \rangle, t}$ we denote the average over
all repeated reaches to the same target, and by 
$m_{i,\langle j \rangle,\langle k \rangle, t}$
we indicate averaging over both targets and repetitions. 
The normalized average trace per condition is defined as:
$m_{i,j}(t) \equiv m_{i,j,\langle k \rangle, t} - m_{i,\langle j \rangle,\langle k
\rangle, t}$. Let $I$ stand for the number of units in $M$, $T$ for the number
of time points, and $J$ for the number of targets. Following  
\cite{churchland_neural_2012}, we unroll the set of $m_{i,j}(t)$ values
into a matrix $X \in R^{JT \times I}$, so we may represent the data through a
matrix $M$
that provides the least-squares solution to the problem $\dot{X} = XM$.
This solution comes from the equation $\hat{M} = (X^TX)^{-1}X^T\dot{X}$.
Furthermore, this matrix can be decomposed into symmetric and anti-symmetric
components $M_{symm} = (\hat{M} + \hat{M}^T)/2, M_{skew} = (\hat{M} -
\hat{M}^T)/2$. The jPCA plane comes from the complex conjugate eigenvalues of
$M_{skew}$.

In practice, our source code follows the detailed explanation provided 
in the Supplementary Information of \cite{churchland_neural_2012}, which
reformulates this matrix problem as a vector problem.

\subsection{Parameter search} \label{sub:parameter}
We kept all parameter values in a range where they still made biological sense.
Parameter values that were not constrained by biological data were adjusted
using a genetic algorithm, and particle swarm optimization (PSO). 
We used a separate optimization run for each one of the configurations,
consisting of roughly 30 iterations of the genetic and PSO algorithms, with
populations sizes of 90 and 45 individuals respectively.
After this we manually adjusted the gain of the
control loop by increasing or decreasing the slope of the sigmoidal units in the
$M$ and $S_A$ populations. This is further described in the Appendix
({\it Gain and oscillations} section).

The parameters used can affect the results in the paper. We chose parameters
that minimized either the error during the second half of the 
learning phase, or the error during center-out reaching. Both of these measures
are agnostic to the other results.

\subsection{Preferred Direction vectors}
Next we describe how PD vectors were obtained for the $M$ units.

Let $m_{jk}$ denote the firing rate of the $j$-th $M$ unit when reaching for
the $k$-th target, averaged over 4 seconds, and across
reaches to the same target. We created a function 
$h_j:\mathds{R}^2 \rightarrow \mathds{R}$
that mapped the X,Y coordinates of each target to its corresponding
$m_{jk}$ value, but in the domain of $h_j$ the coordinates were shifted
so the center location was at the origin.

Next we approximated $h_j$ with a plane, using the least squares method,
and obtained a unit vector $u_j$ normal to that plane, starting at the intersection
of the $z$-axis and the plane, and pointing towards the XY plane.
The PD vector was defined as the projection of $u_j$ on the XY plane.

In order to predict the PD vectors, we first obtained for each muscle
the ``direction of maximum contraction,'' verbally described in
panel B of figure \ref{fig:center_out_2A}. More formally, let $l_{ik}$
denote the length of the $i$-th muscle when the hand is at target $k$, 
and let $l^0_i$ denote its length when the
hand is at the center location. With $\bar{r}_k$ we denote the unit
vector with base at the center location, pointing in the direction 
of the $k$-th target. The direction of maximum length change for
the $i$-th muscle comes from the following vector sum:
\begin{equation}
    \bar{v}_i = \sum_{k=1}^8 \left[ \frac{l^0_i - l_{ik}}{l_i^0} \right]_+ \bar{r}_k,
\end{equation}
where $[x]_+ = \max(x, 0)$.

For the $j$-th unit in $M$, its predicted PD vector comes from a linear combination
of the $\bar{v}_i$ vectors. Let the input to this unit be
$\sum_i w_{ji} e_i$,  where $e_i$ is the output of the $i$-th SPF unit (representing
the error in the $i$-th muscle).
The predicted PD vector is:
\begin{equation}
    \bar{d}_j = \sum_{i=0}^5 w_{ji} \bar{v}_i
\end{equation}

To obtain the main axis of the PD distribution, the $i$-th PD vector was
obtained in the polar form $(r_i, \theta_i)$, with $\theta \in (-\pi, \pi]$. We
reflected vectors in the lower half using the rule: $\theta^*_i = \theta_i +
\pi$ if $\theta_i < 0, \theta^*_i = \theta_i$ otherwise. The angle of the main
axis was the angle of the average PD vector using these modified angles:
$\theta_{main} = \arctan \left( \frac{\sum_i r_i \sin\theta^*_i}{\sum_i r_i
\cos\theta^*_i} \right)$.

\subsection{Statistical tests}
To find whether $M$ units were significantly tuned to the reach direction
we used a bootstrap procedure. For each unit we obtained the length of its
PD vector 10,000 times when the identity of the target for each reach was
randomly shuffled. We considered there was significant
tuning when the length of the true PD vector was longer than
99.9\% of these random samples.

To obtain the coefficient of determination for the 
predicted PD angles, let $\theta_{true}^j$ denote the angle of the true PD
for the $j$-th $M$ unit, and $\theta_{pred}^j$ be the angle of its predicted PD.
We obtained residuals for the angles as 
$\epsilon_j = \theta_{true}^j - \theta_{pred}^j$, where this difference is 
actually the angle of the smallest rotation that turns one angle
into the other. Each residual was then scaled by the norm of its corresponding
PD vector, to account for the fact that these were not homogeneous. Denoting
these scaled residuals as $\epsilon^*_j$ the residual sum of squares is
$SS_{res} = \sum_j (\epsilon^*_j)^2$. The total sum of squares was:
$SS_{tot} = \sum_j (\theta_{true}^j - \bar{\theta}_{true})^2$, where
$\bar{\theta}_{true}$ is the mean of the $\theta_{true}^j$ angles. 
The coefficient of determination comes from the usual formula
$R^2 = 1 - \frac{SS_{res}}{SS_{tot}}$.
 
To assess bimodality of the PD distribution we used a version of the Rayleigh
statistic adapted to look for bimodal distributions where the two modes are
oriented at 180 degrees from each other, introduced in 
\cite{lillicrap_preference_2013}. This test consists of finding an modified
Rayleigh $r$ statistic defined as:
\begin{equation}
    r = \frac{1}{N} \left(\left( \sum_{i=1}^N cos(2\phi_i) \right)^2 +
    \left( \sum_{i=1}^N cos(2\phi_i) \right)^2 \right),
\end{equation}
where the $\phi_i$ angles are the angles for the PDs. A bootstrap procedure
is then used, where this $r$ statistic is produced 100,000 times by sampling
from the uniform distribution on the $(0, \pi)$ interval. The PD distribution
was deemed significantly bimodal if its $r$ value was larger than 99.9\% of the
random $r$ values.

We used a bootstrap test to find whether there was statistical significance to the
linear addition of direction fields.
To make this independent of the individual pair of locations stimulated, we
obtained the direction fields for all 15 possible pairs of locations, and for each
pair calculated the mean angle difference between $D(a+b)$ and $D(a)+D(b)$ as described
in the main text. We next obtained the mean of these 15 average angle deviations, to obtain a 
global average angle deviation $\gamma_{global}$.

We then repeated this procedure 400 times when the identities of the stimulation sites
$a,b$ were shuffled, to obtain 400 global average angle deviations $\gamma_{global}^j$.
We declared statistical significance if $\gamma_{global}$ was smaller than 99\% of
the $\gamma_{global}^j$ values.

\section{Acknowledgements}
The authors wish to thank Prof. Kenji Doya for helping revise early versions
of this manuscript.

\bibliography{library}

\begin{thebibliography}{100}
\expandafter\ifx\csname url\endcsname\relax
  \def\url#1{\texttt{#1}}\fi
\expandafter\ifx\csname urlprefix\endcsname\relax\def\urlprefix{URL }\fi
\providecommand{\bibinfo}[2]{#2}
\providecommand{\eprint}[2][]{\url{#2}}

\bibitem{kaas_primate_2003}
\bibinfo{author}{Kaas, J.~H.} \& \bibinfo{author}{Collins, C.~E.}
\newblock \emph{\bibinfo{title}{The {Primate} {Visual} {System}}}
  (\bibinfo{publisher}{CRC Press}, \bibinfo{year}{2003}).
\newblock \bibinfo{note}{Google-Books-ID: rRu38JwnZXoC}.

\bibitem{ballard_hierarchical_2021}
\bibinfo{author}{Ballard, D.~H.} \& \bibinfo{author}{Zhang, R.}
\newblock \bibinfo{title}{The {Hierarchical} {Evolution} in {Human} {Vision}
  {Modeling}}.
\newblock \emph{\bibinfo{journal}{Topics in Cognitive Science}}
  \textbf{\bibinfo{volume}{13}}, \bibinfo{pages}{309--328}
  (\bibinfo{year}{2021}).
\newblock
  \urlprefix\url{https://onlinelibrary.wiley.com/doi/abs/10.1111/tops.12527}.
\newblock \bibinfo{note}{\_eprint:
  https://onlinelibrary.wiley.com/doi/pdf/10.1111/tops.12527}.

\bibitem{chersi_cognitive_2015}
\bibinfo{author}{Chersi, F.} \& \bibinfo{author}{Burgess, N.}
\newblock \bibinfo{title}{The {Cognitive} {Architecture} of {Spatial}
  {Navigation}: {Hippocampal} and {Striatal} {Contributions}}.
\newblock \emph{\bibinfo{journal}{Neuron}} \textbf{\bibinfo{volume}{88}},
  \bibinfo{pages}{64--77} (\bibinfo{year}{2015}).
\newblock
  \urlprefix\url{https://www.sciencedirect.com/science/article/pii/S0896627315007783}.

\bibitem{moser_spatial_2017}
\bibinfo{author}{Moser, E.~I.}, \bibinfo{author}{Moser, M.-B.} \&
  \bibinfo{author}{McNaughton, B.~L.}
\newblock \bibinfo{title}{Spatial representation in the hippocampal formation:
  a history}.
\newblock \emph{\bibinfo{journal}{Nature Neuroscience}}
  \textbf{\bibinfo{volume}{20}}, \bibinfo{pages}{1448--1464}
  (\bibinfo{year}{2017}).
\newblock \urlprefix\url{https://www.nature.com/articles/nn.4653}.
\newblock \bibinfo{note}{Bandiera\_abtest: a Cg\_type: Nature Research Journals
  Number: 11 Primary\_atype: Comments \& Opinion Publisher: Nature Publishing
  Group Subject\_term: Hippocampus;Spatial memory Subject\_term\_id:
  hippocampus;spatial-memory}.

\bibitem{eccles_physiology_1981}
\bibinfo{author}{Eccles, J.~C.}
\newblock \bibinfo{title}{Physiology of {Motor} {Control} in {Man}}.
\newblock \emph{\bibinfo{journal}{Stereotactic and Functional Neurosurgery}}
  \textbf{\bibinfo{volume}{44}}, \bibinfo{pages}{5--15} (\bibinfo{year}{1981}).
\newblock \urlprefix\url{https://www.karger.com/Article/FullText/102178}.
\newblock \bibinfo{note}{Publisher: Karger Publishers}.

\bibitem{loeb_major_2015}
\bibinfo{author}{Loeb, G.~E.} \& \bibinfo{author}{Tsianos, G.~A.}
\newblock \bibinfo{title}{Major remaining gaps in models of sensorimotor
  systems}.
\newblock \emph{\bibinfo{journal}{Frontiers in Computational Neuroscience}}
  \textbf{\bibinfo{volume}{9}} (\bibinfo{year}{2015}).
\newblock
  \urlprefix\url{https://www.frontiersin.org/articles/10.3389/fncom.2015.00070/full}.
\newblock \bibinfo{note}{Publisher: Frontiers}.

\bibitem{shadmehr_computational_2005}
\bibinfo{author}{Shadmehr, R.} \& \bibinfo{author}{Wise, S.~P.}
\newblock \emph{\bibinfo{title}{The {Computational} {Neurobiology} of
  {Reaching} and {Pointing}: {A} {Foundation} for {Motor} {Learning}}}
  (\bibinfo{publisher}{MIT Press}, \bibinfo{year}{2005}).
\newblock \bibinfo{note}{Google-Books-ID: fKeImql1s\_sC}.

\bibitem{arber_connecting_2018}
\bibinfo{author}{Arber, S.} \& \bibinfo{author}{Costa, R.~M.}
\newblock \bibinfo{title}{Connecting neuronal circuits for movement}.
\newblock \emph{\bibinfo{journal}{Science}} \textbf{\bibinfo{volume}{360}},
  \bibinfo{pages}{1403--1404} (\bibinfo{year}{2018}).
\newblock
  \urlprefix\url{https://www.science.org/lookup/doi/10.1126/science.aat5994}.
\newblock \bibinfo{note}{Publisher: American Association for the Advancement of
  Science}.

\bibitem{tanaka_thalamocortical_2018}
\bibinfo{author}{Tanaka, Y.~H.} \emph{et~al.}
\newblock \bibinfo{title}{Thalamocortical {Axonal} {Activity} in {Motor}
  {Cortex} {Exhibits} {Layer}-{Specific} {Dynamics} during {Motor} {Learning}}.
\newblock \emph{\bibinfo{journal}{Neuron}} \textbf{\bibinfo{volume}{100}},
  \bibinfo{pages}{244--258.e12} (\bibinfo{year}{2018}).
\newblock
  \urlprefix\url{https://www.sciencedirect.com/science/article/pii/S0896627318306895}.

\bibitem{hadders-algra_early_2018}
\bibinfo{author}{Hadders-Algra, M.}
\newblock \bibinfo{title}{Early human motor development: {From} variation to
  the ability to vary and adapt}.
\newblock \emph{\bibinfo{journal}{Neuroscience \& Biobehavioral Reviews}}
  \textbf{\bibinfo{volume}{90}}, \bibinfo{pages}{411--427}
  (\bibinfo{year}{2018}).
\newblock
  \urlprefix\url{https://www.sciencedirect.com/science/article/pii/S0149763418300538}.

\bibitem{verduzco-flores_differential_2022}
\bibinfo{author}{Verduzco-Flores, S.}, \bibinfo{author}{Dorrell, W.} \&
  \bibinfo{author}{De~Schutter, E.}
\newblock \bibinfo{title}{A differential {Hebbian} framework for
  biologically-plausible motor control}.
\newblock \emph{\bibinfo{journal}{Neural Networks}}
  \textbf{\bibinfo{volume}{150}}, \bibinfo{pages}{237--258}
  (\bibinfo{year}{2022}).
\newblock
  \urlprefix\url{https://www.sciencedirect.com/science/article/pii/S0893608022000727}.

\bibitem{woods_homeostasis_2007}
\bibinfo{author}{Woods, S.~C.} \& \bibinfo{author}{Ramsay, D.~S.}
\newblock \bibinfo{title}{Homeostasis: {Beyond} {Curt} {Richter}}.
\newblock \emph{\bibinfo{journal}{Appetite}} \textbf{\bibinfo{volume}{49}},
  \bibinfo{pages}{388--398} (\bibinfo{year}{2007}).
\newblock
  \urlprefix\url{https://www.sciencedirect.com/science/article/pii/S019566630700270X}.

\bibitem{bizzi_new_2000}
\bibinfo{author}{Bizzi, E.}, \bibinfo{author}{Tresch, M.~C.},
  \bibinfo{author}{Saltiel, P.} \& \bibinfo{author}{d'Avella, A.}
\newblock \bibinfo{title}{New perspectives on spinal motor systems}.
\newblock \emph{\bibinfo{journal}{Nature Reviews Neuroscience}}
  \textbf{\bibinfo{volume}{1}}, \bibinfo{pages}{101--108}
  (\bibinfo{year}{2000}).
\newblock \urlprefix\url{https://www.nature.com/articles/35039000}.

\bibitem{lemon_descending_2008}
\bibinfo{author}{Lemon, R.~N.}
\newblock \bibinfo{title}{Descending {Pathways} in {Motor} {Control}}.
\newblock \emph{\bibinfo{journal}{Annual Review of Neuroscience}}
  \textbf{\bibinfo{volume}{31}}, \bibinfo{pages}{195--218}
  (\bibinfo{year}{2008}).
\newblock
  \urlprefix\url{https://www.annualreviews.org/doi/10.1146/annurev.neuro.31.060407.125547}.

\bibitem{arber_motor_2012}
\bibinfo{author}{Arber, S.}
\newblock \bibinfo{title}{Motor {Circuits} in {Action}: {Specification},
  {Connectivity}, and {Function}}.
\newblock \emph{\bibinfo{journal}{Neuron}} \textbf{\bibinfo{volume}{74}},
  \bibinfo{pages}{975--989} (\bibinfo{year}{2012}).
\newblock
  \urlprefix\url{http://www.sciencedirect.com/science/article/pii/S0896627312004771}.

\bibitem{asante_differential_2013}
\bibinfo{author}{Asante, C.~O.} \& \bibinfo{author}{Martin, J.~H.}
\newblock \bibinfo{title}{Differential {Joint}-{Specific} {Corticospinal}
  {Tract} {Projections} within the {Cervical} {Enlargement}}.
\newblock \emph{\bibinfo{journal}{PLOS ONE}} \textbf{\bibinfo{volume}{8}},
  \bibinfo{pages}{e74454} (\bibinfo{year}{2013}).
\newblock
  \urlprefix\url{https://journals.plos.org/plosone/article?id=10.1371/journal.pone.0074454}.

\bibitem{alstermark_circuits_2012}
\bibinfo{author}{Alstermark, B.} \& \bibinfo{author}{Isa, T.}
\newblock \bibinfo{title}{Circuits for {Skilled} {Reaching} and {Grasping}}.
\newblock \emph{\bibinfo{journal}{Annual Review of Neuroscience}}
  \textbf{\bibinfo{volume}{35}}, \bibinfo{pages}{559--578}
  (\bibinfo{year}{2012}).
\newblock \urlprefix\url{https://doi.org/10.1146/annurev-neuro-062111-150527}.

\bibitem{jankowska_spinal_2013}
\bibinfo{author}{Jankowska, E.}
\newblock \bibinfo{title}{Spinal {Interneurons}}.
\newblock In \bibinfo{editor}{Pfaff, D.~W.} (ed.)
  \emph{\bibinfo{booktitle}{Neuroscience in the 21st {Century}: {From} {Basic}
  to {Clinical}}}, \bibinfo{pages}{1063--1099} (\bibinfo{publisher}{Springer},
  \bibinfo{address}{New York, NY}, \bibinfo{year}{2013}).
\newblock \urlprefix\url{https://doi.org/10.1007/978-1-4614-1997-634}.

\bibitem{wang_deconstruction_2017}
\bibinfo{author}{Wang, X.} \emph{et~al.}
\newblock \bibinfo{title}{Deconstruction of {Corticospinal} {Circuits} for
  {Goal}-{Directed} {Motor} {Skills}}.
\newblock \emph{\bibinfo{journal}{Cell}} \textbf{\bibinfo{volume}{171}},
  \bibinfo{pages}{440--455.e14} (\bibinfo{year}{2017}).
\newblock
  \urlprefix\url{http://www.sciencedirect.com/science/article/pii/S009286741730939X}.

\bibitem{ueno_corticospinal_2018}
\bibinfo{author}{Ueno, M.} \emph{et~al.}
\newblock \bibinfo{title}{Corticospinal {Circuits} from the {Sensory} and
  {Motor} {Cortices} {Differentially} {Regulate} {Skilled} {Movements} through
  {Distinct} {Spinal} {Interneurons}}.
\newblock \emph{\bibinfo{journal}{Cell Reports}} \textbf{\bibinfo{volume}{23}},
  \bibinfo{pages}{1286--1300.e7} (\bibinfo{year}{2018}).
\newblock
  \urlprefix\url{http://www.sciencedirect.com/science/article/pii/S2211124718305254}.

\bibitem{wolpaw_adaptive_1983}
\bibinfo{author}{Wolpaw, J.~R.}, \bibinfo{author}{Kieffer, V.~A.},
  \bibinfo{author}{Seegal, R.~F.}, \bibinfo{author}{Braitman, D.~J.} \&
  \bibinfo{author}{Sanders, M.~G.}
\newblock \bibinfo{title}{Adaptive plasticity in the spinal stretch reflex}.
\newblock \emph{\bibinfo{journal}{Brain Research}}
  \textbf{\bibinfo{volume}{267}}, \bibinfo{pages}{196--200}
  (\bibinfo{year}{1983}).
\newblock
  \urlprefix\url{https://www.sciencedirect.com/science/article/pii/0006899383910594}.

\bibitem{grau_learning_2014}
\bibinfo{author}{Grau, J.~W.}
\newblock \bibinfo{title}{Learning from the spinal cord: {How} the study of
  spinal cord plasticity informs our view of learning}.
\newblock \emph{\bibinfo{journal}{Neurobiology of Learning and Memory}}
  \textbf{\bibinfo{volume}{108}}, \bibinfo{pages}{155--171}
  (\bibinfo{year}{2014}).
\newblock
  \urlprefix\url{https://www.sciencedirect.com/science/article/pii/S1074742713001445}.

\bibitem{meyer-lohmann_dominance_1986}
\bibinfo{author}{Meyer-Lohmann, J.}, \bibinfo{author}{Christakos, C.~N.} \&
  \bibinfo{author}{Wolf, H.}
\newblock \bibinfo{title}{Dominance of the short-latency component in
  perturbation induced electromyographic responses of long-trained monkeys}.
\newblock \emph{\bibinfo{journal}{Experimental Brain Research}}
  \textbf{\bibinfo{volume}{64}}, \bibinfo{pages}{393--399}
  (\bibinfo{year}{1986}).
\newblock \urlprefix\url{https://doi.org/10.1007/BF00340475}.

\bibitem{wolpaw_complex_1997}
\bibinfo{author}{Wolpaw, J.~R.}
\newblock \bibinfo{title}{The complex structure of a simple memory}.
\newblock \emph{\bibinfo{journal}{Trends in Neurosciences}}
  \textbf{\bibinfo{volume}{20}}, \bibinfo{pages}{588--594}
  (\bibinfo{year}{1997}).
\newblock
  \urlprefix\url{http://www.sciencedirect.com/science/article/pii/S0166223697011338}.

\bibitem{norton_acquisition_2018}
\bibinfo{author}{Norton, J.~J.} \& \bibinfo{author}{Wolpaw, J.~R.}
\newblock \bibinfo{title}{Acquisition, maintenance, and therapeutic use of a
  simple motor skill}.
\newblock \emph{\bibinfo{journal}{Current Opinion in Behavioral Sciences}}
  \textbf{\bibinfo{volume}{20}}, \bibinfo{pages}{138--144}
  (\bibinfo{year}{2018}).
\newblock
  \urlprefix\url{http://www.sciencedirect.com/science/article/pii/S235215461730219X}.

\bibitem{georgopoulos_relations_1982}
\bibinfo{author}{Georgopoulos, A.}, \bibinfo{author}{Kalaska, J.},
  \bibinfo{author}{Caminiti, R.} \& \bibinfo{author}{Massey, J.}
\newblock \bibinfo{title}{On the relations between the direction of
  two-dimensional arm movements and cell discharge in primate motor cortex}.
\newblock \emph{\bibinfo{journal}{The Journal of Neuroscience}}
  \textbf{\bibinfo{volume}{2}}, \bibinfo{pages}{1527--1537}
  (\bibinfo{year}{1982}).
\newblock
  \urlprefix\url{https://www.ncbi.nlm.nih.gov/pmc/articles/PMC6564361/}.

\bibitem{georgopoulos_neuronal_1986}
\bibinfo{author}{Georgopoulos, A.~P.}, \bibinfo{author}{Schwartz, A.~B.} \&
  \bibinfo{author}{Kettner, R.~E.}
\newblock \bibinfo{title}{Neuronal population coding of movement direction}.
\newblock \emph{\bibinfo{journal}{Science}} \textbf{\bibinfo{volume}{233}},
  \bibinfo{pages}{1416--1419} (\bibinfo{year}{1986}).
\newblock \urlprefix\url{https://science.sciencemag.org/content/233/4771/1416}.
\newblock \bibinfo{note}{Publisher: American Association for the Advancement of
  Science Section: Reports}.

\bibitem{kakei_muscle_1999}
\bibinfo{author}{Kakei, S.}, \bibinfo{author}{Hoffman, D.~S.} \&
  \bibinfo{author}{Strick, P.~L.}
\newblock \bibinfo{title}{Muscle and {Movement} {Representations} in the
  {Primary} {Motor} {Cortex}}.
\newblock \emph{\bibinfo{journal}{Science}}  (\bibinfo{year}{1999}).
\newblock
  \urlprefix\url{https://www.science.org/doi/abs/10.1126/science.285.5436.2136}.
\newblock \bibinfo{note}{Publisher: American Association for the Advancement of
  Science}.

\bibitem{truccolo_primary_2008}
\bibinfo{author}{Truccolo, W.}, \bibinfo{author}{Friehs, G.~M.},
  \bibinfo{author}{Donoghue, J.~P.} \& \bibinfo{author}{Hochberg, L.~R.}
\newblock \bibinfo{title}{Primary {Motor} {Cortex} {Tuning} to {Intended}
  {Movement} {Kinematics} in {Humans} with {Tetraplegia}}.
\newblock \emph{\bibinfo{journal}{Journal of Neuroscience}}
  \textbf{\bibinfo{volume}{28}}, \bibinfo{pages}{1163--1178}
  (\bibinfo{year}{2008}).
\newblock \urlprefix\url{https://www.jneurosci.org/content/28/5/1163}.
\newblock \bibinfo{note}{Publisher: Society for Neuroscience Section:
  Articles}.

\bibitem{kalaska_intention_2009}
\bibinfo{author}{Kalaska, J.~F.}
\newblock \bibinfo{title}{From intention to action: motor cortex and the
  control of reaching movements}.
\newblock \emph{\bibinfo{journal}{Advances in Experimental Medicine and
  Biology}} \textbf{\bibinfo{volume}{629}}, \bibinfo{pages}{139--178}
  (\bibinfo{year}{2009}).

\bibitem{georgopoulos_local_2007}
\bibinfo{author}{Georgopoulos, A.~P.} \& \bibinfo{author}{Stefanis, C.~N.}
\newblock \bibinfo{title}{Local shaping of function in the motor cortex:
  {Motor} contrast, directional tuning}.
\newblock \emph{\bibinfo{journal}{Brain Research Reviews}}
  \textbf{\bibinfo{volume}{55}}, \bibinfo{pages}{383--389}
  (\bibinfo{year}{2007}).
\newblock
  \urlprefix\url{https://www.sciencedirect.com/science/article/pii/S0165017307000720}.

\bibitem{harrison_towards_2013}
\bibinfo{author}{Harrison, T.~C.} \& \bibinfo{author}{Murphy, T.~H.}
\newblock \bibinfo{title}{Towards a circuit mechanism for movement tuning in
  motor cortex}.
\newblock \emph{\bibinfo{journal}{Frontiers in Neural Circuits}}
  \textbf{\bibinfo{volume}{6}} (\bibinfo{year}{2013}).
\newblock
  \urlprefix\url{https://www.frontiersin.org/articles/10.3389/fncir.2012.00127/full}.
\newblock \bibinfo{note}{Publisher: Frontiers}.

\bibitem{tanaka_modeling_2016}
\bibinfo{author}{Tanaka, H.}
\newblock \bibinfo{title}{Modeling the motor cortex: {Optimality}, recurrent
  neural networks, and spatial dynamics}.
\newblock \emph{\bibinfo{journal}{Neuroscience Research}}
  \textbf{\bibinfo{volume}{104}}, \bibinfo{pages}{64--71}
  (\bibinfo{year}{2016}).
\newblock
  \urlprefix\url{https://www.sciencedirect.com/science/article/pii/S0168010215002631}.

\bibitem{morrow_prediction_2003}
\bibinfo{author}{Morrow, M.~M.} \& \bibinfo{author}{Miller, L.~E.}
\newblock \bibinfo{title}{Prediction of {Muscle} {Activity} by {Populations} of
  {Sequentially} {Recorded} {Primary} {Motor} {Cortex} {Neurons}}.
\newblock \emph{\bibinfo{journal}{Journal of Neurophysiology}}
  \textbf{\bibinfo{volume}{89}}, \bibinfo{pages}{2279--2288}
  (\bibinfo{year}{2003}).
\newblock
  \urlprefix\url{https://journals.physiology.org/doi/full/10.1152/jn.00632.2002}.
\newblock \bibinfo{note}{Publisher: American Physiological Society}.

\bibitem{todorov_direct_2000}
\bibinfo{author}{Todorov, E.}
\newblock \bibinfo{title}{Direct cortical control of muscle activation in
  voluntary arm movements: a model}.
\newblock \emph{\bibinfo{journal}{Nature Neuroscience}}
  \textbf{\bibinfo{volume}{3}}, \bibinfo{pages}{391--398}
  (\bibinfo{year}{2000}).
\newblock \urlprefix\url{https://www.nature.com/articles/nn0400\_391}.

\bibitem{scott_dissociation_2001}
\bibinfo{author}{Scott, S.~H.}, \bibinfo{author}{Gribble, P.~L.},
  \bibinfo{author}{Graham, K.~M.} \& \bibinfo{author}{Cabel, D.~W.}
\newblock \bibinfo{title}{Dissociation between hand motion and population
  vectors from neural activity in motor cortex}.
\newblock \emph{\bibinfo{journal}{Nature}} \textbf{\bibinfo{volume}{413}},
  \bibinfo{pages}{161--165} (\bibinfo{year}{2001}).
\newblock \urlprefix\url{https://www.nature.com/articles/35093102}.

\bibitem{rokni_motor_2007}
\bibinfo{author}{Rokni, U.}, \bibinfo{author}{Richardson, A.~G.},
  \bibinfo{author}{Bizzi, E.} \& \bibinfo{author}{Seung, H.~S.}
\newblock \bibinfo{title}{Motor {Learning} with {Unstable}
  {Neural} {Representations}}.
\newblock \emph{\bibinfo{journal}{Neuron}} \textbf{\bibinfo{volume}{54}},
  \bibinfo{pages}{653--666} (\bibinfo{year}{2007}).
\newblock
  \urlprefix\url{https://www.sciencedirect.com/science/article/pii/S0896627307003339}.

\bibitem{padoa-schioppa_neuronal_2004}
\bibinfo{author}{Padoa-Schioppa, C.}, \bibinfo{author}{Li, C.-S.~R.} \&
  \bibinfo{author}{Bizzi, E.}
\newblock \bibinfo{title}{Neuronal {Activity} in the {Supplementary} {Motor}
  {Area} of {Monkeys} {Adapting} to a {New} {Dynamic} {Environment}}.
\newblock \emph{\bibinfo{journal}{Journal of Neurophysiology}}
  \textbf{\bibinfo{volume}{91}}, \bibinfo{pages}{449--473}
  (\bibinfo{year}{2004}).
\newblock
  \urlprefix\url{https://journals.physiology.org/doi/full/10.1152/jn.00876.2002}.
\newblock \bibinfo{note}{Publisher: American Physiological Society}.

\bibitem{shenoy_cortical_2013}
\bibinfo{author}{Shenoy, K.~V.}, \bibinfo{author}{Sahani, M.} \&
  \bibinfo{author}{Churchland, M.~M.}
\newblock \bibinfo{title}{Cortical {Control} of {Arm} {Movements}: {A}
  {Dynamical} {Systems} {Perspective}}.
\newblock \emph{\bibinfo{journal}{Annual Review of Neuroscience}}
  \textbf{\bibinfo{volume}{36}}, \bibinfo{pages}{337--359}
  (\bibinfo{year}{2013}).
\newblock
  \urlprefix\url{http://www.annualreviews.org/doi/10.1146/annurev-neuro-062111-150509}.

\bibitem{sussillo_neural_2015}
\bibinfo{author}{Sussillo, D.}, \bibinfo{author}{Churchland, M.~M.},
  \bibinfo{author}{Kaufman, M.~T.} \& \bibinfo{author}{Shenoy, K.~V.}
\newblock \bibinfo{title}{A neural network that finds a naturalistic solution
  for the production of muscle activity}.
\newblock \emph{\bibinfo{journal}{Nature Neuroscience}}
  \textbf{\bibinfo{volume}{18}}, \bibinfo{pages}{1025--1033}
  (\bibinfo{year}{2015}).
\newblock \urlprefix\url{https://www.nature.com/articles/nn.4042}.

\bibitem{churchland_neural_2012}
\bibinfo{author}{Churchland, M.} \emph{et~al.}
\newblock \bibinfo{title}{Neural population dynamics during reaching}.
\newblock \emph{\bibinfo{journal}{Nature}} \textbf{\bibinfo{volume}{487}},
  \bibinfo{pages}{51--56} (\bibinfo{year}{2012}).
\newblock
  \urlprefix\url{https://www.ncbi.nlm.nih.gov/pmc/articles/PMC3393826/}.

\bibitem{schoner_reaching_2018}
\bibinfo{author}{Sch{\"o}ner, G.}, \bibinfo{author}{Tek{\"u}lve, J.} \&
  \bibinfo{author}{Zibner, S.}
\newblock \bibinfo{title}{Reaching for objects: a neural process account in a
  developmental perspective}.
\newblock In \emph{\bibinfo{booktitle}{Reach-to-Grasp Behavior}},
  \bibinfo{pages}{281--318} (\bibinfo{publisher}{Routledge},
  \bibinfo{year}{2018}).

\bibitem{valero-cuevas_mathematical_2009}
\bibinfo{author}{Valero-Cuevas, F.~J.}
\newblock \bibinfo{title}{A mathematical approach to the mechanical
  capabilities of limbs and fingers}.
\newblock \emph{\bibinfo{journal}{Advances in Experimental Medicine and
  Biology}} \textbf{\bibinfo{volume}{629}}, \bibinfo{pages}{619--633}
  (\bibinfo{year}{2009}).

\bibitem{keener_principles_1995}
\bibinfo{author}{Keener, J.~P.}
\newblock \emph{\bibinfo{title}{Principles {Of} {Applied} {Mathematics}:
  {Transformation} {And} {Approximation}}} (\bibinfo{publisher}{Avalon
  Publishing}, \bibinfo{year}{1995}).
\newblock \bibinfo{note}{Google-Books-ID: 5nlQAAAAMAAJ}.

\bibitem{giszter_motor_2015}
\bibinfo{author}{Giszter, S.~F.}
\newblock \bibinfo{title}{Motor primitives—new data and future questions}.
\newblock \emph{\bibinfo{journal}{Current Opinion in Neurobiology}}
  \textbf{\bibinfo{volume}{33}}, \bibinfo{pages}{156--165}
  (\bibinfo{year}{2015}).
\newblock
  \urlprefix\url{http://www.sciencedirect.com/science/article/pii/S095943881500077X}.

\bibitem{mussa-ivaldi_motor_2000}
\bibinfo{author}{Mussa-Ivaldi, F.~A.} \& \bibinfo{author}{Bizzi, E.}
\newblock \bibinfo{title}{Motor {Learning} through the {Combination} of
  {Primitives}}.
\newblock \emph{\bibinfo{journal}{Philosophical Transactions: Biological
  Sciences}} \textbf{\bibinfo{volume}{355}}, \bibinfo{pages}{1755--1769}
  (\bibinfo{year}{2000}).
\newblock \urlprefix\url{https://www.jstor.org/stable/3066920}.
\newblock \bibinfo{note}{Publisher: The Royal Society}.

\bibitem{bizzi_computations_1991}
\bibinfo{author}{Bizzi, E.}, \bibinfo{author}{Mussa-Ivaldi, F.~A.} \&
  \bibinfo{author}{Giszter, S.}
\newblock \bibinfo{title}{Computations underlying the execution of movement: a
  biological perspective}.
\newblock \emph{\bibinfo{journal}{Science}} \textbf{\bibinfo{volume}{253}},
  \bibinfo{pages}{287--291} (\bibinfo{year}{1991}).
\newblock \urlprefix\url{https://science.sciencemag.org/content/253/5017/287}.
\newblock \bibinfo{note}{Publisher: American Association for the Advancement of
  Science Section: Articles}.

\bibitem{kelso_synergies_2009}
\bibinfo{author}{Kelso, J. A.~S.}
\newblock \bibinfo{title}{Synergies: {Atoms} of {Brain} and {Behavior}}.
\newblock In \bibinfo{editor}{Sternad, D.} (ed.)
  \emph{\bibinfo{booktitle}{Progress in {Motor} {Control}: {A}
  {Multidisciplinary} {Perspective}}}, Advances in {Experimental} {Medicine}
  and {Biology}, \bibinfo{pages}{83--91} (\bibinfo{publisher}{Springer US},
  \bibinfo{address}{Boston, MA}, \bibinfo{year}{2009}).
\newblock \urlprefix\url{https://doi.org/10.1007/978-0-387-77064-2\_5}.

\bibitem{bruton_synergies_2018}
\bibinfo{author}{Bruton, M.} \& \bibinfo{author}{O’Dwyer, N.}
\newblock \bibinfo{title}{Synergies in coordination: a comprehensive overview
  of neural, computational, and behavioral approaches}.
\newblock \emph{\bibinfo{journal}{Journal of Neurophysiology}}
  \textbf{\bibinfo{volume}{120}}, \bibinfo{pages}{2761--2774}
  (\bibinfo{year}{2018}).
\newblock
  \urlprefix\url{https://journals.physiology.org/doi/full/10.1152/jn.00052.2018}.
\newblock \bibinfo{note}{Publisher: American Physiological Society}.

\bibitem{giszter_convergent_1993}
\bibinfo{author}{Giszter, S.~F.}, \bibinfo{author}{Mussa-Ivaldi, F.~A.} \&
  \bibinfo{author}{Bizzi, E.}
\newblock \bibinfo{title}{Convergent force fields organized in the frog's
  spinal cord}.
\newblock \emph{\bibinfo{journal}{Journal of Neuroscience}}
  \textbf{\bibinfo{volume}{13}}, \bibinfo{pages}{467--491}
  (\bibinfo{year}{1993}).
\newblock \urlprefix\url{https://www.jneurosci.org/content/13/2/467}.
\newblock \bibinfo{note}{Publisher: Society for Neuroscience Section:
  Articles}.

\bibitem{mussa-ivaldi_linear_1994}
\bibinfo{author}{Mussa-Ivaldi, F.~A.}, \bibinfo{author}{Giszter, S.~F.} \&
  \bibinfo{author}{Bizzi, E.}
\newblock \bibinfo{title}{Linear combinations of primitives in vertebrate motor
  control}.
\newblock \emph{\bibinfo{journal}{Proceedings of the National Academy of
  Sciences}} \textbf{\bibinfo{volume}{91}}, \bibinfo{pages}{7534--7538}
  (\bibinfo{year}{1994}).
\newblock \urlprefix\url{https://www.pnas.org/content/91/16/7534}.
\newblock \bibinfo{note}{Publisher: National Academy of Sciences Section:
  Research Article}.

\bibitem{yaron_forelimb_2020}
\bibinfo{author}{Yaron, A.}, \bibinfo{author}{Kowalski, D.},
  \bibinfo{author}{Yaguchi, H.}, \bibinfo{author}{Takei, T.} \&
  \bibinfo{author}{Seki, K.}
\newblock \bibinfo{title}{Forelimb force direction and magnitude independently
  controlled by spinal modules in the macaque}.
\newblock \emph{\bibinfo{journal}{Proceedings of the National Academy of
  Sciences}} \textbf{\bibinfo{volume}{117}}, \bibinfo{pages}{27655--27666}
  (\bibinfo{year}{2020}).
\newblock
  \urlprefix\url{https://www.pnas.org/doi/full/10.1073/pnas.1919253117}.
\newblock \bibinfo{note}{Publisher: Proceedings of the National Academy of
  Sciences}.

\bibitem{tresch_case_2009}
\bibinfo{author}{Tresch, M.~C.} \& \bibinfo{author}{Jarc, A.}
\newblock \bibinfo{title}{The case for and against muscle synergies}.
\newblock \emph{\bibinfo{journal}{Current Opinion in Neurobiology}}
  \textbf{\bibinfo{volume}{19}}, \bibinfo{pages}{601--607}
  (\bibinfo{year}{2009}).
\newblock
  \urlprefix\url{https://www.sciencedirect.com/science/article/pii/S095943880900124X}.

\bibitem{de_rugy_are_2013}
\bibinfo{author}{de~Rugy, A.}, \bibinfo{author}{Loeb, G.} \&
  \bibinfo{author}{Carroll, T.}
\newblock \bibinfo{title}{Are muscle synergies useful for neural control?}
\newblock \emph{\bibinfo{journal}{Frontiers in Computational Neuroscience}}
  \textbf{\bibinfo{volume}{0}} (\bibinfo{year}{2013}).
\newblock
  \urlprefix\url{https://www.frontiersin.org/articles/10.3389/fncom.2013.00019/full}.
\newblock \bibinfo{note}{Publisher: Frontiers}.

\bibitem{bizzi_neural_2013}
\bibinfo{author}{Bizzi, E.} \& \bibinfo{author}{Cheung, V.~C.}
\newblock \bibinfo{title}{The neural origin of muscle synergies}.
\newblock \emph{\bibinfo{journal}{Frontiers in Computational Neuroscience}}
  \textbf{\bibinfo{volume}{7}} (\bibinfo{year}{2013}).
\newblock
  \urlprefix\url{https://www.frontiersin.org/articles/10.3389/fncom.2013.00051/full}.
\newblock \bibinfo{note}{Publisher: Frontiers}.

\bibitem{levine_identification_2014}
\bibinfo{author}{Levine, A.~J.} \emph{et~al.}
\newblock \bibinfo{title}{Identification of a cellular node for motor control
  pathways}.
\newblock \emph{\bibinfo{journal}{Nature neuroscience}}
  \textbf{\bibinfo{volume}{17}}, \bibinfo{pages}{586--593}
  (\bibinfo{year}{2014}).
\newblock
  \urlprefix\url{https://www.ncbi.nlm.nih.gov/pmc/articles/PMC4569558/}.

\bibitem{takei_neural_2017}
\bibinfo{author}{Takei, T.}, \bibinfo{author}{Confais, J.},
  \bibinfo{author}{Tomatsu, S.}, \bibinfo{author}{Oya, T.} \&
  \bibinfo{author}{Seki, K.}
\newblock \bibinfo{title}{Neural basis for hand muscle synergies in the primate
  spinal cord}.
\newblock \emph{\bibinfo{journal}{Proceedings of the National Academy of
  Sciences}} \textbf{\bibinfo{volume}{114}}, \bibinfo{pages}{8643--8648}
  (\bibinfo{year}{2017}).
\newblock \urlprefix\url{https://www.pnas.org/content/114/32/8643}.
\newblock \bibinfo{note}{Publisher: National Academy of Sciences Section:
  Biological Sciences}.

\bibitem{pulvermuller_biological_2021}
\bibinfo{author}{Pulvermüller, F.}, \bibinfo{author}{Tomasello, R.},
  \bibinfo{author}{Henningsen-Schomers, M.~R.} \& \bibinfo{author}{Wennekers,
  T.}
\newblock \bibinfo{title}{Biological constraints on neural network models of
  cognitive function}.
\newblock \emph{\bibinfo{journal}{Nature Reviews Neuroscience}}
  \bibinfo{pages}{1--15} (\bibinfo{year}{2021}).
\newblock \urlprefix\url{https://www.nature.com/articles/s41583-021-00473-5}.
\newblock \bibinfo{note}{Bandiera\_abtest: a Cg\_type: Nature Research Journals
  Primary\_atype: Reviews Publisher: Nature Publishing Group Subject\_term:
  Cognitive neuroscience;Human behaviour;Network models Subject\_term\_id:
  cognitive-neuroscience;human-behaviour;network-models}.

\bibitem{oreilly_six_1998}
\bibinfo{author}{O'Reilly, R.~C.}
\newblock \bibinfo{title}{Six principles for biologically based computational
  models of cortical cognition}.
\newblock \emph{\bibinfo{journal}{Trends in Cognitive Sciences}}
  \textbf{\bibinfo{volume}{2}}, \bibinfo{pages}{455--462}
  (\bibinfo{year}{1998}).
\newblock
  \urlprefix\url{http://www.sciencedirect.com/science/article/pii/S1364661398012418}.

\bibitem{richards_deep_2019}
\bibinfo{author}{Richards, B.~A.} \emph{et~al.}
\newblock \bibinfo{title}{A deep learning framework for neuroscience}.
\newblock \emph{\bibinfo{journal}{Nature Neuroscience}}
  \textbf{\bibinfo{volume}{22}}, \bibinfo{pages}{1761--1770}
  (\bibinfo{year}{2019}).
\newblock \urlprefix\url{https://www.nature.com/articles/s41593-019-0520-2}.
\newblock \bibinfo{note}{Bandiera\_abtest: a Cg\_type: Nature Research Journals
  Number: 11 Primary\_atype: Reviews Publisher: Nature Publishing Group
  Subject\_term: Learning algorithms;Machine learning;Neural circuits
  Subject\_term\_id: learning-algorithms;machine-learning;neural-circuit}.

\bibitem{powers_feedback:_1973}
\bibinfo{author}{Powers, W.~T.}
\newblock \bibinfo{title}{Feedback: {Beyond} {Behaviorism} {Stimulus}-response
  laws are wholly predictable within a control-system model of behavioral
  organization}.
\newblock \emph{\bibinfo{journal}{Science}} \textbf{\bibinfo{volume}{179}},
  \bibinfo{pages}{351--356} (\bibinfo{year}{1973}).
\newblock \urlprefix\url{http://www.sciencemag.org/content/179/4071/351}.

\bibitem{powers_behavior:_2005}
\bibinfo{author}{Powers, W.~T.}
\newblock \emph{\bibinfo{title}{Behavior: {The} {Control} of {Perception} (2nd
  ed. rev. \& exp.)}}, vol. \bibinfo{volume}{xiv}
  (\bibinfo{publisher}{Benchmark Press}, \bibinfo{address}{New Canaan, CT, US},
  \bibinfo{year}{2005}).

\bibitem{adams_predictions_2013}
\bibinfo{author}{Adams, R.~A.}, \bibinfo{author}{Shipp, S.} \&
  \bibinfo{author}{Friston, K.~J.}
\newblock \bibinfo{title}{Predictions not commands: active inference in the
  motor system}.
\newblock \emph{\bibinfo{journal}{Brain Structure and Function}}
  \textbf{\bibinfo{volume}{218}}, \bibinfo{pages}{611--643}
  (\bibinfo{year}{2013}).
\newblock \urlprefix\url{http://link.springer.com/10.1007/s00429-012-0475-5}.

\bibitem{mileusnic_mathematical_2006}
\bibinfo{author}{Mileusnic, M.~P.}, \bibinfo{author}{Brown, I.~E.},
  \bibinfo{author}{Lan, N.} \& \bibinfo{author}{Loeb, G.~E.}
\newblock \bibinfo{title}{Mathematical {Models} of {Proprioceptors}. {I}.
  {Control} and {Transduction} in the {Muscle} {Spindle}}.
\newblock \emph{\bibinfo{journal}{Journal of Neurophysiology}}
  \textbf{\bibinfo{volume}{96}}, \bibinfo{pages}{1772--1788}
  (\bibinfo{year}{2006}).
\newblock \urlprefix\url{https://www.physiology.org/doi/10.1152/jn.00868.2005}.

\bibitem{mountcastle_columnar_1997}
\bibinfo{author}{Mountcastle, V.~B.}
\newblock \bibinfo{title}{The columnar organization of the neocortex.}
\newblock \emph{\bibinfo{journal}{Brain}} \textbf{\bibinfo{volume}{120}},
  \bibinfo{pages}{701--722} (\bibinfo{year}{1997}).
\newblock
  \urlprefix\url{https://academic.oup.com/brain/article/120/4/701/372118}.

\bibitem{porr_strongly_2006}
\bibinfo{author}{Porr, B.} \& \bibinfo{author}{W\"{o}rg\"{o}tter, F.}
\newblock \bibinfo{title}{Strongly {Improved} {Stability} and {Faster}
  {Convergence} of {Temporal} {Sequence} {Learning} by {Using} {Input}
  {Correlations} {Only}}.
\newblock \emph{\bibinfo{journal}{Neural Computation}}
  \textbf{\bibinfo{volume}{18}}, \bibinfo{pages}{1380--1412}
  (\bibinfo{year}{2006}).
\newblock \urlprefix\url{https://doi.org/10.1162/neco.2006.18.6.1380}.

\bibitem{shafi_variability_2007}
\bibinfo{author}{Shafi, M.} \emph{et~al.}
\newblock \bibinfo{title}{Variability in neuronal activity in primate cortex
  during working memory tasks}.
\newblock \emph{\bibinfo{journal}{Neuroscience}}
  \textbf{\bibinfo{volume}{146}}, \bibinfo{pages}{1082--1108}
  (\bibinfo{year}{2007}).
\newblock
  \urlprefix\url{http://www.sciencedirect.com/science/article/pii/S0306452206017593}.

\bibitem{steinmetz_distributed_2019}
\bibinfo{author}{Steinmetz, N.~A.}, \bibinfo{author}{Zatka-Haas, P.},
  \bibinfo{author}{Carandini, M.} \& \bibinfo{author}{Harris, K.~D.}
\newblock \bibinfo{title}{Distributed coding of choice, action and engagement
  across the mouse brain}.
\newblock \emph{\bibinfo{journal}{Nature}} \textbf{\bibinfo{volume}{576}},
  \bibinfo{pages}{266--273} (\bibinfo{year}{2019}).
\newblock \urlprefix\url{https://www.nature.com/articles/s41586-019-1787-x}.
\newblock \bibinfo{note}{Number: 7786 Publisher: Nature Publishing Group}.

\bibitem{najafi_excitatory_2020}
\bibinfo{author}{Najafi, F.} \emph{et~al.}
\newblock \bibinfo{title}{Excitatory and {Inhibitory} {Subnetworks} {Are}
  {Equally} {Selective} during {Decision}-{Making} and {Emerge}
  {Simultaneously} during {Learning}}.
\newblock \emph{\bibinfo{journal}{Neuron}} \textbf{\bibinfo{volume}{105}},
  \bibinfo{pages}{165--179.e8} (\bibinfo{year}{2020}).
\newblock
  \urlprefix\url{https://www.sciencedirect.com/science/article/pii/S0896627319308487}.

\bibitem{berg_balanced_2007}
\bibinfo{author}{Berg, R.~W.}, \bibinfo{author}{Alaburda, A.} \&
  \bibinfo{author}{Hounsgaard, J.}
\newblock \bibinfo{title}{Balanced {Inhibition} and {Excitation} {Drive}
  {Spike} {Activity} in {Spinal} {Half}-{Centers}}.
\newblock \emph{\bibinfo{journal}{Science}} \textbf{\bibinfo{volume}{315}},
  \bibinfo{pages}{390--393} (\bibinfo{year}{2007}).
\newblock \urlprefix\url{https://science.sciencemag.org/content/315/5810/390}.

\bibitem{berg_when_2019}
\bibinfo{author}{Berg, R.~W.}, \bibinfo{author}{Willumsen, A.} \&
  \bibinfo{author}{Lindén, H.}
\newblock \bibinfo{title}{When networks walk a fine line: balance of excitation
  and inhibition in spinal motor circuits}.
\newblock \emph{\bibinfo{journal}{Current Opinion in Physiology}}
  \textbf{\bibinfo{volume}{8}}, \bibinfo{pages}{76--83} (\bibinfo{year}{2019}).
\newblock
  \urlprefix\url{https://www.sciencedirect.com/science/article/pii/S2468867319300069}.

\bibitem{goulding_inhibition_2014}
\bibinfo{author}{Goulding, M.}, \bibinfo{author}{Bourane, S.},
  \bibinfo{author}{Garcia-Campmany, L.}, \bibinfo{author}{Dalet, A.} \&
  \bibinfo{author}{Koch, S.}
\newblock \bibinfo{title}{Inhibition downunder: an update from the spinal
  cord}.
\newblock \emph{\bibinfo{journal}{Current Opinion in Neurobiology}}
  \textbf{\bibinfo{volume}{26}}, \bibinfo{pages}{161--166}
  (\bibinfo{year}{2014}).
\newblock
  \urlprefix\url{http://www.sciencedirect.com/science/article/pii/S0959438814000567}.

\bibitem{cowan_wilsoncowan_2016}
\bibinfo{author}{Cowan, J.~D.}, \bibinfo{author}{Neuman, J.} \&
  \bibinfo{author}{van Drongelen, W.}
\newblock \bibinfo{title}{Wilson–{Cowan} {Equations} for {Neocortical}
  {Dynamics}}.
\newblock \emph{\bibinfo{journal}{The Journal of Mathematical Neuroscience}}
  \textbf{\bibinfo{volume}{6}}, \bibinfo{pages}{1} (\bibinfo{year}{2016}).
\newblock \urlprefix\url{https://doi.org/10.1186/s13408-015-0034-5}.

\bibitem{petersen_premotor_2014}
\bibinfo{author}{Petersen, P.~C.}, \bibinfo{author}{Vestergaard, M.},
  \bibinfo{author}{Jensen, K. H.~R.} \& \bibinfo{author}{Berg, R.~W.}
\newblock \bibinfo{title}{Premotor spinal network with balanced excitation and
  inhibition during motor patterns has high resilience to structural division}.
\newblock \emph{\bibinfo{journal}{The Journal of Neuroscience: The Official
  Journal of the Society for Neuroscience}} \textbf{\bibinfo{volume}{34}},
  \bibinfo{pages}{2774--2784} (\bibinfo{year}{2014}).

\bibitem{pierrot-deseilligny_circuitry_2005}
\bibinfo{author}{Pierrot-Deseilligny, E.} \& \bibinfo{author}{Burke, D.}
\newblock \emph{\bibinfo{title}{The {Circuitry} of the {Human} {Spinal} {Cord}:
  {Its} {Role} in {Motor} {Control} and {Movement} {Disorders}}}
  (\bibinfo{publisher}{Cambridge University Press},
  \bibinfo{address}{Cambridge}, \bibinfo{year}{2005}).
\newblock
  \urlprefix\url{https://www.cambridge.org/core/books/circuitry-of-the-human-spinal-cord/BBDFC31B44453B61DD5FD801B34CFFFF}.

\bibitem{nishimura_spike-timing-dependent_2013}
\bibinfo{author}{Nishimura, Y.}, \bibinfo{author}{Perlmutter, S.~I.},
  \bibinfo{author}{Eaton, R.~W.} \& \bibinfo{author}{Fetz, E.~E.}
\newblock \bibinfo{title}{Spike-{Timing}-{Dependent} {Plasticity} in {Primate}
  {Corticospinal} {Connections} {Induced} during {Free} {Behavior}}.
\newblock \emph{\bibinfo{journal}{Neuron}} \textbf{\bibinfo{volume}{80}},
  \bibinfo{pages}{1301--1309} (\bibinfo{year}{2013}).
\newblock
  \urlprefix\url{https://www.sciencedirect.com/science/article/pii/S0896627313007629}.

\bibitem{kaufman_cortical_2014}
\bibinfo{author}{Kaufman, M.~T.}, \bibinfo{author}{Churchland, M.~M.},
  \bibinfo{author}{Ryu, S.~I.} \& \bibinfo{author}{Shenoy, K.~V.}
\newblock \bibinfo{title}{Cortical activity in the null space: permitting
  preparation without movement}.
\newblock \emph{\bibinfo{journal}{Nature Neuroscience}}
  \textbf{\bibinfo{volume}{17}}, \bibinfo{pages}{440--448}
  (\bibinfo{year}{2014}).
\newblock \urlprefix\url{https://www.nature.com/articles/nn.3643}.
\newblock \bibinfo{note}{Bandiera\_abtest: a Cg\_type: Nature Research Journals
  Number: 3 Primary\_atype: Research Publisher: Nature Publishing Group
  Subject\_term: Motor cortex;Premotor cortex Subject\_term\_id:
  motor-cortex;premotor-cortex}.

\bibitem{bastian_cerebellar_1996}
\bibinfo{author}{Bastian, A.~J.}, \bibinfo{author}{Martin, T.~A.},
  \bibinfo{author}{Keating, J.~G.} \& \bibinfo{author}{Thach, W.~T.}
\newblock \bibinfo{title}{Cerebellar ataxia: abnormal control of interaction
  torques across multiple joints}.
\newblock \emph{\bibinfo{journal}{Journal of Neurophysiology}}
  \textbf{\bibinfo{volume}{76}}, \bibinfo{pages}{492--509}
  (\bibinfo{year}{1996}).
\newblock
  \urlprefix\url{https://journals.physiology.org/doi/abs/10.1152/jn.1996.76.1.492}.
\newblock \bibinfo{note}{Publisher: American Physiological Society}.

\bibitem{becker_multi-joint_1991}
\bibinfo{author}{Becker, W.~J.}, \bibinfo{author}{Morrice, B.~L.},
  \bibinfo{author}{Clark, A.~W.} \& \bibinfo{author}{Lee, R.~G.}
\newblock \bibinfo{title}{Multi-{Joint} {Reaching} {Movements} and {Eye}-{Hand}
  {Tracking} in {Cerebellar} {Incoordination}: {Investigation} of a {Patient}
  with {Complete} {Loss} of {Purkinje} {Cells}}.
\newblock \emph{\bibinfo{journal}{Canadian Journal of Neurological Sciences}}
  \textbf{\bibinfo{volume}{18}}, \bibinfo{pages}{476--487}
  (\bibinfo{year}{1991}).
\newblock
  \urlprefix\url{https://www.cambridge.org/core/journals/canadian-journal-of-neurological-sciences/article/div-classtitlemulti-joint-reaching-movements-and-eye-hand-tracking-in-cerebellar-incoordination-investigation-of-a-patient-with-complete-loss-of-purkinje-cellsdiv/10DE187987FC9A31457C01168A1E8382}.
\newblock \bibinfo{note}{Publisher: Cambridge University Press}.

\bibitem{gilman_kinematic_1976}
\bibinfo{author}{Gilman, S.}, \bibinfo{author}{Carr, D.} \&
  \bibinfo{author}{Hollenberg, J.}
\newblock \bibinfo{title}{{Kinematic} {effects} {of} {deafferentiation} {and}
  {cerebelar} {ablation}}.
\newblock \emph{\bibinfo{journal}{Brain}} \textbf{\bibinfo{volume}{99}},
  \bibinfo{pages}{311--330} (\bibinfo{year}{1976}).
\newblock \urlprefix\url{https://doi.org/10.1093/brain/99.2.311}.

\bibitem{day_influence_1998}
\bibinfo{author}{Day, B.~L.}, \bibinfo{author}{Thompson, P.~D.},
  \bibinfo{author}{Harding, A.~E.} \& \bibinfo{author}{Marsden, C.~D.}
\newblock \bibinfo{title}{Influence of vision on upper limb reaching movements
  in patients with cerebellar ataxia.}
\newblock \emph{\bibinfo{journal}{Brain}} \textbf{\bibinfo{volume}{121}},
  \bibinfo{pages}{357--372} (\bibinfo{year}{1998}).
\newblock \urlprefix\url{https://doi.org/10.1093/brain/121.2.357}.

\bibitem{lillicrap_preference_2013}
\bibinfo{author}{Lillicrap, T.} \& \bibinfo{author}{Scott, S.}
\newblock \bibinfo{title}{Preference {Distributions} of {Primary} {Motor}
  {Cortex} {Neurons} {Reflect} {Control} {Solutions} {Optimized} for {Limb}
  {Biomechanics}}.
\newblock \emph{\bibinfo{journal}{Neuron}} \textbf{\bibinfo{volume}{77}},
  \bibinfo{pages}{168--179} (\bibinfo{year}{2013}).
\newblock
  \urlprefix\url{http://www.sciencedirect.com/science/article/pii/S0896627312009920}.

\bibitem{kurtzer_primate_2006}
\bibinfo{author}{Kurtzer, I.}, \bibinfo{author}{Pruszynski, J.~A.},
  \bibinfo{author}{Herter, T.~M.} \& \bibinfo{author}{Scott, S.~H.}
\newblock \bibinfo{title}{Primate {Upper} {Limb} {Muscles} {Exhibit} {Activity}
  {Patterns} {That} {Differ} {From} {Their} {Anatomical} {Action} {During} a
  {Postural} {Task}}.
\newblock \emph{\bibinfo{journal}{Journal of Neurophysiology}}
  \textbf{\bibinfo{volume}{95}}, \bibinfo{pages}{493--504}
  (\bibinfo{year}{2006}).
\newblock
  \urlprefix\url{https://journals.physiology.org/doi/full/10.1152/jn.00706.2005}.
\newblock \bibinfo{note}{Publisher: American Physiological Society}.

\bibitem{chambers_stable_2017}
\bibinfo{author}{Chambers, A.~R.} \& \bibinfo{author}{Rumpel, S.}
\newblock \bibinfo{title}{A stable brain from unstable components: {Emerging}
  concepts and implications for neural computation}.
\newblock \emph{\bibinfo{journal}{Neuroscience}}
  \textbf{\bibinfo{volume}{357}}, \bibinfo{pages}{172--184}
  (\bibinfo{year}{2017}).
\newblock
  \urlprefix\url{http://www.sciencedirect.com/science/article/pii/S0306452217304013}.

\bibitem{kalidindi_rotational_2021}
\bibinfo{author}{Kalidindi, H.~T.} \emph{et~al.}
\newblock \bibinfo{title}{Rotational dynamics in motor cortex are consistent
  with a feedback controller}.
\newblock \emph{\bibinfo{journal}{eLife}} \textbf{\bibinfo{volume}{10}},
  \bibinfo{pages}{e67256} (\bibinfo{year}{2021}).
\newblock \urlprefix\url{https://doi.org/10.7554/eLife.67256}.
\newblock \bibinfo{note}{Publisher: eLife Sciences Publications, Ltd}.

\bibitem{dewolf_spiking_2016}
\bibinfo{author}{DeWolf, T.}, \bibinfo{author}{Stewart, T.~C.},
  \bibinfo{author}{Slotine, J.-J.} \& \bibinfo{author}{Eliasmith, C.}
\newblock \bibinfo{title}{A spiking neural model of adaptive arm control}.
\newblock \emph{\bibinfo{journal}{Proceedings of the Royal Society B:
  Biological Sciences}}  (\bibinfo{year}{2016}).
\newblock
  \urlprefix\url{https://royalsocietypublishing.org/doi/abs/10.1098/rspb.2016.2134}.
\newblock \bibinfo{note}{Publisher: The Royal Society}.

\bibitem{edelman_degeneracy_2001}
\bibinfo{author}{Edelman, G.~M.} \& \bibinfo{author}{Gally, J.~A.}
\newblock \bibinfo{title}{Degeneracy and complexity in biological systems}.
\newblock \emph{\bibinfo{journal}{Proceedings of the National Academy of
  Sciences}} \textbf{\bibinfo{volume}{98}}, \bibinfo{pages}{13763--13768}
  (\bibinfo{year}{2001}).
\newblock \urlprefix\url{https://www.pnas.org/doi/full/10.1073/pnas.231499798}.
\newblock \bibinfo{note}{Publisher: Proceedings of the National Academy of
  Sciences}.

\bibitem{sanguineti_cerebellar_2003}
\bibinfo{author}{Sanguineti, V.} \emph{et~al.}
\newblock \bibinfo{title}{Cerebellar ataxia: {Quantitative} assessment and
  cybernetic interpretation}.
\newblock \emph{\bibinfo{journal}{Human Movement Science}}
  \textbf{\bibinfo{volume}{22}}, \bibinfo{pages}{189--205}
  (\bibinfo{year}{2003}).
\newblock
  \urlprefix\url{https://www.sciencedirect.com/science/article/pii/S0167945702001598}.

\bibitem{richter_adaptive_2004}
\bibinfo{author}{Richter, S.} \emph{et~al.}
\newblock \bibinfo{title}{Adaptive {Motor} {Behavior} of {Cerebellar}
  {Patients} {During} {Exposure} to {Unfamiliar} {External} {Forces}}.
\newblock \emph{\bibinfo{journal}{Journal of Motor Behavior}}
  \textbf{\bibinfo{volume}{36}}, \bibinfo{pages}{28--38}
  (\bibinfo{year}{2004}).
\newblock \urlprefix\url{https://doi.org/10.3200/JMBR.36.1.28-38}.
\newblock \bibinfo{note}{Publisher: Routledge \_eprint:
  https://doi.org/10.3200/JMBR.36.1.28-38}.

\bibitem{bizzi_motor_2020}
\bibinfo{author}{Bizzi, E.} \& \bibinfo{author}{Ajemian, R.}
\newblock \bibinfo{title}{From motor planning to execution: a sensorimotor loop
  perspective}.
\newblock \emph{\bibinfo{journal}{Journal of Neurophysiology}}
  \textbf{\bibinfo{volume}{124}}, \bibinfo{pages}{1815--1823}
  (\bibinfo{year}{2020}).
\newblock
  \urlprefix\url{https://journals.physiology.org/doi/full/10.1152/jn.00715.2019}.
\newblock \bibinfo{note}{Publisher: American Physiological Society}.

\bibitem{verduzco-flores_draculab:_2019}
\bibinfo{author}{Verduzco-Flores, S.} \& \bibinfo{author}{De~Schutter, E.}
\newblock \bibinfo{title}{Draculab: {A} {Python} {Simulator} for {Firing}
  {Rate} {Neural} {Networks} {With} {Delayed} {Adaptive} {Connections}}.
\newblock \emph{\bibinfo{journal}{Frontiers in Neuroinformatics}}
  \textbf{\bibinfo{volume}{13}} (\bibinfo{year}{2019}).
\newblock
  \urlprefix\url{https://www.frontiersin.org/articles/10.3389/fninf.2019.00018/full}.

\bibitem{lin_neural_2002}
\bibinfo{author}{Lin, C.-C.~K.} \& \bibinfo{author}{Crago, P.~E.}
\newblock \bibinfo{title}{Neural and {Mechanical} {Contributions} to the
  {Stretch} {Reflex}: {A} {Model} {Synthesis}}.
\newblock \emph{\bibinfo{journal}{Annals of Biomedical Engineering}}
  \textbf{\bibinfo{volume}{30}}, \bibinfo{pages}{54--67}
  (\bibinfo{year}{2002}).
\newblock \urlprefix\url{https://doi.org/10.1114/1.1432692}.

\bibitem{cheah_adaptive_2006}
\bibinfo{author}{Cheah, C.~C.}, \bibinfo{author}{Liu, C.} \&
  \bibinfo{author}{Slotine, J. J.~E.}
\newblock \bibinfo{title}{Adaptive {Tracking} {Control} for {Robots} with
  {Unknown} {Kinematic} and {Dynamic} {Properties}}.
\newblock \emph{\bibinfo{journal}{The International Journal of Robotics
  Research}} \textbf{\bibinfo{volume}{25}}, \bibinfo{pages}{283--296}
  (\bibinfo{year}{2006}).
\newblock \urlprefix\url{https://doi.org/10.1177/0278364906063830}.
\newblock \bibinfo{note}{Publisher: SAGE Publications Ltd STM}.

\bibitem{sanner_gaussian_1992}
\bibinfo{author}{Sanner, R.} \& \bibinfo{author}{Slotine, J.-J.}
\newblock \bibinfo{title}{Gaussian networks for direct adaptive control}.
\newblock \emph{\bibinfo{journal}{IEEE Transactions on Neural Networks}}
  \textbf{\bibinfo{volume}{3}}, \bibinfo{pages}{837--863}
  (\bibinfo{year}{1992}).
\newblock \bibinfo{note}{Conference Name: IEEE Transactions on Neural
  Networks}.

\bibitem{bekolay_nengo:_2014}
\bibinfo{author}{Bekolay, T.} \emph{et~al.}
\newblock \bibinfo{title}{Nengo: a {Python} tool for building large-scale
  functional brain models}.
\newblock \emph{\bibinfo{journal}{Frontiers in Neuroinformatics}}
  \textbf{\bibinfo{volume}{7}} (\bibinfo{year}{2014}).
\newblock
  \urlprefix\url{https://www.frontiersin.org/articles/10.3389/fninf.2013.00048/full}.

\bibitem{dura-bernal_cortical_2015}
\bibinfo{author}{Dura-Bernal, S.} \emph{et~al.}
\newblock \bibinfo{title}{Cortical {Spiking} {Network} {Interfaced} with
  {Virtual} {Musculoskeletal} {Arm} and {Robotic} {Arm}}.
\newblock \emph{\bibinfo{journal}{Frontiers in Neurorobotics}}
  \textbf{\bibinfo{volume}{9}}, \bibinfo{pages}{13} (\bibinfo{year}{2015}).
\newblock
  \urlprefix\url{https://www.frontiersin.org/article/10.3389/fnbot.2015.00013}.

\bibitem{li_hierarchical_2005}
\bibinfo{author}{Li, W.}, \bibinfo{author}{Todorov, E.} \&
  \bibinfo{author}{Pan, X.}
\newblock \bibinfo{title}{Hierarchical {Feedback} and {Learning} for
  {Multi}-joint {Arm} {Movement} {Control}}.
\newblock In \emph{\bibinfo{booktitle}{2005 {IEEE} {Engineering} in {Medicine}
  and {Biology} 27th {Annual} {Conference}}}, \bibinfo{pages}{4400--4403}
  (\bibinfo{year}{2005}).
\newblock \bibinfo{note}{ISSN: 1558-4615}.

\bibitem{martin_redundancy_2009}
\bibinfo{author}{Martin, V.}, \bibinfo{author}{Scholz, J.~P.} \&
  \bibinfo{author}{Schöner, G.}
\newblock \bibinfo{title}{Redundancy, {Self}-{Motion}, and {Motor} {Control}}.
\newblock \emph{\bibinfo{journal}{Neural Computation}}
  \textbf{\bibinfo{volume}{21}}, \bibinfo{pages}{1371--1414}
  (\bibinfo{year}{2009}).
\newblock \urlprefix\url{https://doi.org/10.1162/neco.2008.01-08-698}.

\bibitem{caligiore_integrating_2014}
\bibinfo{author}{Caligiore, D.}, \bibinfo{author}{Parisi, D.} \&
  \bibinfo{author}{Baldassarre, G.}
\newblock \bibinfo{title}{Integrating reinforcement learning, equilibrium
  points, and minimum variance to understand the development of reaching: {A}
  computational model}.
\newblock \emph{\bibinfo{journal}{Psychological Review}}
  \textbf{\bibinfo{volume}{121}}, \bibinfo{pages}{389--421}
  (\bibinfo{year}{2014}).
\newblock \bibinfo{note}{Place: US Publisher: American Psychological
  Association}.

\bibitem{sutton_reinforcement_2018}
\bibinfo{author}{Sutton, R.~S.} \& \bibinfo{author}{Barto, A.~G.}
\newblock \emph{\bibinfo{title}{Reinforcement {Learning}: {An} {Introduction}}}
  (\bibinfo{publisher}{MIT Press}, \bibinfo{year}{2018}).
\newblock \bibinfo{note}{Google-Books-ID: sWV0DwAAQBAJ}.

\bibitem{feldman_once_1986}
\bibinfo{author}{Feldman, A.~G.}
\newblock \bibinfo{title}{Once {More} on the {Equilibrium}-{Point} {Hypothesis}
  (λ {Model}) for {Motor} {Control}}.
\newblock \emph{\bibinfo{journal}{Journal of Motor Behavior}}
  \textbf{\bibinfo{volume}{18}}, \bibinfo{pages}{17--54}
  (\bibinfo{year}{1986}).
\newblock \urlprefix\url{https://doi.org/10.1080/00222895.1986.10735369}.
\newblock \bibinfo{note}{Publisher: Routledge \_eprint:
  https://doi.org/10.1080/00222895.1986.10735369}.

\bibitem{izawa_biological_2004}
\bibinfo{author}{Izawa, J.}, \bibinfo{author}{Kondo, T.} \&
  \bibinfo{author}{Ito, K.}
\newblock \bibinfo{title}{Biological arm motion through reinforcement
  learning}.
\newblock \emph{\bibinfo{journal}{Biological Cybernetics}}
  \textbf{\bibinfo{volume}{91}}, \bibinfo{pages}{10--22}
  (\bibinfo{year}{2004}).
\newblock \urlprefix\url{https://doi.org/10.1007/s00422-004-0485-3}.

\bibitem{mici_incremental_2018}
\bibinfo{author}{Mici, L.}, \bibinfo{author}{Parisi, G.~I.} \&
  \bibinfo{author}{Wermter, S.}
\newblock \bibinfo{title}{An {Incremental} {Self}-{Organizing} {Architecture}
  for {Sensorimotor} {Learning} and {Prediction}}.
\newblock \emph{\bibinfo{journal}{IEEE Transactions on Cognitive and
  Developmental Systems}} \textbf{\bibinfo{volume}{10}},
  \bibinfo{pages}{918--928} (\bibinfo{year}{2018}).
\newblock \bibinfo{note}{Conference Name: IEEE Transactions on Cognitive and
  Developmental Systems}.

\bibitem{tsianos_useful_2014}
\bibinfo{author}{Tsianos, G.~A.}, \bibinfo{author}{Goodner, J.} \&
  \bibinfo{author}{Loeb, G.~E.}
\newblock \bibinfo{title}{Useful properties of spinal circuits for learning and
  performing planar reaches}.
\newblock \emph{\bibinfo{journal}{Journal of Neural Engineering}}
  \textbf{\bibinfo{volume}{11}}, \bibinfo{pages}{056006}
  (\bibinfo{year}{2014}).
\newblock
  \urlprefix\url{https://doi.org/10.1088\%2F1741-2560\%2F11\%2F5\%2F056006}.

\bibitem{sussillo_generating_2009}
\bibinfo{author}{Sussillo, D.} \& \bibinfo{author}{Abbott, L.~F.}
\newblock \bibinfo{title}{Generating {Coherent} {Patterns} of {Activity} from
  {Chaotic} {Neural} {Networks}}.
\newblock \emph{\bibinfo{journal}{Neuron}} \textbf{\bibinfo{volume}{63}},
  \bibinfo{pages}{544--557} (\bibinfo{year}{2009}).
\newblock
  \urlprefix\url{https://www.sciencedirect.com/science/article/pii/S0896627309005479}.

\bibitem{bashor_large-scale_1998}
\bibinfo{author}{Bashor, D.~P.}
\newblock \bibinfo{title}{A large-scale model of some spinal reflex circuits}.
\newblock \emph{\bibinfo{journal}{Biological Cybernetics}}
  \textbf{\bibinfo{volume}{78}}, \bibinfo{pages}{147--157}
  (\bibinfo{year}{1998}).
\newblock \urlprefix\url{https://doi.org/10.1007/s004220050421}.

\bibitem{stienen_analysis_2007}
\bibinfo{author}{Stienen, A. H.~A.}, \bibinfo{author}{Schouten, A.~C.},
  \bibinfo{author}{Schuurmans, J.} \& \bibinfo{author}{van~der Helm, F. C.~T.}
\newblock \bibinfo{title}{Analysis of reflex modulation with a biologically
  realistic neural network}.
\newblock \emph{\bibinfo{journal}{Journal of Computational Neuroscience}}
  \textbf{\bibinfo{volume}{23}}, \bibinfo{pages}{333} (\bibinfo{year}{2007}).
\newblock \urlprefix\url{https://doi.org/10.1007/s10827-007-0037-7}.

\bibitem{cutsuridis_does_2007}
\bibinfo{author}{Cutsuridis, V.}
\newblock \bibinfo{title}{Does abnormal spinal reciprocal inhibition lead to
  co-contraction of antagonist motor units? a modeling study}.
\newblock \emph{\bibinfo{journal}{International Journal of Neural Systems}}
  \textbf{\bibinfo{volume}{17}}, \bibinfo{pages}{319--327}
  (\bibinfo{year}{2007}).
\newblock
  \urlprefix\url{https://www.worldscientific.com/doi/abs/10.1142/S0129065707001160}.
\newblock \bibinfo{note}{Publisher: World Scientific Publishing Co.}

\bibitem{cisi_simulation_2008}
\bibinfo{author}{Cisi, R. R.~L.} \& \bibinfo{author}{Kohn, A.~F.}
\newblock \bibinfo{title}{Simulation system of spinal cord motor nuclei and
  associated nerves and muscles, in a {Web}-based architecture}.
\newblock \emph{\bibinfo{journal}{Journal of Computational Neuroscience}}
  \textbf{\bibinfo{volume}{25}}, \bibinfo{pages}{520--542}
  (\bibinfo{year}{2008}).
\newblock \urlprefix\url{https://doi.org/10.1007/s10827-008-0092-8}.

\bibitem{farina_effective_2014}
\bibinfo{author}{Farina, D.}, \bibinfo{author}{Negro, F.} \&
  \bibinfo{author}{Dideriksen, J.~L.}
\newblock \bibinfo{title}{The effective neural drive to muscles is the common
  synaptic input to motor neurons}.
\newblock \emph{\bibinfo{journal}{The Journal of Physiology}}
  \textbf{\bibinfo{volume}{592}}, \bibinfo{pages}{3427--3441}
  (\bibinfo{year}{2014}).
\newblock
  \urlprefix\url{https://physoc.onlinelibrary.wiley.com/doi/abs/10.1113/jphysiol.2014.273581}.
\newblock \bibinfo{note}{\_eprint:
  https://physoc.onlinelibrary.wiley.com/doi/pdf/10.1113/jphysiol.2014.273581}.

\bibitem{shevtsova_organization_2015}
\bibinfo{author}{Shevtsova, N.~A.} \emph{et~al.}
\newblock \bibinfo{title}{Organization of left–right coordination of neuronal
  activity in the mammalian spinal cord: {Insights} from computational
  modelling}.
\newblock \emph{\bibinfo{journal}{The Journal of Physiology}}
  \textbf{\bibinfo{volume}{593}}, \bibinfo{pages}{2403--2426}
  (\bibinfo{year}{2015}).
\newblock
  \urlprefix\url{https://physoc.onlinelibrary.wiley.com/doi/abs/10.1113/JP270121}.
\newblock \bibinfo{note}{\_eprint:
  https://physoc.onlinelibrary.wiley.com/doi/pdf/10.1113/JP270121}.

\bibitem{shevtsova_organization_2016}
\bibinfo{author}{Shevtsova, N.~A.} \& \bibinfo{author}{Rybak, I.~A.}
\newblock \bibinfo{title}{Organization of flexor–extensor interactions in the
  mammalian spinal cord: insights from computational modelling}.
\newblock \emph{\bibinfo{journal}{The Journal of Physiology}}
  \textbf{\bibinfo{volume}{594}}, \bibinfo{pages}{6117--6131}
  (\bibinfo{year}{2016}).
\newblock
  \urlprefix\url{https://www.ncbi.nlm.nih.gov/pmc/articles/PMC5088238/}.

\bibitem{dannerComputationalModelingSpinal2017}
\bibinfo{author}{Danner, S.~M.}, \bibinfo{author}{Shevtsova, N.~A.},
  \bibinfo{author}{Frigon, A.} \& \bibinfo{author}{Rybak, I.~A.}
\newblock \bibinfo{title}{Computational modeling of spinal circuits controlling
  limb coordination and gaits in quadrupeds}.
\newblock \emph{\bibinfo{journal}{eLife}} \textbf{\bibinfo{volume}{6}},
  \bibinfo{pages}{e31050} (\bibinfo{year}{2017}).
\newblock \urlprefix\url{https://doi.org/10.7554/eLife.31050}.
\newblock \bibinfo{note}{Publisher: eLife Sciences Publications, Ltd}.

\bibitem{zelenin_differential_2021}
\bibinfo{author}{Zelenin, P.~V.} \emph{et~al.}
\newblock \bibinfo{title}{Differential {Contribution} of {V0} {Interneurons} to
  {Execution} of {Rhythmic} and {Nonrhythmic} {Motor} {Behaviors}}.
\newblock \emph{\bibinfo{journal}{The Journal of Neuroscience: The Official
  Journal of the Society for Neuroscience}} \textbf{\bibinfo{volume}{41}},
  \bibinfo{pages}{3432--3445} (\bibinfo{year}{2021}).

\bibitem{stachowski_spinal_2021}
\bibinfo{author}{Stachowski, N.~J.} \& \bibinfo{author}{Dougherty, K.~J.}
\newblock \bibinfo{title}{Spinal {Inhibitory} {Interneurons}: {Gatekeepers} of
  {Sensorimotor} {Pathways}}.
\newblock \emph{\bibinfo{journal}{International Journal of Molecular Sciences}}
  \textbf{\bibinfo{volume}{22}}, \bibinfo{pages}{2667} (\bibinfo{year}{2021}).
\newblock
  \urlprefix\url{https://www.ncbi.nlm.nih.gov/pmc/articles/PMC7961554/}.

\bibitem{borisyukModelingConnectomeSimple2011}
\bibinfo{author}{Borisyuk, R.}, \bibinfo{author}{Azad, A.},
  \bibinfo{author}{Conte, D.}, \bibinfo{author}{Roberts, A.} \&
  \bibinfo{author}{Soffe, S.}
\newblock \bibinfo{title}{Modeling the {Connectome} of a {Simple} {Spinal}
  {Cord}}.
\newblock \emph{\bibinfo{journal}{Frontiers in Neuroinformatics}}
  \textbf{\bibinfo{volume}{5}}, \bibinfo{pages}{20} (\bibinfo{year}{2011}).
\newblock
  \urlprefix\url{https://www.frontiersin.org/article/10.3389/fninf.2011.00020}.

\bibitem{cangianoMechanismsRhythmGeneration2005}
\bibinfo{author}{Cangiano, L.} \& \bibinfo{author}{Grillner, S.}
\newblock \bibinfo{title}{Mechanisms of {Rhythm} {Generation} in a {Spinal}
  {Locomotor} {Network} {Deprived} of {Crossed} {Connections}: {The} {Lamprey}
  {Hemicord}}.
\newblock \emph{\bibinfo{journal}{Journal of Neuroscience}}
  \textbf{\bibinfo{volume}{25}}, \bibinfo{pages}{923--935}
  (\bibinfo{year}{2005}).
\newblock \urlprefix\url{https://www.jneurosci.org/content/25/4/923}.
\newblock \bibinfo{note}{Publisher: Society for Neuroscience Section:
  Behavioral/Systems/Cognitive}.

\bibitem{zappacosta_general_2018}
\bibinfo{author}{Zappacosta, S.}, \bibinfo{author}{Mannella, F.},
  \bibinfo{author}{Mirolli, M.} \& \bibinfo{author}{Baldassarre, G.}
\newblock \bibinfo{title}{General differential {Hebbian} learning: {Capturing}
  temporal relations between events in neural networks and the brain}.
\newblock \emph{\bibinfo{journal}{PLOS Computational Biology}}
  \textbf{\bibinfo{volume}{14}}, \bibinfo{pages}{e1006227}
  (\bibinfo{year}{2018}).
\newblock
  \urlprefix\url{https://journals.plos.org/ploscompbiol/article?id=10.1371/journal.pcbi.1006227}.
\newblock \bibinfo{note}{Publisher: Public Library of Science}.

\bibitem{fiete_spike-time-dependent_2010}
\bibinfo{author}{Fiete, I.~R.}, \bibinfo{author}{Senn, W.},
  \bibinfo{author}{Wang, C. Z.~H.} \& \bibinfo{author}{Hahnloser, R. H.~R.}
\newblock \bibinfo{title}{Spike-{Time}-{Dependent} {Plasticity} and
  {Heterosynaptic} {Competition} {Organize} {Networks} to {Produce} {Long}
  {Scale}-{Free} {Sequences} of {Neural} {Activity}}.
\newblock \emph{\bibinfo{journal}{Neuron}} \textbf{\bibinfo{volume}{65}},
  \bibinfo{pages}{563--576} (\bibinfo{year}{2010}).
\newblock
  \urlprefix\url{https://www.sciencedirect.com/science/article/pii/S0896627310000917}.

\bibitem{kaleb_network-centered_2021}
\bibinfo{author}{Kaleb, K.}, \bibinfo{author}{Pedrosa, V.} \&
  \bibinfo{author}{Clopath, C.}
\newblock \bibinfo{title}{Network-centered homeostasis through inhibition
  maintains hippocampal spatial map and cortical circuit function}.
\newblock \emph{\bibinfo{journal}{Cell Reports}} \textbf{\bibinfo{volume}{36}},
  \bibinfo{pages}{109577} (\bibinfo{year}{2021}).
\newblock
  \urlprefix\url{https://www.sciencedirect.com/science/article/pii/S2211124721010111}.

\bibitem{lim_balanced_2013}
\bibinfo{author}{Lim, S.} \& \bibinfo{author}{Goldman, M.~S.}
\newblock \bibinfo{title}{Balanced cortical microcircuitry for maintaining
  information in working memory}.
\newblock \emph{\bibinfo{journal}{Nature Neuroscience}}
  \textbf{\bibinfo{volume}{16}}, \bibinfo{pages}{1306--1314}
  (\bibinfo{year}{2013}).
\newblock \urlprefix\url{https://www.nature.com/articles/nn.3492}.
\newblock \bibinfo{note}{Bandiera\_abtest: a Cg\_type: Nature Research Journals
  Number: 9 Primary\_atype: Research Publisher: Nature Publishing Group
  Subject\_term: Network models Subject\_term\_id: network-models}.

\bibitem{helmchen_dendrites_1999}
\bibinfo{author}{Helmchen, F.}
\newblock \bibinfo{title}{Dendrites as biochemical compartments}.
\newblock In \emph{\bibinfo{booktitle}{Dendrites}}, \bibinfo{pages}{376}
  (\bibinfo{publisher}{Oxford University Press}, \bibinfo{address}{Oxford,
  England}, \bibinfo{year}{1999}).
\newblock
  \urlprefix\url{https://pure.mpg.de/pubman/faces/ViewItemOverviewPage.jsp?itemId=item\_2537436}.

\bibitem{kawato_50_2020}
\bibinfo{author}{Kawato, M.}, \bibinfo{author}{Ohmae, S.},
  \bibinfo{author}{Hoang, H.} \& \bibinfo{author}{Sanger, T.}
\newblock \bibinfo{title}{50 years since the {Marr}, {Ito}, and {Albus} models
  of the cerebellum}.
\newblock \emph{\bibinfo{journal}{arXiv:2003.05647 [q-bio]}}
  (\bibinfo{year}{2020}).
\newblock \urlprefix\url{http://arxiv.org/abs/2003.05647}.
\newblock \bibinfo{note}{ArXiv: 2003.05647}.

\end{thebibliography}
\bibliographystyle{naturemag}

\appendix

\section{Supplementary Discussion}

\subsection{Comparison with previous models} \label{subsec:comparison}

There are many other models of reaching and motor control. Most of them have
one or more of the following limitations:
\begin{enumerate}
    \item They use non-neural systems to produce motor commands.
    \item They control a single degree of freedom, sidestepping the problem
    of controller configuration, since the error is one-dimensional.
    \item They do not model a biologically plausible form of synaptic learning.
\end{enumerate}
We will contrast our model with some of the work that does not strongly present these
limitations, and with a few others. Due to space constraints the contributions of
many models will not be addressed, and for those mentioned we will limit ourselves
to explain some of their limitations.

The model in \cite{dewolf_spiking_2016} is similar has similar goals to our model, 
but with very different assumptions. 
In their model, motor cortex receives a location error vector $x$, and
transforms it into a vector of joint torques. So that this transformation implements
adaptive kinematics, it must approximate a Jacobian matrix that includes the effects
of the arm's inertia matrix, using location errors and
joint velocities as training signals. This is accomplished by adapting an algorithm 
taken from the robotics literature \cite{cheah_adaptive_2006},
implementing it in a spiking
neural network. Additionally a second algorithm from robotics 
\cite{sanner_gaussian_1992} is used
to provide an adaptive dynamics component, which is interpreted as the
cerebellar contributions.

In order to implement vector functions in spiking neural networks 
\cite{dewolf_spiking_2016} uses
the Neural Engineering Framework \cite{bekolay_nengo:_2014}. 
The essence of this approach is to represent
values in populations of neurons with cosine-like tuning functions. These
populations implement expansive recoding, becoming a massively overcomplete basis
of the input space. Implementing a function using this population as the input
is akin to using a linear decoder to extract the desired function values 
from the population activity. This can be done through standard methods, such as
least-squares minimization, or random gradient descent. The parameters of the
linear decoder then become weights of a feedforward neural layer implementing the
function.

The model in \cite{dewolf_spiking_2016} has therefore a rather different approach.
They use
engineering techniques to create a powerful motor control system, using algorithms
from robotics, and 30,000 neurons to approximate their computations, which are then
ascribed to sensory, motor, and premotor cortices,
as well as the cerebellum. In contrast, we use 74 firing rate units, and unlike
\cite{dewolf_spiking_2016} we include muscles, muscle afferents, transmission delays,
and a spinal cord.

There is nothing intrinsically wrong with using an engineering approach to try
to understand a biological function. The crucial part is which model will be
experimentally validated. Some differences between the models that may be able
to separate them experimentally are: 1) In \cite{dewolf_spiking_2016} 
premotor cortex is required
to produce the error signal, whereas we ascribed this to sensory cortex. 2) In
\cite{dewolf_spiking_2016} direct afferent connections to motor cortex are not considered,
whereas in our model they are important to maintain stability during learning
(in the absence of a cerebellum). 3) In \cite{dewolf_spiking_2016} spinal cord 
adaptation is not necessary to implement adaptive kinematics.
In contrast, spinal cord adaptation is important in one of the
interpretations of our model.

The model in \cite{dura-bernal_cortical_2015} uses spiking neurons, and a 
realistic neuromechanical
model in order to perform 2D reaching. The feedback is in term of
muscle lengths, rather than muscle afferent signals. There is no mechanism to
stop the arm, or hold it on target. Most importantly, learning relies on a
critic, sending rewarding or punishing signals depending on whether the hand was
approaching or getting away from the target. This is implicitly reducing the
error dimension using a hidden mechanism. Furthermore, each single target
must be trained individually, and it is not discussed how this can lead to a
flexible reaching mechanism without suffering from catastrophic interference.

The model in \cite{todorov_direct_2000} is used to obtain characteristics of M1 activity given the required muscle forces to produce a movement. It is an open-loop, locally-linear model, where all connections from M1 directly stimulate a linear motoneuron. Among other things, it showed that representations of kinematic parameters can appear when the viscoelastic properties of the muscles are taken into account,  giving credence to the hypothesis that M1 directly activates muscle groups. Outside of its scope are neural implementation, learning, or the role of spinal cord.

\cite{li_hierarchical_2005} proposes a 2-level hierarchical controller for a 2-DOF 
arm. Since this 
model is based on optimal control theory, it is given a cost function, and
proceeds by iteratively approaching a solution of the associated Jacobi-Bellman
equation. There is no neural implementation of these computations, nor a description
of learning.

\cite{martin_redundancy_2009} is not properly a neural model, but it is rooted in 
Dynamic Field Theory, which assumes that the population of neuronal activities encodes
"activation variables". For this model activation variables represent kinematic
parameters as a virtual joint location vector $\lambda(t)$, which is 
an internal representation of the arm's configuration. 

The innovative part of \cite{martin_redundancy_2009} is in describing how
$\lambda(t)$ is updated and used to control a redundant manipulator.
In particular, the kinematic Jacobian, its null-space matrix, their derivatives, and the Moore-Penrose pseudoinverse are all computed analytically in order to obtain differential equations where the joint motions that move the end effector decouple from those which don't.

Encapsulating the muscle commands into a virtual joint location whose dynamics are decoupled for motions that don't affect the end-effector location is a very interesting concept. 
Still, this is far from a neural implementation, and learning is not considered.

The model in \cite{caligiore_integrating_2014} studies the long-term development
of infant reaching
using a PD controller, and an actor critic mechanism implementing a neural version
of TD-learning \cite{sutton_reinforcement_2018}.
The 2-dimensional PD controller receives two desired joint angles (interpreted as an
equilibrium position), producing appropriate torques.
Since it uses the Equilibrium Point (EP) Hypothesis \cite{feldman_once_1986}, 
the reaching portion of this model is tantamount to associating states with 
equilibrium positions. This model thus performs at a higher level of analysis.
Our model could operate at the
same level if we added a reinforcement learning component to
learn $S_P$ values allowing the hand to {\it touch} a target whose distance
is known. \cite{caligiore_integrating_2014} does not consider separate neuronal 
populations (e.g. spinal cord, sensory cortex), propagation delays, or low-level
learning.

The model in \cite{izawa_biological_2004} shows how reinforcement learning can
be applied to redundant actuators, such as biological arms. It is, however, not 
a neural model.

In \cite{mici_incremental_2018}, a neural network produces predictions of visual
input in order to deal with temporal delays in a sensorimotor system. The network
used for this study uses non-local learning, and adds or destroys nodes as
required during its operation. It is thus not biologically-plausible.

\cite{tsianos_useful_2014} is a reaching model that also considers the spinal cord
as a highly-configurable controller. Corticospinal inputs are assumed to be step
commands, which means that motor cortex operates in an open-loop configuration. In
order to produce reaching based on these constant commands,
a biologically-implausible gradient descent mechanism is required, where the same
reach is performed for various values of a synaptic weight, keeping the value that
led to the best performance. Furthermore, the model learns to reach one target at
a time, which would require learning anew when the new target is more than 45 degrees
apart.

As mentioned in the main text, in the context of rotational dynamics,
the model in \cite{sussillo_neural_2015} was used to produce desired muscle
forces using a recurrent neural network. This model uses the FORCE algorithm
\cite{sussillo_generating_2009} to adjust the weights of
a neural network with activity vector $\mathbf{r}(t)$ so it can produce
experimentally observed muscle activity $\mathbf{m}(t)$. Oscillatory dynamics
arise when the model is constrained to be simple.

Although very insightful, this model is limited by the fact that equations
\ref{eq:downstream} and \ref{eq:dynamical} represent an open-loop configuration,
where only the M1 dynamics are considered. In essence, the model is doing a
function approximation with the FORCE algorithm. The question of how the 
training data $\mathbf{m}(t)$ is produced is not addressed, nor is the role 
of spinal cord or sensory cortex (but see their Supplementary Figure 1).

Other than the aforementioned model in \cite{tsianos_useful_2014}, we are
unaware of spinal cord models addressing arm reaching. When these models are
coupled with a musculoskeletal system, it is usually for the control of one
degree of freedom using antagonistic neural populations. We mention some
examples next.

The spinal cord model in \cite{bashor_large-scale_1998}
inspired much of the subsequent work,
organizing the spinal circuit into six pairs of populations controlling
two antagonistic muscle groups at one joint. With this model, the effect
of Ia and Ib afferent input was studied in various neuronal populations.

\cite{stienen_analysis_2007} used a model similar to that in 
\cite{bashor_large-scale_1998} in conjunction with
a one DOF musculoskeletal model to study the effect of Ia afferents in the
modulation of reflexes. \cite{cutsuridis_does_2007} also adapted a similar model,
and used it to inquire whether Parkinsonian rigidity arose from alterations in reciprocal
inhibition mediated by reduced dopamine.

The model in \cite{cisi_simulation_2008} has several features not found in 
\cite{bashor_large-scale_1998}, including
a heterogeneous motoneuron population, and a mechanism to generate electromyograms.
This was used to study the generation of the H-reflex, and response properties
of the motor neuron pool \cite{farina_effective_2014}.

The model in \cite{shevtsova_organization_2015} 
goes beyond models such as \cite{bashor_large-scale_1998}
by using populations of genetically-identified interneurons. This model is used to
study rhythm generating abilities of the spinal circuit, as well as
coordination between flexors and extensors
\cite{shevtsova_organization_2016,dannerComputationalModelingSpinal2017}.
Knowledge regarding
the role of genetically identified spinal interneurons in movement generation
is still evolving \cite[e.g.]{zelenin_differential_2021,stachowski_spinal_2021}.

This paper focuses on mammals, but the spinal cord of simpler organisms
is better characterized
\cite[e.g.]{borisyukModelingConnectomeSimple2011,cangianoMechanismsRhythmGeneration2005},
and may lead to the first realistic models producing ethological behavior.

\subsection{Possible implementations of the learning rule}
\label{sub:possible}
The learning rule in equation \ref{eq:rga21} is a Hebbian rule that also
presents derivatives, heterosynaptic competition, and normalization (e.g.
removing the mean) of its terms. 
None of these is new in a model claiming biological plausibility 
\cite[e.g.]{porr_strongly_2006,zappacosta_general_2018,fiete_spike-time-dependent_2010,kaleb_network-centered_2021}.
We nevertheless mention possible ways for derivatives and normalized terms to 
appear.

Formally, the 
time derivative of a function $f:\mathds{R} \rightarrow \mathds{R}$ evaluated at time 
$t$ is the limit $\frac{f(t+\Delta t) - f(t)}{\Delta t}$ as
$\Delta t \rightarrow 0$. If $f$ 
represents the firing rate of a cell, a measure of change roughly 
proportional to the derivative can come from $f(t) - f(t- \Delta t)$ for 
some small value $\Delta t$.
The most obvious way that such a difference may arise is through 
feedforward inhibition (for the $e_j$ signal), and feedback 
inhibition (for the $c_i$ signal). Feedforward and feedback inhibition are
common motifs in spinal circuits \cite{pierrot-deseilligny_circuitry_2005}. 

A somewhat different way to approach a time derivative is by using two 
low-pass filters with different time constants: 
$$\frac{df}{dt} \approx f_ {fast}(t) - f_{slow}(t),$$
where 
\begin{align*}
\tau_{1} \dot{f}_{fast}(t) &= f(t) - f_{ fast}(t), \\ 
\tau_{2} \dot{f}_{slow}(t) &= f(t) - f_{slow}(t), \\
\tau_{2} &\gg \tau_{1}.
\end{align*}

These principles are illustrated in \cite{lim_balanced_2013} , where they 
are used to explain negative-derivative feedback. 
    
Low-pass filtering can also arise in the biochemical cascades following 
synaptic depolarization. The most salient case is intracellular calcium 
concentration, which has been described as an indicator of firing rate with 
leaky integrator dynamics \cite{helmchen_dendrites_1999}. Although the 
physiology of spinal interneurons has not been characterized with sufficient 
detail to make specific hypotheses, it is clearly possible that feedback 
inhibition and low-pass filtering are enough to approximate a second-order 
derivative.

The $e_j, c_i$ terms in our learning rule are mean-centered. The most 
straightforward way to subtract a mean is to have inhibitory units with 
converging inputs (e.g. receiving all the $e_j$ signals) providing
input to the $c_i$ units. The Ib interneurons \cite{pierrot-deseilligny_circuitry_2005}
are one possibility for mediating this. Another possibility is that
the mean-subtraction happens at the single unit level when the input 
($e_j$) and lateral ($c_i$)
connections are located at different parts of the dendritic tree. In particular,
a larger level of overall input activation $\langle \ddot{e} \rangle$ 
could produce a scarcity of postsynaptic resources flowing from the main 
branches of the dendritic tree into the individual dendritic spines, 
resulting in reduced plasticity.

\subsection{Limitations of the model} \label{sec:limitations}
A model as simple as ours will undoubtedly be wrong in many details. 
The main simplifying assumptions of our model are:
\begin{itemize}
    \item {\bf Firing rate encoding.} Each unit in this model captures
    the mean-field activity of a neuronal population. This may occlude
    computations depending on spike timing, as well as fine-grained
    computations at the neuronal and dendritic level.
    \item {\bf Trivial sensory cortex}. We assumed that sensory cortex
    conveyed errors directly based on static gamma afferent activity.
    Sensory cortex may instead build complex representations of afferent
    activity. It would be interesting to test how the requirement of
    controllability could guide learning of these representations.
    \item {\bf No visual information.} Should a visual error be available, it could be treated by $M$ in a similar way to somatosensory errors. 
    If the visual error holds a monotonic relation with the afferent 
    signals, then it should be possible to adjust the
    long-loop reflex in order to reduce it. When the relation between the visual
    error and the afferent signals is not monotonic (e.g. in some context the
    afferent signals correlate positively, and in other negatively), an alternative
    approach involving reinforcement learning can be pursued
    \cite{verduzco-flores_differential_2022}.
    \item {\bf Very gross anatomical detail.} The detailed anatomical
    organization of cortex and spinal cord is not considered. 
    Moreover, the proportions for different types of cells are not considered.
    \item {\bf Errors must be monotonic.} Muscle activation may not
    monotonically reduce visual errors. Moreover, the haptic perception
    of contact is dependent on the environment, so it would not make
    an appropriate error signal.
    \item {\bf No cerebellum, basal ganglia, brainstem, 
    or premotor cortex.}
\end{itemize}
Each of these omissions is also a possible route for improving the model.
We aim to grow a more sophisticated model, but each new component must
integrate with the rest, improve functionality and
agree with biological literature.

A final limitation concerns proofs of convergence. Many factors complicate
them for this model: transmission delays, noise, synaptic learning,
fully neural implementation, as well as complex muscle and afferent models.
We tested our results for many initial conditions,
but this of course is no guarantee.

50 years ago Marr's model of the cerebellum became a stepping stone
to further theoretical and experimental progress, despite all
its shortcomings \cite{kawato_50_2020}. 
We aspire our model to be the next step towards a complete model of motor control.

\section{Variations of the spinal learning model} \label{sec:variations}

The main text mentions two variations of the spinal learning network that
emphasize the robustness and potential of the learning mechanism. We will
explain the rationale behind those two variations.

There is evidence for interneurons that drive a
set of muscles, possibly setting the circuit foundation for motor synergies 
\cite{giszter_motor_2015,levine_identification_2014,bizzi_neural_2013}.
To explore whether our ideas were compatible with interneurons activating
multiple muscles, we explored 
whether reaching can be learned when the $CE$ and
$CI$ units send projections to more than one motoneuron. To achieve this
we modified the architecture of figure \ref{fig:architecture} so that for every
combination of two different muscles there was a pair of $CE, CI$ 
units that stimulated both of them. 

As illustrated in Supplementary figure \ref{fig:synergy_pairs}, from the set of
6 muscles there are 15 combinations of 2 muscles, but 3 of them consist of
antagonist pairs. Removing these we are left with 12 pairs of muscles, and for
each muscle pair we had a Wilson-Cowan-like $CE, CI$ pair sending projections
to the alpha motoneurons of both muscles. Furthermore, for each pair of muscles,
there is another pair
that contains both their antagonists, and we can use this fact to
generalize the concept of antagonists when interneurons project to several
motoneurons. The $CE$ units sent projections to the $CI$ units of their
antagonists. This organization allowed us to maintain the balance between
excitation and inhibition in the network, along with the
connectivity motifs used previously.

Because this model could be considered a proof-of-concept for the
compatibility of our learning mechanisms
with this particular type of synergies, we refer to this model as the
``synergistic'' network.

To introduce the second variation of the spinal learning network, we may notice
that in all configurations so far the projections from $S_{PA}$ to  $M$ use 
one-to-one connectivity (each $M$ unit controls the length error of one 
muscle).  Interestingly, this is not necessary.
In a second variation of the spinal learning network, dubbed the
``mixed errors'' network, each unit in $M$ can be driven by a linear 
combination of $S_{PA}$ errors.

To ensure that the information about the $S_{PA}$ activity was transmitted to
$M$, we based our $S_{PA}$ to $M$ connections
on the following 6-dimensional orthogonal matrix:
\begin{equation}
    R = 
    \begin{bmatrix}
    1 & 1 & 1 & 1 & 1 & 1 \\
    1 & 1 & 1 &-1 &-1 &-1 \\
    1 &-2 & 1 &-1 & 2 &-1 \\
    -1&-1 & 2 & 2 &-1 &-1 \\
    -1& 1 & 0 & 0 & 1 &-1 \\
    -1& 0 & 1 &-1 & 0 & 1
    \end{bmatrix}.
\end{equation}
The rows of this matrix form an orthogonal basis in $\mathds{R}^6$, and normalizing each
row we obtain a matrix $R^*$ whose rows form an orthonormal basis.
Connections from the 12 $S_{PA}$ units to the 12 $M$ units used this 12$\times$12
matrix:
\begin{equation}
    \begin{bmatrix}
    R^* & - R^* \\
    -R^* & R^*
    \end{bmatrix}.
\end{equation}

The one-to-one connections from $S_{PA}$ to $M$ used in our models are 
unrealistic, but the mixed errors network shows that this simplification
can be overcome, since all the results of the paper also apply to this
variation.

All numerical tests applied to the 3 configurations in the main text of the
paper were also applied to the two variations of the spinal learning model.
Results can be seen in this Appendix, in the Comparison of the 5 configurations.
Supplementary figures \ref{tp_c4} and \ref{tp_c5} illustrate the training phase
for the synergistic and mixed error networks.

We also tested whether the stimulation of an isolated spinal cord produced
convergent direction fields in the synergistic network, as was done in the main
text for the spinal network $C$ common to the other four configurations.
We found that the mean angle difference $\gamma_{a,b}$ between the direction
fields $D(a+b)$ and $D(a) + D(b)$, averaged over the 144 possible $(a,b)$ pairs,
was 19.8 degrees.

Randomly choosing pairs $(a,b)$ and $(c,d)$ for the stimulation locations lead
to a $\gamma_{a,b}$ angle of 72.3 degrees. As before, a bootstrap test
showed that this $\gamma_{a,b}$ value is significantly different ($p < 0.0001$).
Removing the resting field does not alter this result. Moreover, we found no
evidence for supralinear summation of force fields in the synergistic network.

\section{The model fails when elements are removed}
\label{sec:fail}

Due to the larger number of tests,
we only used 5 trials for each configuration in this section. 
The p values reported in this section come from the one-sided t-test.
For brevity, the different configurations of the model will be denoted by 
numbers in the rest of this Appendix:
1 = spinal learning, 2 = cortical learning, 3 = static network,
4 = synergistic network, 5 = mixed errors network.

A model with fully random connectivity and no plasticity has an exceedingly
low probability of having an input-output structure leading it to reduce errors.
The configurations of the model with plasticity (configurations 1,2,4,5), 
however, only have
random connections at one of the projections in the sensorimotor loop (either
from $M$ to $C$, or from $S_{PA}$ to $M$). This may increase the chance of
randomly obtaining a good input-output structure, which could throw 
the usefulness of the learning rules into question.

Removing both types of plasticity in configurations 1,2,4,5 impaired reaching in
all 5 tests for each configuration, as reflected by the inability to reduce the
average error below 10 centimeters in any of the last 4 reaches of the learning
phase. This was also true when removing 
only the plasticity in the connections from $M$ to $C$
(configurations 1, 4, 5) or from $S_{PA}$ to $M$ (configuration 2).
In each one of the plastic configurations (1,2,4,5) the average error for the last
4 training targets was $(22 \pm 10| 22 \pm 4| 24 \pm 9| 22 \pm 5)$ centimeters,
which was significantly higher than the case with normal plasticity
($p<0.001$ for configurations 2,4,5, $p=0.028$ for configuration
1, where the failed trials were not discarded).

Removing plasticity in the connections from $A$ to $C$ or from $A$ to $M$
individually had for the most part no statistically significant effects.
Removing plasticity in both connections simultaneously, however, roughly
duplicated the error in configurations 2 and 4 during center-out reaching
(one-sided t-test, $p<0.001$). Error may be slightly increased for configuration 1
(the small sample size allowed no strong conclusions), whereas configuration 5
was seemingly unaffected. 

Configurations (1,2,4,5) could still learn to reach after removing the $ACT$
unit. Configuration 4 roughly duplicated its center-out reaching error
($p < 0.001$) and its time to initially reach ($p=.016$). Configuration 1
increased its time to initially reach about 3 times ($p < 0.001$), and the other
two configurations were seemingly unaffected. The $ACT$ unit was essential for
previous, less robust versions of the model.

Removing noise made learning too slow, to the point where mean error in the last
4 training reaches could not be reduced below 10 cm in any trial for configurations 
4, and 5. It was reduced below 10 cm in a single trial for configuration 2.
Configuration 1 managed t learn normally. Center-out reaches
were not possible in configurations 2, 4, and 5 with mean errors of
$(15 \pm 8| 17 \pm 11| 9 \pm 3)$ centimeters respectively, at least 3 times
larger than the models with noise ($p < 0.001$). Center-out reaches were normal
in configuration 1, but the first reach with mean error below 10 centimeters 
took significantly longer to happen (from 2.5 to 6.4 attempts in average, 
$p=0.001$).

Removing Ia and Ib afferents, and instead sending the output of II afferents to
$C$, $M$, and $S_A$ prevented reaching in configurations 1, 2, 3, and 4 (except
for a single trial in configuration 1). Configuration 5 could still learn to
reach, but the mean error in the last 4 training reaches and during the
center-out reaching was significantly higher ($p < 0.001$).

Removing the agonist-antagonist connections in $C$ prevented reaching in 
(5| 0| 0| 3| 2) trials for configurations (1,2,3,4,5) respectively. Error in
center-out reaching was significantly increased for configurations 4 and 5, and
it did not increase significantly for configurations 2 and 3.

Figure \ref{fig:fail1} is the same as figure \ref{fig:training_phase}, but in
this case the noise and the ACT units were both removed. The average distance
from the hand to the target was roughly 18 cm. A video illustrating this failure
to learn is included in the supplementary materials (section \ref{sec:videos}).

Configuration 1 was resilient to removal of noise and the ACT unit individually.
The simulation of figure \ref{fig:fail1} suggests that removing more than one
element will have larger consequences on the performance of the model.

\section{Comparison of the 5 configurations}
\label{sec:config_compare}
Once again, the 5 configurations in this paper are represented by a number:
1 = spinal learning, 2 = cortical learning, 3 = static network,
4 = synergistic network, 5 = mixed errors network.
The following table shows the connectivity in each one. Abbreviations:
A2A: all-to-all, O2O: one-to-one, DH: the differential Hebbian learning rule
from \cite{verduzco-flores_differential_2022}, IC: the Input Correlation
learning rule, S: static connections (see section \ref{sub:connections} for
details on the weights of static connections).

\begin{tabular}{|c|c|c|c|c|}
\hline
{\bf Configuration} & {\bf $S_{PA}$ to $M$} & {\bf $M$ to $C$} & {\bf $A$ to
$C$,$M$} \\
\hline
\hline
1 & O2O, S & A2A, DH & A2A, IC \\ 
\hline
2 & A2A, DH & S & A2A, IC \\
\hline
3 & O2O, S & S & S \\
\hline
4 & O2O, S & A2A, DH & A2A, IC \\
\hline
5 & S & A2A, DH & A2A, IC \\
\hline
\end{tabular}

\begin{itemize}
  \item The spinal learning model (configuration 1)
    is a ``basic'' network where the input-output structure of the
    control loop happens in the spinal cord, in the connections from $M$ to $C$.
  \item The cortical learning model (configuration 2)
    is also a ``basic'' network, but the input-output structure is
    resolved in the intracortical connections from $S_{PA}$ to $M$.
  \item The static network (configuration 3) uses only
     static connections, and is meant to show that the
     results in the paper appear in a close to optimal configuration of feedback control,
     rather than being some sophisticated product of the plasticity rules.
  \item The synergistic network (configuration 4)
     is an extension of configuration 1, where the spinal cord has 12
     $CE,CI$ pairs rather than 6, and each pair stimulates 2 $\alpha$ motoneurons.
  \item The mixed errors network (configuration 5)
    is a different variation of configuration 1, where the
    connections from $S_{PA}$ to $M$ are not one-to-one, but instead come from an
    orthogonal matrix.
\end{itemize}

The following table summarizes the numerical results for the 5 configurations.

\begin{tabular}{|c|c|c|c|c|c|}
\hline
{\bf Measurement } & 1 & 2 & 3 & 4 & 5 \\
\hline
Failed reaches$^1$ & 
    $1.8 \pm 2$ &
    $1.2 \pm .9$ &
    $0 \pm 0$ &
    $1.6 \pm 1.3$ &
    $4 \pm 2.5$ \\
\hline
Center-out error$^2$ &
    $3.3 \pm .01$ &
    $2.9 \pm .001$ &
    $2.9 \pm .0003$ &
    $3 \pm .008$ &
    $2.8 \pm .0007$ \\
\hline
$M$ units tuned & & & & & \\ 
to direction & 
    $11.8 \pm .4$ &
    $12 \pm 0$ &
    $12 \pm 0$ &
    $12 \pm 0$ &
    $12 \pm 0$ \\
\hline
$R^2$ for & & & & & \\
predicted PD &
    $.74 \pm .18$ &
    $.88 \pm .14$ &
    $.86 \pm .01$ &
    $.89 \pm .06$ &
    $.82 \pm .03$ \\
\hline
PD distribution & & & & & \\
main axis (deg) &
    $59 \pm 7$ &
    $52 \pm 2$ &
    $54 \pm .5$ &
    $60 \pm 3$ &
    $58 \pm 1$ \\
\hline
PD drift & & & & & \\
angle (deg) &
    $3.3 \pm 2.4$ &
    $4.9 \pm 2.1$ &
    $.3 \pm .2$ &
    $1.8 \pm 1.3$ &
    $7 \pm 6$ \\
\hline 
Muscle PD & & & & & \\
drift angle (deg) &
    $3.8 \pm 2.1$ &
    $6.4 \pm 2.9$ &
    $.2 \pm .2$ &
    $11.4 \pm 15.2$ &
    $27.7 \pm 34.5$ \\
\hline 
Center-out error & & & & & \\
(10 targets) & 3 & 3.6 & 2.9 & 2.6 & 4.5 \\
\hline
Variance in & & & & & \\
first jPCA &
    $.42 \pm .04$ &
    $.42 \pm .04$ &
    $.46 \pm .03$ &
    $.45 \pm .04$ &
    $.47 \pm .07$ \\
\hline
Center-out error & & & & & \\
(light arm) & 2.5 & 3.2 & 3 & 6.1 & 2.9 \\
\hline
Center-out error & & & & & \\
(heavy arm) & 2.4 & 3.3 & 2.9 & 5.6 & 3.2 \\
\hline
\end{tabular} \\
{\small $^1$ Average number of reaches before the first reach when the mean error
was below 10 cm. \\
$^2$ Average distance (in centimeters) between the hand and the target during 
center-out reaching.}

\section{Gain and oscillations}
\label{sec:gain}
The gain of a feedback loop describes how the amplitude of the error signal
increases as it gets transformed into a control signal sent to the plant.
As described in the Methods (section \ref{sub:parameter}), we used a relatively low
number of iterations of an optimization algorithm to find suitable parameters
for each configuration of the model.  This led to configurations with 
gains that were coarsely tuned. Supplementary figure 
\ref{fig:center_out_1A_raw}
is analogous to figure \ref{fig:center_out_1A}, and shows the hand trajectories
right after the optimization algorithm was finished. It can be observed that
configurations 2 and 3 were particularly prone to oscillations, and
configuration 1 would undershoot many targets.

To improve performance, as well as to facilitate comparison of the 5
configurations, we adjusted their gains. This involved manually adjusting the
slope of the sigmoidal units in populations $M$ and $S_A$, until they appeared
stable, but on the verge of oscillating (so reaching would be faster).
This required from 1 to 3 attempts. The gain of configuration 1 was slightly
increased, whereas the gain of configurations 2,3,4 was reduced. Configuration 5
was left with the same parameters.

The trajectories in panels C and D of figure \ref{fig:center_out_1A_raw} are
reminiscent of terminal tremors in cerebellar ataxia. An animation showing the
movement of the arm for the 5 configurations before gain adjustment is included
among the Supplementary Videos. In addition, supplementary figures
\ref{fig:training_phase3} and \ref{fig:training_phase4} show the error and
activity of several units during the training reaches for configurations 3 and
4, analogous to figure \ref{fig:training_phase}. It can be observed that the
oscillations are present in the whole network, suggesting that the control
signals are trying to catch up with an error that keeps reversing direction.

\section{Supplementary videos}
\label{sec:videos}
To help visualization of the arm's learning and performance under different
conditions, 4 videos were produced.
The videos indicate the model configuration using the enumeration from this
Appendix: 
1 = spinal learning, 2 = cortical learning, 3 = static network,
4 = synergistic network, 5 = mixed errors network.

To download these videos, please visit \\
\verb+https://gitlab.com/sergio.verduzco/public_materials/-/tree/master/adaptive_plasticity+.

The videos' content is as follows:
\begin{itemize}
  \item \verb+Video 1.mp4+: Visualization of the arm and the muscles
  during the learning phase for configuration 1. Data comes from the simulation
  shown in figure \ref{fig:training_phase}. Speed is roughly 4X.
  \item \verb+video 2.mp4+: Arm animation of the first ~180 seconds
  of center-out reaching for the 5 configurations. Speed is roughly 4X.
  \item \verb+Video 3.mp4+: The learning phase for a simulation with
  configuration 1. Both noise and the $ACT$ unit were removed, reducing
  exploration and disrupting learning. Data comes from the same simulation in
  figure \ref{fig:fail1}. Speed is ~4X.
  \item \verb+Video 4.mp4+: The first 180 seconds of center-out
  reaching for the 5 configurations before the gains were adjusted. Speed is
  roughly ~4X. Configuration 2 shows target-dependent oscillations after ~70
  seconds. 
\end{itemize}

\section{Parameter values}
\label{sec:parameters}

Values appear in order for configurations 1-5. A single number means that all
configurations use that same value.
\subsection{Unit parameters}
The superscript $^x$ on a population name indicates that a parameter has 
heterogeneous values. This means that a random value was added to the parameter
for each individual unit.
This random value comes from a uniform distribution centered at zero, with a 
width equal to 1\% of the reported parameter value.

\begin{tabular}{|c|c|c|c|}
\hline
{\bf Parameter} & {\bf Equation} &  {\bf Population} & {\bf Value} \\
\hline
$\tau_u$ & \ref{eq:unit} & $CHG, A, ACT$ & 10 [ms] \\ \cline{3-4}
& & $\alpha, S_A, S_{PA}^x, CI$ & 20 [ms] \\ \cline{3-4}
& & $CE$ &  140, 70, 150, 180, 110 [ms]  \\ \cline{3-4}
& & $M^x$ & 50 [ms]  \\
\hline
$\beta$ & \ref{eq:sig}  & $\alpha^x, ACT$ & 2 \\ \cline{3-4}
&   & $CE$ & 1.63, 1.72, 1.70, 3.38, 1.44 \\ \cline{3-4}
&   & $CI$ & 4.0, 3.38, 3.44, 2.46, 3.63 \\ \cline{3-4}
&   & $M^x$ & 1.5, 1.5, 2.0, 2.0, 1.17  \\ \cline{3-4}
&   & $S_A$ & 3.0, 2.2, 2.0, 2.3, 3.0  \\ \cline{3-4}
&   & $CHG, S_{PA}^x$ & $9$ \\ 
\hline
$\eta $ & \ref{eq:sig}  & $\alpha^x$ & 1.1 \\ \cline{3-4}
&   & $CE$ &  2.0, 1.93, 2.13, 2.31, 1.67  \\ \cline{3-4}
&   & $CI$ & 1.5, 1.41, 1.63, 1.72, 1.7 \\ \cline{3-4}
&   & $M_x$ & 1.3, 1.96, 0.68, 1.19, 1.38 \\ \cline{3-4}
&   & $ACT$ & 1 \\ \cline{3-4}
&   & $S_A$ & 0.75 \\ 
&   & units 0,3 & \\ \cline{3-4}
&   & $S_A$ & .4 \\ 
&   & units 1,2,5 & \\ \cline{3-4}
&   & $S_A$ & .3 \\ 
&   & unit 4 & \\ \cline{3-4}
&   & $CHG$ & .25 \\ \cline{3-4}
&   & $S_{PA}^x$ & .1 \\ \cline{3-4}
\hline
$\varsigma$ & \ref{eq:langevin} & $CE,CI$ & 0.63, 0, 0, 0.69, 0.72 \\
\cline{3-4}
& & $M$ & 0, 0.62, 0, 0, 0 \\
\hline
$\tau_a$ & \ref{eq:log} & $A$ & 10 [ms] \\
\hline
$T$ & \ref{eq:log} & $A$ & .2  \\
& & $I_b,II$ afferents & \\ \cline{3-4}
& & $A$ & 0  \\
& & $I_a$ afferents & \\ 
\hline
$\tau_{slow}$ & \ref{eq:low-pass}, \ref{eq:decay} & $CE, CI$ & 11 [s] \\
\hline
$\theta_{ACT}$ & \ref{eq:act1}, \ref{eq:act2} & $ACT$ & .31 \\
\hline
$\tau_{ACT}$ & \ref{eq:act2} & $ACT$ & 10 [ms] \\
\hline
$\gamma$ & \ref{eq:act2} & $ACT$ & 8 \\
\hline
$\alpha$ & \ref{eq:chg_syn} & $CHG$ (synapse) & 20 \\
\hline
\end{tabular}
\\ \\ \\

\subsection{Learning rules}
\begin{tabular}{|c|c|c|}
    \hline
{\bf Parameter} & {\bf Equation} & {\bf Value} \\ \hline
$\Delta t$ & \ref{eq:rga21} & 0.33, 0.37, 0.15, 0.36, 0.32 [s] \\
\hline
$\alpha (M$ to $C)$& \ref{eq:w_norm} & 500, 0, 0, 500, 500 \\
\hline
$\alpha (S_{PA}$ to $M)$ & \ref{eq:w_norm} & 0, 527, 0, 0, 0 \\
\hline
$\alpha (M$ to $\alpha)$& \ref{eq:w_norm} & 300, 0, 0, 300, 300 \\
\hline
$\lambda$ & \ref{eq:w_norm} & .03 \\
\hline
$\omega_{sa} (M$ to $C,\alpha)^1$& \ref{eq:w_norm} & 2.52 \\
\hline
$\omega_{sa} (S_{PA}$ to $M)$& \ref{eq:w_norm} & 3.23, 3.19, 2.98, 3.23, 3.23 \\
\hline
$\omega_{sb} (M$ to $C)$ & \ref{eq:w_norm} & 3.29, 2.14, 1.50, 3.69, 0.57 \\
\hline
$\omega_{sb} (M$ to $\alpha)$ & \ref{eq:w_norm} & 2.86, 2.86, 1.50, 2.86, 2.86 \\
\hline
$\alpha_{IC} (A$ to $M)$& \ref{eq:inp_corr} & 26.17 \\
\hline
$\alpha_{IC} (A$ to $C)$& \ref{eq:inp_corr} & 22.5 \\
\hline
$\omega_s (A$ to $M)$& \ref{eq:inp_corr} & 0.85, 1.14, 1, 0.53, 0.53 \\
\hline
$\omega_s (A$ to $C)$& \ref{eq:inp_corr} & 1.68, 2, 2, 1.55, 2.88 \\
\hline
$\omega_{max} (A$ to $M)$ & \ref{eq:inp_corr} & .48, .22, .2, .25, .33 \\
\hline
$\omega_{max} (A$ to $C)$ & \ref{eq:inp_corr} & .3, .64, .64, .28, .59 \\
\hline
\end{tabular}

$^1$ {\small Constraints in the sum of weights are also used with static
connections.}

\subsection{Muscles and afferents}

\begin{tabular}{|c|c|c|}
\hline
{\bf Parameter} & {\bf Equation} & {\bf Value} \\ \hline
$K_{SE}$ & \ref{eq:hill1} & 20 [N/m] \\
\hline
$K_{PE}$ & \ref{eq:hill1} & 20 [N/m] \\
\hline
$b$ & \ref{eq:hill1} & 1 [N $\cdot$ s / m] \\
\hline
$g$ & \ref{eq:hill1} & 67.11 [N] \\
& muscles 0,3 &   \\
\hline
$g$ & \ref{eq:hill1} & .75 [N] \\
& muscles 1,2,4,5 &   \\
\hline
$K_{SE}^s$ & \ref{eq:hill2},\ref{eq:Ia},\ref{eq:II} & 2 [N/m] \\
\hline
$K_{PE}^s$ & \ref{eq:hill2},\ref{eq:Ia},\ref{eq:II} & 2 [N/m] \\
\hline
$b^s$ & \ref{eq:hill2}, \ref{eq:II} & $.5$ [N $\cdot$ s / m] \\
\hline
$K_{SE}^d$ & \ref{eq:hill2},\ref{eq:Ia},\ref{eq:II} & 1 [N/m] \\
\hline
$K_{PE}^d$ & \ref{eq:hill2} & .2 [N/m] \\
\hline
$b^d$ & \ref{eq:hill2} & $2$ [N $\cdot$ s / m] \\
\hline
$l_0^s$ & \ref{eq:hill2} & .7  \\
\hline
$l_0^d$ & \ref{eq:hill2} & .8  \\
\hline
$g_{I_a}$ & \ref{eq:Ia} & [7.5, 25, 25, 7.5, 25, 25] $[m^{-1}]$ \\
& muscles 0-5 &  \\
\hline
$f_s^{I_a}$ & \ref{eq:Ia} & 0.1 \\
\hline
$g_{II}$ & \ref{eq:II} & $[5.46, 8, 8, 5.46, 8, 8]^*$, \\
& muscles 0-5 & $[5.8, 8, 8, 5.8, 8, 8]^{**}$ $[m^{-1}]$ \\
\hline
$f_s^{II}$ & \ref{eq:II} & 0.5 \\
\hline
$g_{I_b}$ & \ref{eq:gto1} & 1 \\
\hline
$T_0$ & \ref{eq:gto1} & 10 [N] \\
\hline
$\tau_g$ & \ref{eq:gto2} & 50 [ms] \\
\hline
\end{tabular}

\subsection{Connection delays and weights}
Connections considered "local" used a delay of 10 [ms], unless those
those connections implied a not-modeled disynaptic inhibition. All other
connections had a delay of 20 [ms].

\begin{tabular}{|c|c|c|}
\hline
{\bf Source } & {\bf Target } & {\bf Delay } \\ \hline
$A$ & $M, S_A$ & 20 [ms] \\ \cline{1-2}
$ACT$ & $CE,CI$  & \\ \cline{1-2}
$\alpha$ & muscle & \\ \cline{1-2}
$M$ & $CE,CI, \alpha$  & \\ \cline{1-2}
$M$ & $M$  & \\ \cline{1-2}
afferents & $A$ & \\ \cline{1-2}
$CHG, S_{PA}$ & $ACT$ & \\ \cline{1-2}
$S_{PA}$ & $M, S_{PA}$ & \\ \cline{1-2}
\hline
$A$ & $M$ & 10 [ms] \\ \cline{1-2}
$CE,CI$ & $\alpha$  &  \\ \cline{1-2}
$CE,CI$ & $CE,CI$ & \\ \cline{1-2}
$CI$ & $CE$ & \\ \cline{1-2}
$S_A, S_P$ & $S_{PA}$ & \\ \cline{1-2}
$S_P$ & $CHG$ & \\ \cline{1-2}
\hline
\end{tabular}

The next table shows the value of fixed synaptic weights not specified
in section \ref{sub:connections}. 
Columns indicate the source of the connection,
rows indicate the target. "Aff" stands for the muscle afferents.
Potentially plastic connections are marked as "+". 
If a connection marked ``+'' is static in
one of the configurations, its weight is determined by the $\omega_{sa},
\omega_{sb}, \omega_s$ parameters of equations \ref{eq:w_norm}, \ref{eq:inp_corr}.

\begin{tabular}{|c|c|c|c|c|c|c|c|c|c|} \hline
& $\alpha$ & $A$ & $CE$ & $CI$ & $M$ & Aff. & $S_A$ & $S_P$ & $S_{PA}$ \\
\hline
$\alpha$ &   &   & 1 & -1 & + &   &   &   & \\ \hline 
$A$ &   &   &   &   &   &  2 ($I_a,I_b$), &   &   & \\ 
&   &   &   &   &   &  4 ($II$) &   &   & \\ \hline
$ACT$ &   &   &   &   &   &   &   &   & 1 \\ \hline
$CE$ &   & + & $.5^a, .18^b$ &$-1.8^c$ & + &   &   &   & \\ \hline
$CHG$ &   &   &   &   &   &   &   & + & \\ \hline
$CI$ &   & + & $.5^c$ &   & + &   &   &   & \\ 
&   & & $1.83^d, .16^e$ &   &   &   &   &   & \\ \hline
$M$ &   & + &   &   & * &   &   &   & + \\ \hline
muscles & 1 &   &   &   &   &   &   &   & \\ \hline
$S_A$ &   & 1 &   &   &   &   &   &   & \\ \hline
$S_{PA}$ &   &   &   &   &   &   & 1 or -1 & 1 or -1  & -1.77 \\ \hline
 \hline
\end{tabular}

$^a$ {\small Agonist connections}. 
$^b$ {\small Partial agonist connections}. 
$^c$ {\small Withing the same $(CE, CI, \alpha)$ triplet}.
$^d$ {\small Antagonist connections}. 
$^e$ {\small Partial antagonist connections}. 

$^*$ {\small $M$ units inhibited their duals with weights
that depended on the configuration: -0.93, -0.74, -1.00, -1.14, 0.0}.

All connections whose source is $CHG$ or $ACT$ have a weight of 1.

\section{Supplementary figures}
\label{sec:sup_figs}

\begin{figure}
    \centering
    \includegraphics[width=0.9\columnwidth]{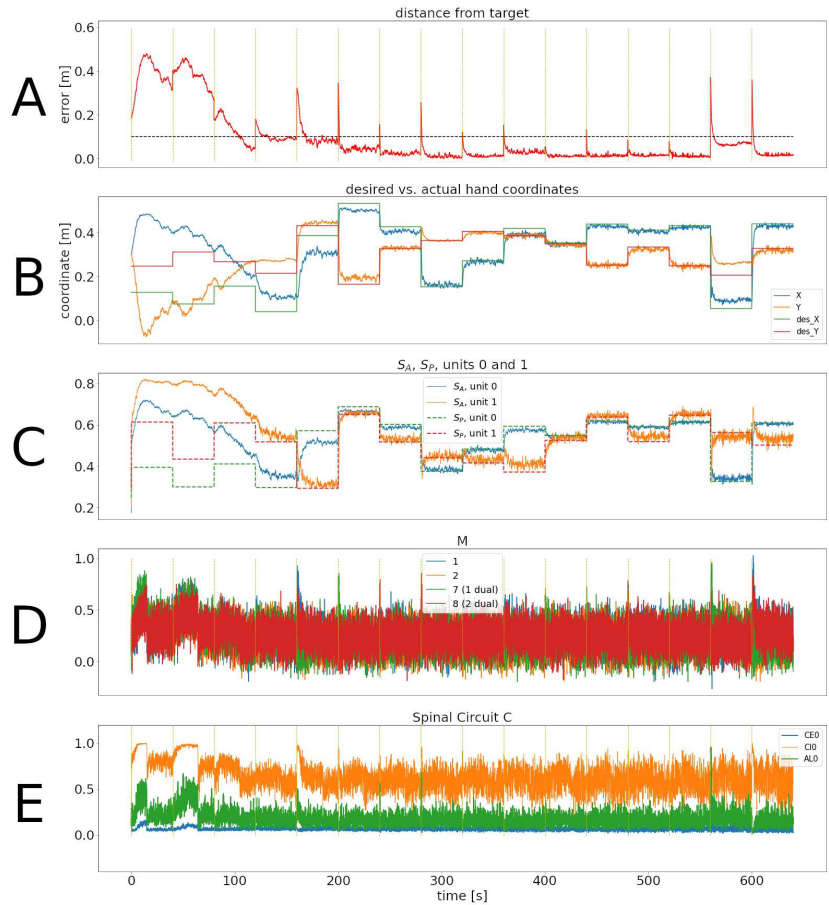}
    \caption{Representative training phase of the simulation for
    configuration 2 (cortical learning). 
    Panels are as in figure \ref{fig:training_phase}.
    }
    \label{tp_c2}
\end{figure}

\begin{figure}
    \centering
    \includegraphics[width=0.9\columnwidth]{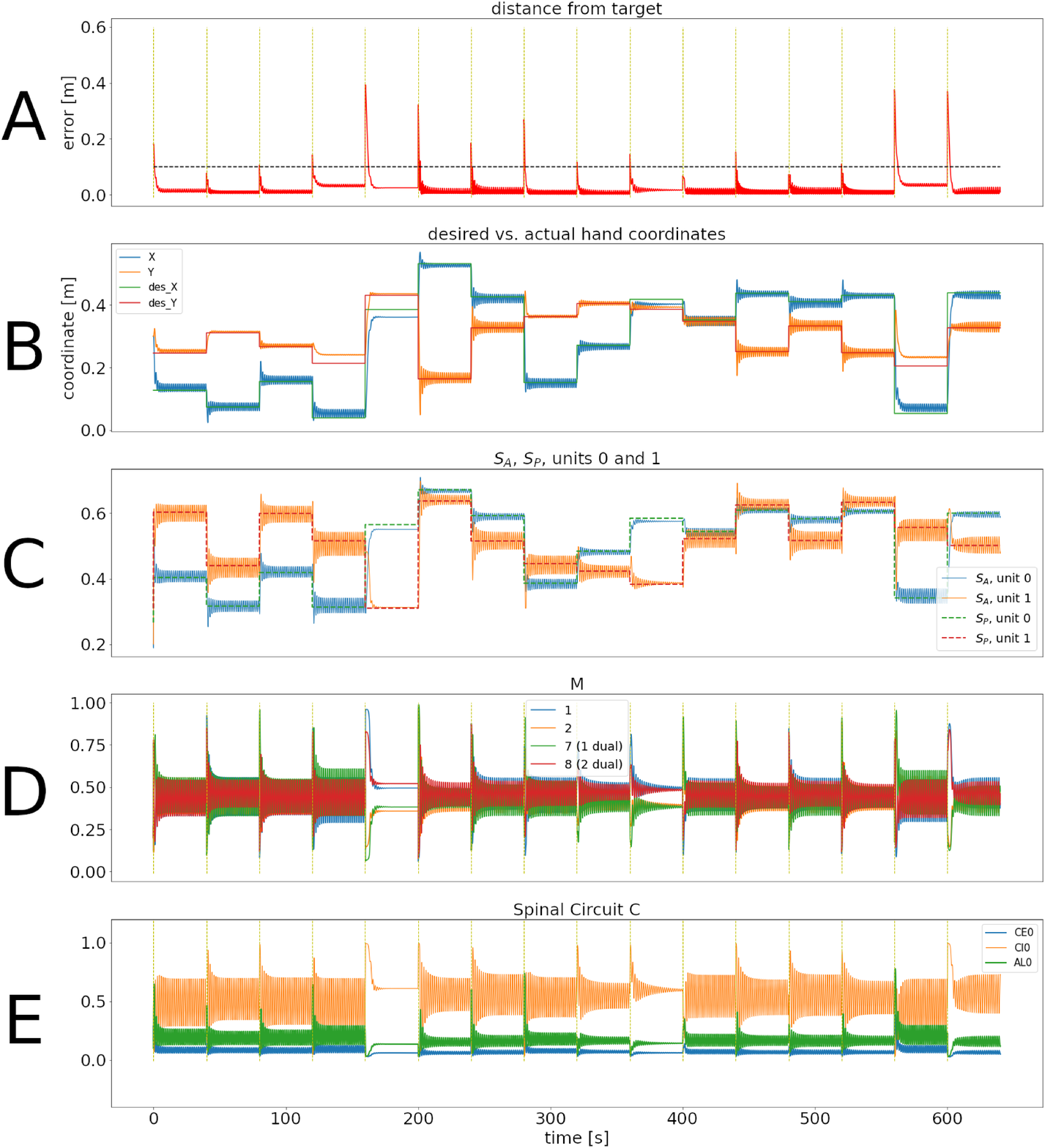}
    \caption{Representative training phase of the simulation for
    configuration 3 (static network). 
    Panels are as in figure \ref{fig:training_phase}.
    }
    \label{tp_c3}
\end{figure}

\begin{figure}
    \centering
    \includegraphics[width=0.9\columnwidth]{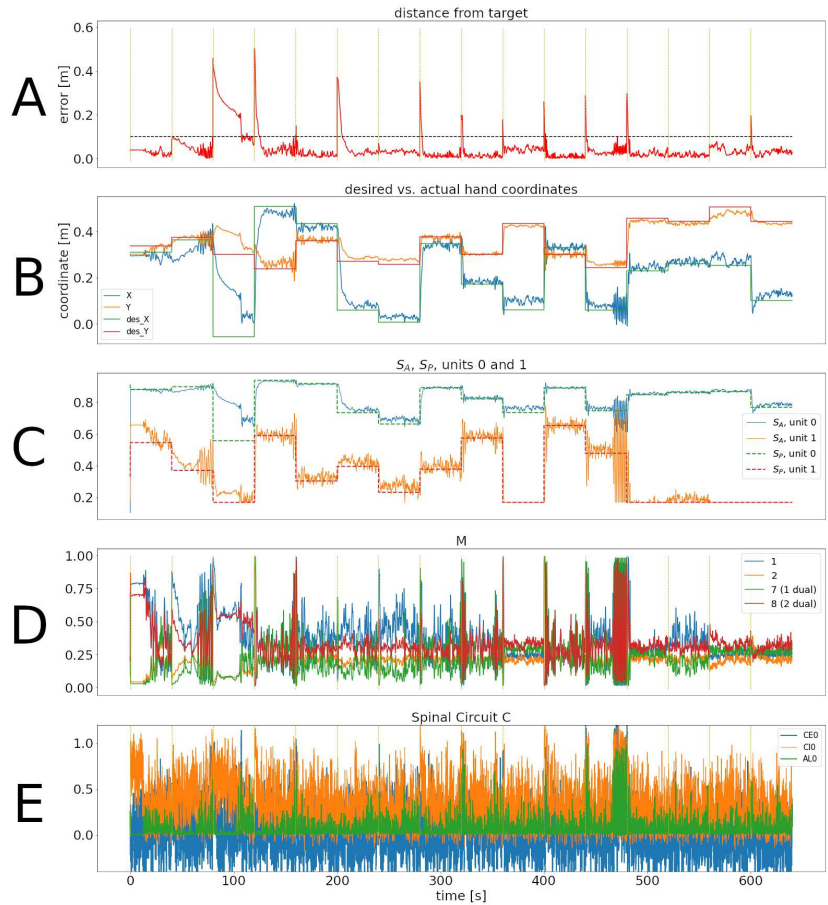}
    \caption{Representative training phase of the simulation for
    configuration 4 (synergistic network). 
    Panels are as in figure \ref{fig:training_phase}.
    }
    \label{tp_c4}
\end{figure}

\begin{figure}
    \centering
    \includegraphics[width=0.9\columnwidth]{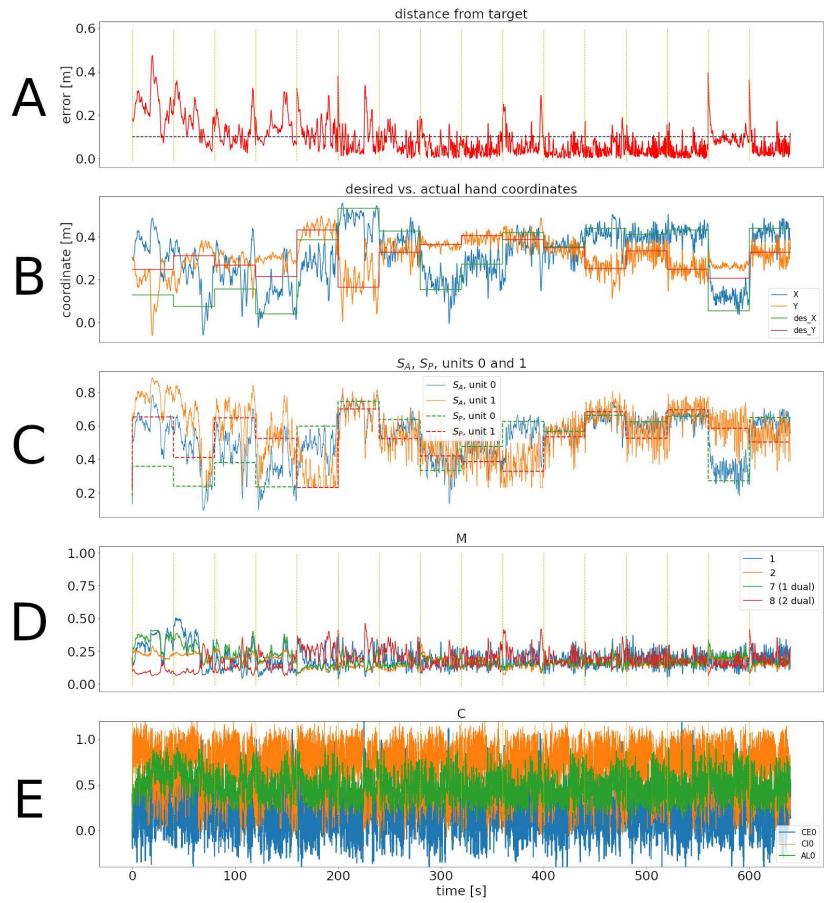}
    \caption{Representative training phase of the simulation for
    configuration 5 (mixed errors network).
    Panels are as in figure \ref{fig:training_phase}.
    }
    \label{tp_c5}
\end{figure}

\begin{figure}
    \centering
    \includegraphics[width=0.7\columnwidth]{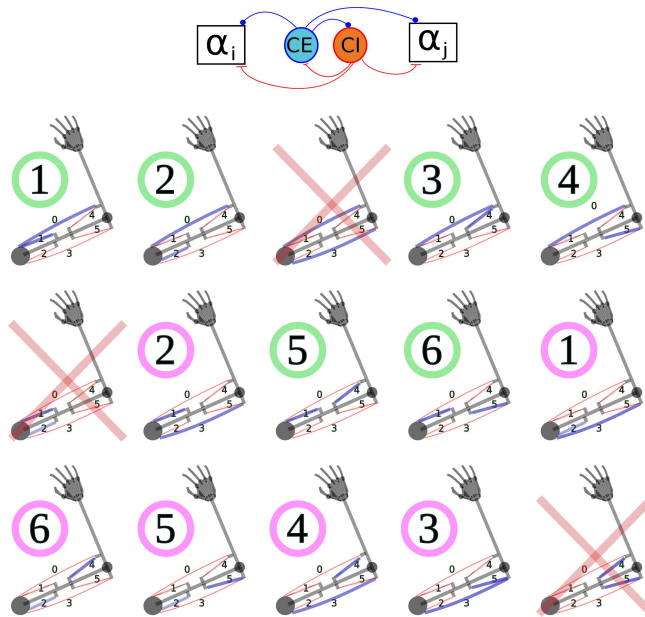}
    \caption{Modified architecture of the $C$ population.
    Each pair of $CE, CI$ units projects to two $\alpha$ motoneurons (top). There are 15 possible
    pairs of muscles, corresponding to the blue lines for each arm in the figure. Three of the pairs
    (marked with red crosses) contain antagonist muscles, and are
    not included. The remaining 12 pairs can be arranged into 2 groups of 6 units each. The units in
    the group marked with green circles are the antagonists of the units with the same number, marked
    with pink circles.}
    \label{fig:synergy_pairs}
\end{figure}

\begin{figure}
    \centering
    \includegraphics[width=0.8\columnwidth]{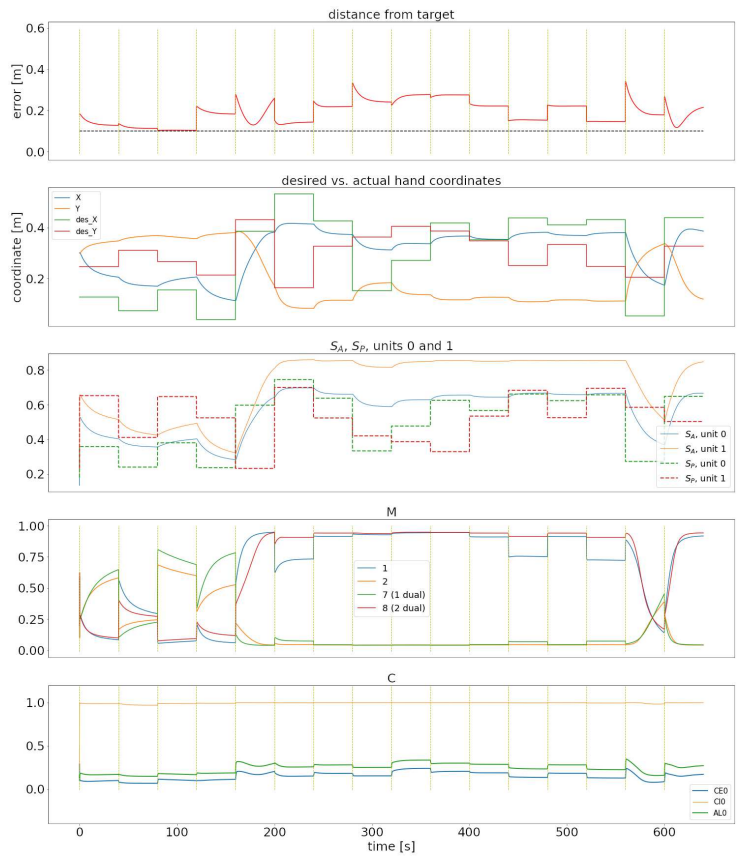}
    \caption{A failure to learn for a simulation of configuration 1 
    (spinal learning) with no noise
    and no ACT unit. Captions are as in figure \ref{fig:training_phase} of the
    main text.
    }
    \label{fig:fail1}
\end{figure}

\begin{figure}
    \centering
    \includegraphics[width=0.9\columnwidth]{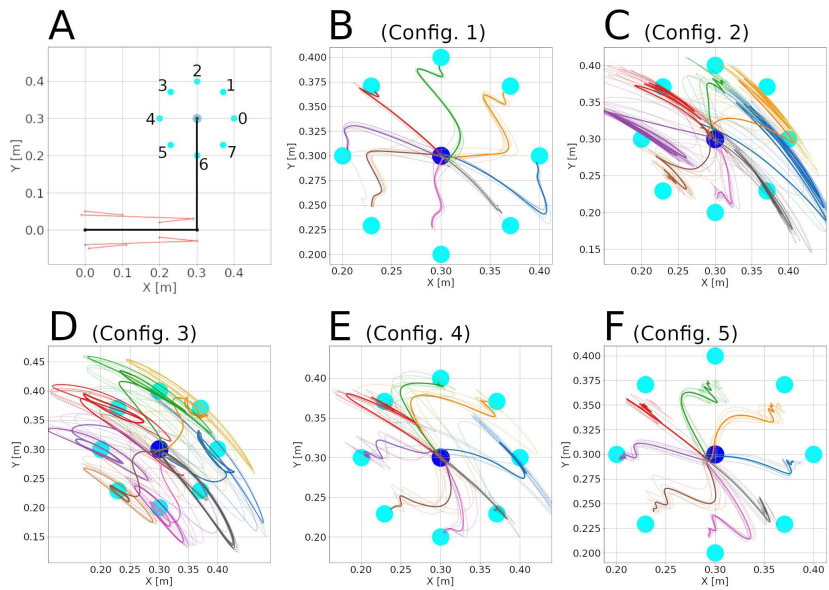}
    \caption{Center-out reaching before the control loop gain was adjusted.
    Panels are as in figure \ref{fig:center_out_1A}, with the two extra
    configurations (4 and 5) presented in this Appendix.
    }
    \label{fig:center_out_1A_raw}
\end{figure}

\begin{figure}
    \centering
    \includegraphics[width=0.9\columnwidth]{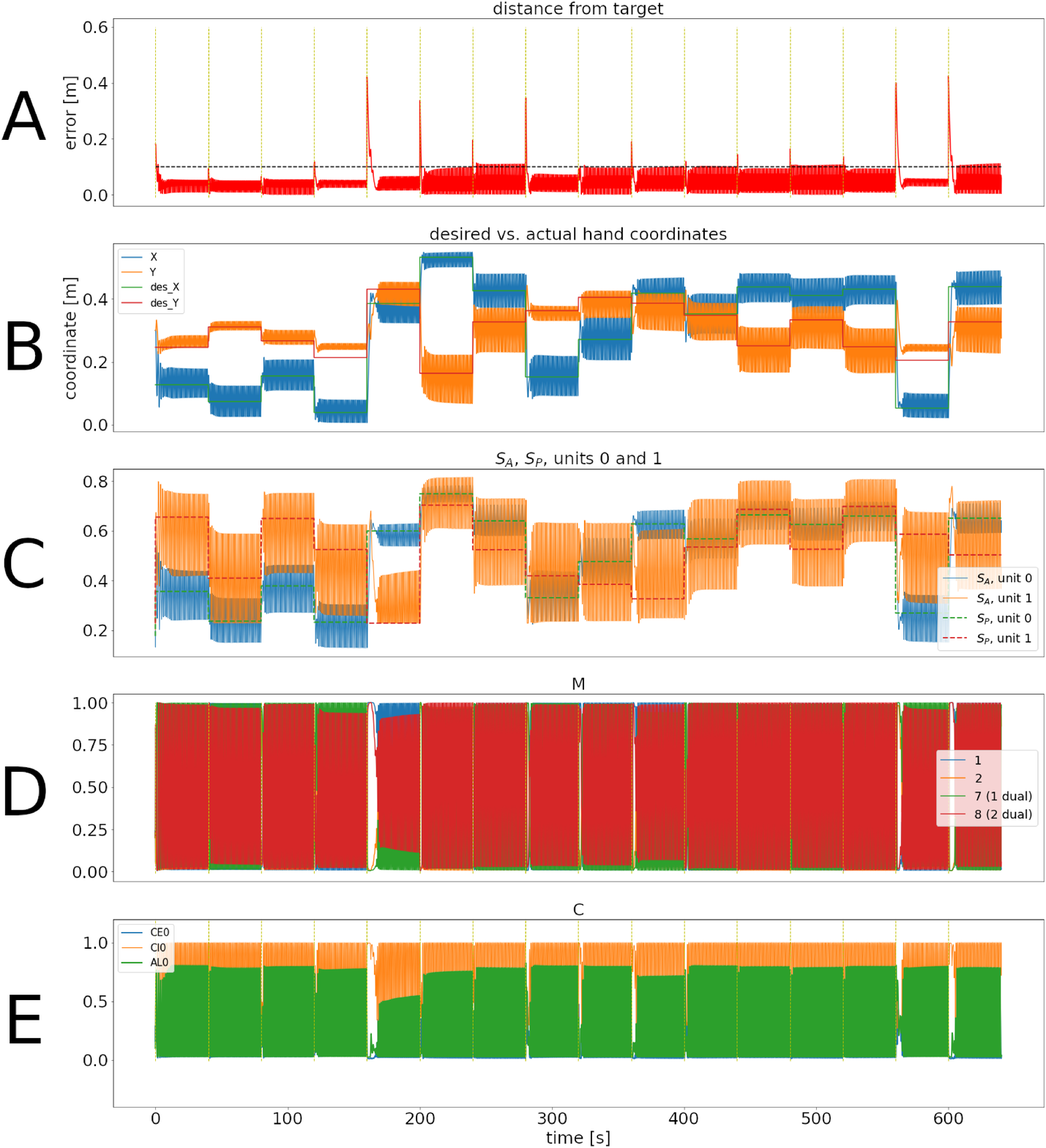}

    \caption{Training phase of the simulation for
    configuration 3 before gain was reduced. 
    Panels are as in figure \ref{fig:training_phase}.
    } 
    \label{fig:training_phase3}
\end{figure}

\begin{figure}
    \centering
    \includegraphics[width=0.9\columnwidth]{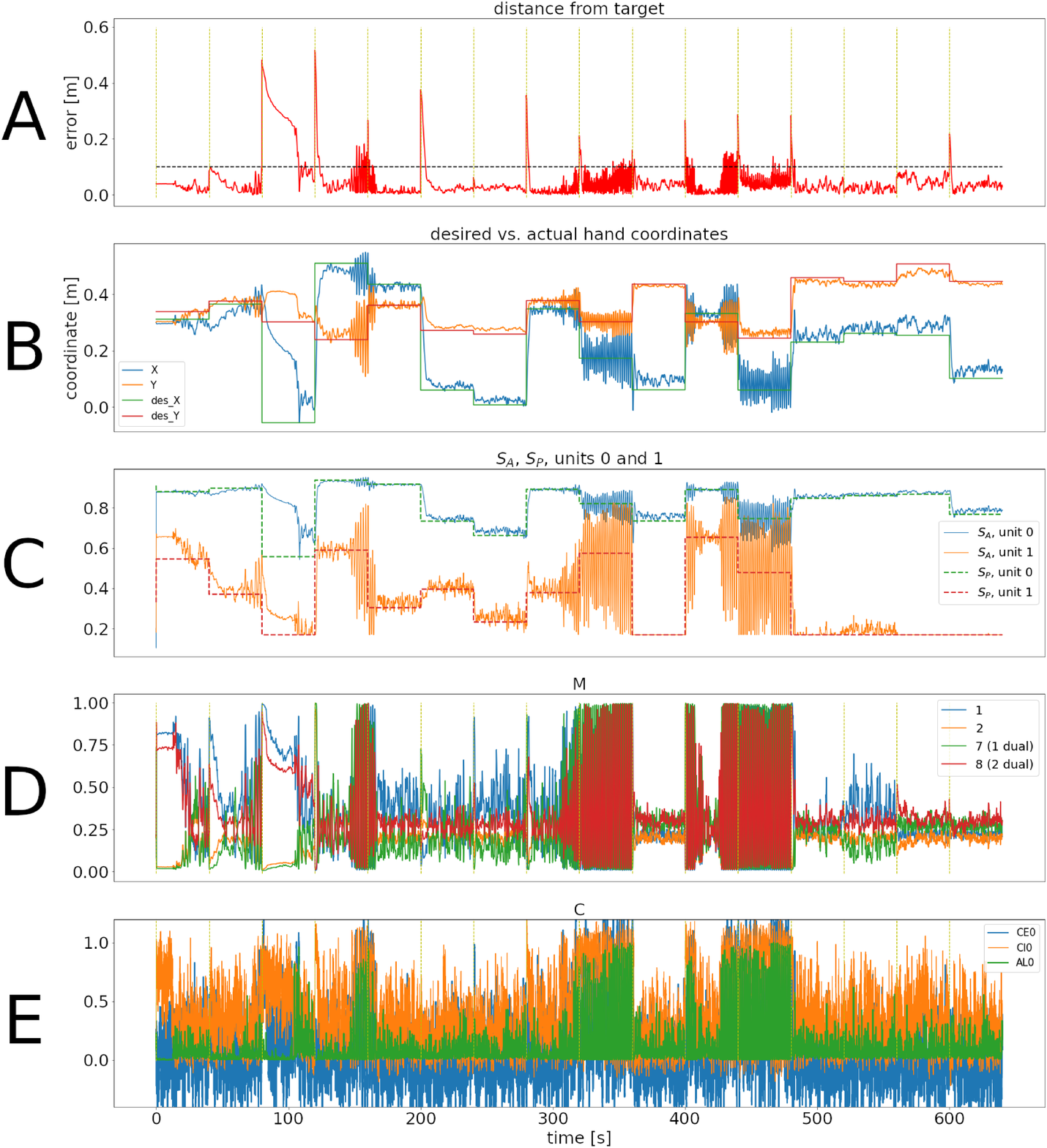}

    \caption{Training phase of the simulation for
    configuration 4 before gain was reduced. 
    Panels are as in figure \ref{fig:training_phase}.
    } 
    \label{fig:training_phase4}
\end{figure}

\clearpage

\end{document}